\documentclass{aa}
\usepackage{graphicx}
\usepackage{longtable}
\usepackage{color}

\def\gtrsim{\mathrel{\hbox{\rlap{\hbox{\lower4pt\hbox{$\sim$}}}\hbox{$>$}}}}
\def\ltsim{\mathrel{\hbox{\rlap{\hbox{\lower4pt\hbox{$\sim$}}}\hbox{$<$}}}}

\begin{document}

\title{ 
The Magnetic Fields at the Surface of Active Single G-K Giants\thanks{Based on observations  obtained at  the T\'elescope Bernard Lyot (TBL) at Observatoire du Pic du Midi, CNRS/INSU and Universit\'e de Toulouse, France, and at the Canada-France-Hawaii Telescope (CFHT) which is operated by the National Research Council of Canada, CNRS/INSU and the University of Hawaii.}}
%\subtitle{}
\author{M. Auri\`ere\inst{1,2}, R. Konstantinova-Antova\inst{3,1}, C. Charbonnel\inst{4,2}, G.A. Wade \inst{5}, S. Tsvetkova\inst{3}, P. Petit\inst{1,2}, B. Dintrans\inst{1,2}, N. A. Drake \inst{6,7}, T. Decressin\inst{8,4}, N. Lagarde \inst{9}, J.-F. Donati\inst{1,2}, T. Roudier\inst{1,2},  F. Ligni\`eres\inst{1,2}, K.-P. Schr\"oder\inst{10},  J.D.~Landstreet\inst{11,12}, A. L\`ebre\inst{13}, W.W. Weiss\inst{14}, J-P Zahn\inst{15}}
\offprints{M. Auri\`ere, {\tt michel.auriere@irap.omp.eu}}
\institute{Universit\'e de Toulouse, UPS-OMP, Institut de Recherche en Astrophysique et Plan\'etologie, Toulouse, France
\and
CNRS, UMR 5277, Institut de Recherche en Astrophysique et Plan\'etologie, 14 Avenue Edouard Belin, 31400 Toulouse, France
\and
Institute of Astronomy and NAO, Bulgarian Academy of Sciences, 72 Tsarigradsko shose, 1784 Sofia, Bulgaria
\and
Geneva Observatory, University of Geneva, 51 Chemin des Maillettes, 1290 Versoix, Switzerland
\and
Department of Physics, Royal Military College of Canada,
  PO Box 17000, Station 'Forces', Kingston, Ontario, Canada K7K 4B4
\and
Sobolev Astronomical Institute, St. Petersburg State University, Universitetski pr.28, St. Petersburg 198504, Russia
\and
Observat\'orio Nacional/MCTI, Rua Jos\' e Cristino 77, CEP 20921-400, Rio de Janeiro-RJ, Brazil 
\and
INAF - Osservatorio Astronomico di Roma, Via Frascati 33, I-00040 Monte Porzio Catone (RM), Italy
\and
  School of Physics and Astronomy, University of Birmingham, Edgbaston, Birmingham, B15 2TT, UK
\and
Departamento de Astronomia, Universitad de Guanajuato, A.P. 144, C.P. 36000, GTO, Mexico
\and
Physics \& Astronomy Department, The University of Western Ontario, London, Ontario, Canada N6A 3K7
\and
Armagh Observatory, College Hill, Armagh, BT619DG, Northern Ireland, UK
\and
LUPM - UMR 5299 - CNRS and Universit\'e Montpellier II - Place E. Bataillon, 34090 Montpellier, France
\and 
Institut f\"ur Astronomie, Universit\"at Wien, T\"urkenschanzstrasse 17, A-1180 Wien, Austria
\and
LUTH, CNRS UMR 8102, Universit\'e Paris Diderot, 5 place Jules Janssen, 92195 Meudon, France}

 \date{Received ??; accepted ??}

\abstract
 {}
{We investigate the magnetic field at the surface of 48 red giants selected as promising for detection of Stokes $V$ Zeeman signatures in their spectral lines. In our sample, 24 stars are identified from the literature as presenting moderate to strong signs of magnetic activity. An additional 7 stars are identified as those in which thermohaline mixing appears not to have occured, which could be due to hosting a strong magnetic field. Finally, we observed 17 additional very bright stars which enable a sensitive search to be performed with the spectropolarimetric technique.}
{We use the spectropolarimeters Narval and ESPaDOnS to detect circular polarization within the photospheric absorption lines of our targets.  We treat the spectropolarimetric data using the least-squares deconvolution (LSD) method to create high signal-to-noise ratio mean Stokes $V$ profiles. We also measure the classical $S$-index activity indicator for the Ca~{\sc ii} $H \& K$ lines, and the stellar radial velocity. To infer the evolutionary status of our giants and to interpret our results, we use state-of-the-art stellar evolutionary models with predictions of convective turnover times.}
{We unambiguously detect magnetic fields via Zeeman signatures in 29 of the 48 red giants in our sample. Zeeman signatures are found in all but one of the 24 red giants exhibiting signs of activity, as well as 6 out of 17 bright giant stars. The majority of the magnetically detected giants are either in the first dredge up phase or at the beginning of core He burning, i.e. phases when the convective turnover time is at a maximum: this corresponds to a 'magnetic strip' for red giants in the Hertzsprung-Russell diagram. A close study of the 16 giants with known rotational periods shows that the measured magnetic field strength is tightly correlated with the rotational properties, namely to the rotational period and to the Rossby number $Ro$. 
Our results show that the magnetic fields of these giants are produced by a dynamo, possibly of $\alpha$-$\omega$ origin since $Ro$ is in general smaller than unity. Four stars for which the magnetic field is measured to be outstandingly strong with respect to that expected from the rotational period/magnetic field relation or their evolutionary status are interpreted as being probable descendants of magnetic Ap stars. In addition to the weak-field giant Pollux, 4 bright giants (Aldebaran, Alphard, Arcturus, $\eta$ Psc) are detected with magnetic field strength at the sub-gauss level. %Besides Arcturus, these stars were not considered to be active giants before this study and are very similar in other respects to ordinary giants, with $S$-index indicating consistency with basal chromospheric flux.
}
{}

   \authorrunning {M. Auri\`ere et al.}
   \titlerunning {Magnetic fields of Active Red Giants}

\maketitle

\section{Introduction}
  
 Ordinary G and K giants are expected to harbour mainly weak surface magnetic fields because of their large radii and their slow rotation (e.g. Landstreet, 2004). However, activity (in the form of, e.g., emission in the cores of strong chromospheric lines, photometric variability, X-ray emission) is a feature which occurs among these stars, and which has been observed for several decades (e.g. reviews by Korhonen 2014, Konstantinova-Antova et al. 2013).  Magnetic fields have been detected via Zeeman signatures revealed by the spectropolarimetric method in the case of rapidly rotating giants situated in synchronized binaries (RS CVn stars, e.g. HR 1099, Donati et al. 1990) or supposed to originate from coalesced binaries (FK Com type stars, e.g. Petit et al. 2004). For the slower rotators, in spite of some early investigations (Hubrig et al. 1994, Tarasova 2002), reliable detection of surface magnetic fields with the Zeeman effect was not obtained before the introduction of the twin spectropolarimeters ESPaDOnS at the Canada-France-Hawaii telescope (CFHT) and Narval at t\'elescope Bernard Lyot (TBL, Pic du Midi Observatory, Konstantinova-Antova et al., 2008). In this paper we report on observations obtained with Narval and ESPaDOnS of a sample of 48 single G-K red giants (or wide binaries in which synchronization play no role) including fast to slow rotators. The stars were selected as appearing as good candidates for leading to magnetic field detection, in particular on the basis of activity signatures and/or fast rotation. However, well known FK Com type stars were not included in our sample.

 The aim of this pilot study was to investigate if the activity signatures of single giants were due to magnetic fields, if a dynamo operates in these stars (or if there is another possible origin for their magnetic field and activity, e.g. that they are descendants of magnetic Ap stars) and how the magnetic fields in giant stars and potential dynamo depend on their rotation and evolutionary status. 

Section 2 presents the sample of selected giants and some of their properties, and \S~3 presents our observations. Section 4 presents the criteria related to the detection of magnetic field via Zeeman signatures, as well as the measurement of the longitudinal magnetic field ($B_l$), the $S$-index, and radial velocities (RVs). In  \S~5, evolutionary models of Lagarde et al. (2012) and Charbonnel et al. (in prep.) are used to locate the giants in the Hertzsprung-Russell diagram (HRD) and to compute the Rossby number ($Ro$) for the giants with measured rotational periods. Section 6 presents the analysis of the data for our sample stars.  Section 7 is our discussion and \S~8 reports the conclusions. We also provide 3 appendices. In Appendix A we report complementary results for 3 stars which have been followed up during several seasons. Properties and results for the giants, when not available in Tables or in the main text, are presented in Appendix B.  CNO abundances and $^{12}$C/$^{13}$C ratios for 12 stars with ambiguous evolutionary status, including our own determinations for 3 stars, are presented in Appendix C.

\section{The observed sample}

\subsection{Selection of the sample of red giants}

The main aim of our study was to detect and measure the magnetic field at the surface of single red giant stars (or wide binaries in which synchronization plays no role in their fast rotation and magnetic activity), beginning our investigation with those already known to exhibit activity signatures or appearing as good candidates for Zeeman detection.  
Our sample was constructed from three subsamples of promising stars with respect to Zeeman detection. The observations of the 2 first subsamples were initiated with Narval, the third one was initially a snapshot program with ESPaDOnS, then observations were followed up with Narval:

Sample 1) This sample consists of the giants reported in the literature to present evidence for activity (hereafter the 'Active Giants' subsample of 24 stars):
We first included the G K giants selected by Fekel \& Balachandran (1993), consisting mainly of fast rotating objects with respect to other giants (De Medeiros \& Mayor 1999) with strong Ca~{\sc ii} H\&K activity. These stars are generally also outstanding X-ray emitters ($L_{\rm x} > 10^{30}$\ erg s$^{-1}$) in the sample studied by Gondoin (1999, 2005b). Although highly active, we explicitly excluded FK Com stars from our observational sample, since it is believed that their fast rotation originates from merger events (which are not included in the evolutionary models). Consequently, we do not present here our results concerning the FK Com candidate HD 232862 (L\`ebre et al. 2009, Auri\`ere et al. in preparation). 
We also included giants with slower apparent rotation, but for which strong emission has been detected in X-rays (Gondoin 1999, Schr\"oder et al. 1998), or for which variations in Ca~{\sc ii} $H \& K$ emission cores have been measured (Choi et al., 1995).

Sample 2) The 'Thermohaline deviants' subsample (hereafter the 'THD' subsample of 7 stars):
This sample is composed of red giant stars which may have escaped thermohaline mixing (Charbonnel \& Zahn 2007a). Such 'thermohaline deviants' have been proposed to host strong and deeply buried magnetic fields and to be descendants of magnetic Ap stars (Charbonnel \& Zahn 2007b). We selected this subsample based on their high $^{12}C / ^{13}C$ ratio and/or anomalous Li abundance: one is a deviant object of Charbonnel \& do Nascimento (1998); six stars were part of a large ESO/OHP/McDonald spectroscopic survey by Charbonnel et al. (in prep.). These stars have been observed both with Narval and ESPaDOnS. 

Sample 3) The 'CFHT snapshot' subsample and miscellaneous stars (hereafter the 'CFHT \& miscellaneous' subsample of 15 + 2 stars):
The CFHT snapshot program was designed for execution even during the worst sky conditions at CFHT. This subsample is composed of 15 very bright ($V< 4$) red and yellow giants, in which dynamo-driven magnetic fields may occur. 
 
This subsample contained some of the stars already in the active giants subsample (namely: $\kappa$ Her A, $\beta$ Boo, $\rho$ Cyg, $\beta$ Ceti). This program led to the detection of a magnetic field at the surface of 4 red giants which were followed up with Narval, namely Pollux, $\epsilon$ Taurus, Aldebaran, and Alphard. We added to this subsample two stars ($\eta$ Psc and $\mu$ Peg) selected from the list of possibly magnetic late giants of Tarasova (2002).

\subsection{Some properties of the stars in the sample: description of Table 1}
\label{subsect:stellarproperties}

The properties of the red giants included in our sample are summarized in Table 1. Columns 1 and 2 give the HD number and the name of the stars. The $V$ magnitude comes from the Hipparcos catalog (ESA 1997), and the spectral class is from SIMBAD database. 

For $T_{\rm eff}$, $\log g$, [Fe/H], and $v \sin i$, we tried to use a homogeneous set of data. With the increasing number of works devoted to searching for stars hosting exoplanets,  recent measurements of fundamental parameters, rotation and metallicity of red giant stars have become available. The compilation of Massarotti et al. (2008), who studied stars nearer than 100 pc, includes a large part (64\%) of the 'Active Giants' and 'CFHT \& miscellaneous' subsamples and we used the results whenever possible. However, some of the most active giants of our sample, as well as the 'THD' stars, are too distant to be included in this work: the references for the parameters adopted for these stars are given in the Appendix B. Rather large uncertainties of several tens of K are present for $T_{\rm eff}$, as illustrated in our work on Pollux (Auri\`ere et al. 2009). This is in particular seen when data obtained using different techniques are used. For example, Massarotti et al. (2008) use photometry, Fekel \& Balachandran (1993) use spectroscopy, and Wright et al. (2003) derive $T_{\rm eff}$ from the spectral type. As to $v \sin i$, when not available from Massarotti et al. (2008), the quantity is taken from de Medeiros \& Mayor (1999) or from other sources (given in Appendix B.) Each value of these 4 parameters which is not taken from Massarotti et al. (2008) is marked with an asterisk in Table 1.

Luminosity values are computed using the stellar parallaxes from the New Reduction Hipparcos catalog by van Leeuwen (2007), the $V$ magnitudes from the 1997 Hipparcos catalog (ESA 1997), and applying the bolometric correction relation of Flower (1996)\footnote{We use M$_{bol,\odot}$=4.75}. Since the stars are rather close to the Sun (most of them are nearer than 100 pc), we did not apply a correction for interstellar extinction. The adopted luminosity values are shown in Fig. 5 (see \S~5) %\S~\ref{sect:evolutionstatus})
where the error bars on luminosity reflect only the uncertainties on the parallaxes which generally correspond to the dominant ones. 
We also provide in Table 1 the stellar radius $R_*$ obtained from the Stefan-Boltzmann law using the adopted values for the stellar effective temperature and luminosity.

The X-ray luminosities come from the catalogs of H\"unsch et al. (1998a,b), the pointed observations of H\"unsch et al. (1996) and Collura et al.  (1993), or the works of Schr\"oder et al. (1998) and Gondoin (1999). The luminosities $L_{\rm x}$ quoted in Table~1 are computed using the Hipparcos distances (van Leeuwen 2007).

The last column of Table 1 indicates if the star is detected as magnetic via the detection of a significant Stokes $V$ signature in its LSD profiles (described in \S~4). Useful information about the individual stars, when not included in the Tables, is given in the Appendix B.

The origin of the rotational periods given in Table 1 is described in the next subsection. The discussion about the evolutionary status of our sample stars is presented in \S~5 (see also Appendix C).

\begin{table*}
\caption{Properties of Red Giants Stars in the sample: Active Giants (selected from literature), Thermohaline Deviants, CFHT snapshot and miscellaneous stars}
\begin{tabular}{llclccccccccc}
\hline
\hline
HD       &Name     &$V_{ \rm mag}$&  Sp type& $T_{\rm eff}$  &$\log g$&[Fe/H]& $\log L$    & $R_*$          &$v \sin i$    & $P_{\rm rot}$&$L_{\rm x}$                 & Det.\\
         &         &        &         &   K       &      &      &$L_{\odot}$   & $R_{\odot}$      &km $s^{-1}$    & day     &$10^{27}$ erg $s^{-1}$ &\\
\hline
\hline
\multicolumn {13} {c} {Active  Giants} \\
\hline
3229    &14 Cet    &5.84    &F5 IV    &6495*       &3.8*  & 0.0* &  1.03        & 2.6           &   5*       &         & 336                 &DD\\
4128    &$\beta$ Cet  &2.04 &K0 III   &4797        &2.7   &-0.09 &  2.19        & 18.0          &   5.8      & 215     & 1585                &DD \\
9746    &OP And    &6.20    &K1 III   &4420*       &2.3*  & 0.0* &  2.08        & 18.6          &   9.0*     &  76     & 25300               &DD \\
27536   &EK Eri    &6.15    &G8 III-IV&5058        &3.2   & 0.02*&  1.12        & 4.7           &   1.4      & 308.8   & 1000                &DD\\
28307   &77 Tau    &3.84    &K0 IIIb  &4955        &2.90  & 0.04 &  1.84        & 11.3          &   4.2      & 140     & 1996                &DD\\
31993   &V1192 Ori &7.48    &K2 III   & 4500*     &3.0*   &0.10* &  1.84        & 13.6          &   33*      & 25.3    & 23500               &DD\\
33798   &V390 Aur  &6.91    &G8 III   &  4970*     &3.0*  &-0.05*&  1.38        & 6.6           &   25*      &  9.8    & 5040                &DD\\
47442   &nu3 CMa   &4.42    &K0 II-III&4510        &2.34  & -0.24&  2.60        & 32.7          &   4.3      & 183     & 624                 &DD\\
68290   &19 Pup    &4.72    & K0 III  &4932        &2.9   &-0.03 &  1.62        & 8.8           &   2.2      & 159     & 586                 &DD\\
72146   &FI Cnc    &7.45    &G8 III   & 5150*      &      &      &  1.49        & 6.9           &            &  28.5   & 26000               &DD \\
82210   &24 UMa    &4.54    &G4 III-IV&5253*       &3.43* & -0.34*& 1.18        & 4.7           &   5.5*     &         & 901                 &DD\\
85444   &39 Hya    &4.11    &G7 III   &4977        &2.7   & -0.14&  2.20        & 16.8          &   4.9      &         & 4690                &DD\\
111812  &31 Com    &4.93    &G0 III   & 5660*      &3.51* &-0.15*&  1.87        & 8.9           &   67*      &  6.8    & 6325                &DD \\
112989  &37 Com    &4.88    &G8 II-III& 4600*      &2.3*  &-0.05*&  2.77        & 38.2          &   11*      & 111     & 5200                &DD \\
121107  &7 Boo     &5.71    &G5 III   & 5150*      &      & 0.08 &  2.36        & 19.0          &   14.5*    &         & 3720                &MD \\
133208  &$\beta$ Boo&3.49   &G8 IIIa  &4932        &2.8   &-0.13 &  2.32        & 19.8          &   2.5*     &         & 153                 &nd\\
141714  &$\delta$ CrB&4.59  &G3.5 III &5248        &3.2   &-0.32 &  1.58        & 7.4           &   7.2      & 59      & 1456                &DD\\
145001  &$\kappa$ HerA&5.00 &G8 III   & 4990*      &      &-0.26*&  2.13        & 15.5          &   9.9*     &         & 2980                &DD \\
150997  &$\eta$ Her&3.50    &G7.5 IIIb&4943        &2.8   &-0.37 &  1.69        & 9.5           &   2.2      &         & 63                  &DD\\
163993  &$\xi$ Her&3.70     &G8 III   &4966        &2.8   & -0.1 &  1.79        & 10.7          &   2.8      &         & 3000                &DD\\
203387  &$\iota$ Cap&4.28   &G8 III   &5012        &2.7   &-0.23 &  1.87        & 11.4          &   7.1      & 68      & 4482                &DD \\
205435  &$\rho$ Cyg&3.98    &G8 III   &5012        &3.    &-0.31 &  1.59        & 8.2           &   3.9      &         & 1072                &DD \\
218153  &KU Peg    &7.64    &G8 II    & 5000*      &3.0*   &-0.15*& 1.58        & 8.2           &   27.1*    & 25      & 11800               &DD\\
223460  &OU And    &5.86    &G1 III   & 5360*      &2.8*   &      & 1.81        & 7.8           &   21.5*    & 24.2    & 8203                &DD\\
\hline
\hline
\multicolumn {13} {c} {THD} \\
\hline
50885   &          & 5.69   & K4 III  &4750*       &      &      & 2.18         & 18.3          &            &         &                     &nd\\
95689   &$\alpha$ UMa& 1.81 &K0 Iab   &4655*       &2.2*  &-0.19*& 2.52         &   28.2        &            &         &                     &nd\\
150580  &          &6.07    &K2       &4420*       &      &      & 1.98         &   16.7        &            &         &                     &nd\\
178208  &          &6.45    &K3 III   &3950*       &      &      & 2.34         &   31.6        &            &         &                     &nd\\
186619  &          &5.86    &M0 III   &3690*       &1.63* &      & 2.73         &   57.0        &            &         &                     &nd\\
199101  &          &5.47    &K5 III   &3940*       &1.65* &-0.36*& 2.73         &   49.6        &            &         &                     &nd\\
218452  &4 And     &5.30    &K5 III   &4100*       &1.91* &-0.02*& 2.23         &   25.9        &            &         &                     &nd\\
\hline
\hline
\multicolumn {13} {c} {CFHT \& miscellaneous} \\
\hline
9270    &$\eta$ Psc   &3.62 &G7 IIa   &4898        &2.44  &-0.14 & 2.65         &  29.5         &  8.4       &         &                     &DD\\
9927    &$\upsilon$ Per&3.59&K3 III   &4325        &2.2   &0.    & 2.23         &  23.2         &  5.9       &         &                     &nd\\ 
12929   &$\alpha$ Ari  &2.0&K2 III    &4498        &2.4   &-0.25 & 1.95         &  15.6         & 4.2        &         &                     &nd\\
28305   &$\epsilon$ Tau&3.53& G9.5 III&4797        &2.6   & 0.04 & 1.95         & 13.7          & 4.4        &         & 21                  &DD\\ 
29139   &Aldebaran     &0.87&K5 III   &3936        &1.    & -0.34& 2.66         & 45.7          & 4.3        &         &                     &DD\\
32887   &$\epsilon$ Lep&3.19& K4 III  &4150*       &1.8*  &      & 2.62         & 39.7          & 4.3*       &         &                     &nd\\
62509   & Pollux       &1.16&K0 III   &4842        &2.9   &-0.07 & 1.64         & 9.3           & 2.8        & 590     & 5                   &DD\\
76294   &$\zeta$ Hya   &3.12&G9 II-III&4819        &2.6   &-0.21 & 2.23         & 18.8          & 2.5        &         &                     &nd\\ 
81797   & Alphard      &2.0&K3 II-III &4027        &1.8   &-0.12 & 3.02         & 66.1          & 2.3*       &         &                     &DD\\
89484   &$\gamma$ Leo A&2.12&K1 IIIb  &4365        &2.3   &-0.49 & 2.58         & 34.2          & 4.3        &         &                     &nd\\
93813   &$\nu$ Hya     & 3.11& K0/K1 III&4335      &2.3   &-0.3  & 2.24         & 23.4          & 5.3        &         & 71                  &nd\\
105707  &$\epsilon$ Crv&3.02&K2 III   &4320*       &2.16* &0.13* & 2.97         & 54.7          & 2.6*       &         &                     &nd\\
124897  & Arcturus     &-0.05&K1.5 III&4325        &2.1   &-0.6  & 2.36         & 26.9          & 4.2        &         &                     &MD\\
129989  &$\epsilon$ Boo A&2.70&K0 II-III&4550      &2.2   &-0.13 & 2.77         & 39.2          & 10.9       &         &                     &nd\\
131873  &$\beta$ UMi   &2.07&K4 III   & 4085*      &1.6*  &-0.15*& 2.68         & 43.8          & 1.7*       &         &                     &nd\\ 
163917  &$\nu$ Oph     &3.32 &G9 III  & 4831       &2.7   &0.02  & 2.06         & 15.3          & 2.1        &         &                     &nd\\
216131  &$\mu$ Peg &3.51    &G8 III   &4943        &2.8  &-0.16  & 1.65         & 9.2           &  4.0       &         & 1.1                 &nd\\
\hline
%\hline 
\end{tabular} 
\tablefoot{Columns are described in \S~2.2. $T_{\rm eff}$, $\log g$, [Fe/H], $v \sin i$ are from Massarotti (2008) when available. An asterisk indicates data that are from another reference given in Appendix B. $log(L)$ is computed using relevant data from the Hipparcos catalogs (1997, 2007) and the bolometric correction of Flower (1996) and is used in Fig. 5;  $R_*$ is the corresponding radius obtained from the Stefan-Boltzmann law.}  
%\end{tabular}
\end{table*} 
%\end{longtable}

\subsection{Rotational periods for our sample red giant stars}

The rotational period is a fundamental parameter of magnetic stars and is required to learn about the origin of the magnetic field, e.g. via the Rossby number. 
Unfortunately, rotational periods have not been previously determined for all stars for which we detect and measure the magnetic field". From the literature we have collected rotational periods of some giants of the 'Active Giants' subsample. These were derived from optical photometry or variations of chromospheric Ca~{\sc ii} $H \& K$ lines. We inferred a few additional rotational periods from the variations of the measured magnetic field.
 To extend the number of active stars with known rotational periods, it might be tempting to use predictions obtained from the chromospheric flux as made by Young et al. (1989) and used by Gondoin (2005b). Nevertheless, in this case since the obtained rotational periods are only representations of the chromospheric fluxes and not of its variations, they can be far from the real period and therefore misleading.

\noindent Optical photometry:

Chromospherically active stars have been the subject of several large photometric surveys carried out with automatic ground-based telescopes (e.g. Strassmeier et al. 1990, 1999, Henry et al. 1995, 2000). Space laboratories, as MOST, COROT and KEPLER (and soon BRITE constellation) are also providing (or provided) huge amounts of rotational data for evolved stars, but for the COROT and KEPLER their targets are generally fainter than those in the present study. Henry et al. (2000) obtained photometric measurements of 187 G and K giants. One of their conclusions is that the light variations in the vast majority of G and K variables are most likely due to pulsation. However, in  the case of chromospherically active stars with fast rotation,  they suggest that rotational modulation of active regions is the principal variability mechanism. Eight stars from our sample have their rotational period derived from photometry.  V390 Aur ($P_{\rm rot}$ = 9.825 d., Hooten \& Hall 1990), V1192 Ori ($P_{\rm rot}$ = 25.3 $\pm$ 0.3 d; Strassmeier et al. 2003), EK Eri ($P_{\rm rot}$ = 308.8 d., Dall et al. 2010), OP And ($P_{\rm rot}$ = 76 d., Strassmeier \& Hall 1988, Konstantinova-Antova et al. 2005), FI Cnc ($P_{\rm rot}$ = 29 d, Henry et al. 1995, Strassmeier et al. 2000, Erdem et al. 2009), 31 Com ($P_{\rm rot}$ = 6.8 d, Strassmeier et al. 2010), OU And ($P_{\rm rot}$ = 24.2 d., Strassmeier et al. 1999), KU Peg ($P_{\rm rot}$ = 24.96 d., Weber \& Strassmeier 2001).

\noindent Chromospheric emission flux:

The study of Choi et al. (1995) gives rotational periods for 5 stars of our sample: 77 Tau ($P_{\rm rot}$ = 140 d., in agreement with the value of Beck et al. 2014), $\nu$3 CMa ($P_{\rm rot}$ = 183 d.), 19 Pup ($P_{\rm rot}$ = 159 d.), $\delta$ CrB ($P_{\rm rot}$ = 59 d., confirmed by photometry, Fernie 1999), $\iota$ Cap ($P_{\rm rot}$ = 68 d., confirmed by photometry, Henry et al. 1995).

\noindent Magnetic measurements:

In our sample, periods are inferred from magnetic measurements for 3 stars: $\beta$ Ceti ($P_{\rm rot}$ = 215 d., Tsvetkova et al., 2013), Pollux ($P_{\rm rot}$ = 590 d, Auri\`ere et al. 2014a and in prep.), 37 Com ($P_{\rm rot}$ = 110 d., Tsvetkova et al., 2014 and in prep.). 

For 5 giants (V390 Aur, OP And, EK Eri, OU And, 31 Com) our Zeeman Doppler imaging is consistent with the photometric period being the rotational one, excluding the possible effect of active longitudes (starspots concentrating on two active longitudes about 180$^\circ$  apart, in which case the measured $P_{\rm rot}$ could be half the real value (e.g. Berdyugina, 2005)). For the 3 remaining giants among the 8 ones with $P_{\rm rot}$ determined by photometry, in the cases of V1192 Ori (Strassmeier et al. 2003) and KU Peg (Weber and Strassmeier 2001)  Doppler imaging was obtained which also  reduces significantly the possibility of such an error. As to the last giant, FI Cnc, the  $P_{\rm rot}$ of 29 d was derived by 3 different authors in different seasons spanning more than 10 years (Henry et al. 1995, Strassmeier et al. 2000, Erdem et al. 2009).

Figure 1 shows the histogram of the known rotational periods for 16 stars of our sample.  The periods determined by photometric or $S$-index variations are expected to correspond to strongly and moderate active stars; they represent the stars with $P_{\rm rot} < 200$ d. EK Eri and Pollux ($P_{\rm rot} > 300$ d), located in the tail of the histogram, are special cases that merit further discussion later in the paper.

\begin{figure}
\centering
\includegraphics[width=9 cm,angle=0] {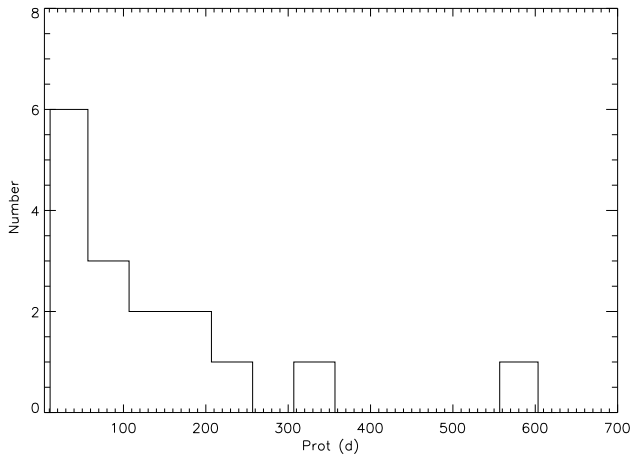}
\caption{Distribution of the rotational periods for 16 of our sample stars with known periods.}
%\label{f1}
\end{figure}  

%\clearpage

\section{Observations with Narval and ESPaDOnS}

 ESPaDOnS at the CFHT (Donati et al., 2006a) and Narval at the TBL are twin spectropolarimeters. Each instrument consists of a Cassegrain polarimetric module connected by optical fibres to an echelle spectrometer. In polarimetric mode, the instrument simultaneously acquires two orthogonally polarized spectra covering the spectral range from 370 nm to 1000 nm in a single exposure, with a resolving power of about 65,000. 

A standard circular polarization observation consists of a series of 4 sub-exposures between which the 
 half-wave retarders (Fresnel rhombs) are rotated in order to exchange the paths of the orthogonally polarized beams within the whole instrument (and therefore the positions of the two spectra on the CCD), thereby reducing spurious polarization signatures. The extraction of the spectra, including wavelength calibration, correction to the heliocentric frame and continuum normalization, was performed using Libre-ESpRIT (Donati et al. 1997), a dedicated and automatic reduction package installed both at CFHT and at TBL. The extracted spectra are output in ASCII format, and consist of the normalised Stokes $I$ ($I/I_{\rm c}$) and Stokes $V$ ($V/I_{\rm c}$) parameters as a function of wavelength , along with their associated Stokes $V$ uncertainty $\sigma_V$ (where $I_{\rm c}$ represents the continuum intensity). Also included in the output are 'diagnostic null' spectra $N$, which are in principle featureless, and therefore serve to diagnose the presence of spurious contributions to the Stokes $V$ spectrum. Observing red giants suspected to host weak magnetic fields required rather long exposures. To avoid saturation of the CCD, we made concurrent series of 4, 8 or 16 Stokes $V$ sequences which were then averaged.

To obtain a high-precision diagnosis of the spectral line circular polarization, least-squares deconvolution 
(LSD, Donati et al. 1997) was applied to each reduced Stokes $I$ and $V$ spectrum. LSD is a multi-line technique, similar to cross correlation, which assumes that all spectral lines have the same profile shape, scaled by a certain factor, and expressed using line masks summarizing the relevant atomic data. The masks were constructed using appropriate temperature and gravity for each star, from ATLAS9 models of solar abundance (Kurucz 1993) or from data provided by the Vienna Atomic Line Database VALD (Kupka et al. 1999). The selected lines have a lower limit for intrinsic depth between 0.1 and 0.25. The number of lines included in each mask is mainly temperature dependant and generally comprised between about 6500 and 14000. At the end, the signal to noise ratio (S/N) of the LSD Stokes $V$ profile is about 30 times higher than the S/N in the original spectrum.

\section{Magnetic field detection, and derivation of its strength, $S$-index and radial velocity}

\subsection{Magnetic field detection and measurement}

\subsubsection{The Stokes $V$ detection criterion}

The output of the LSD procedure contains the mean Stokes $V$ and $I$ profiles, as well as a diagnostic null ($N$) profile (Donati et al. 1997). To diagnose a Stokes $V$ Zeeman detection we performed a statistical test outside and inside the spectral line to infer detection probabilities (as described by Donati et al. 1997). We consider that an observation displays a 'definite detection' (DD) if the signal detection probability inside the line is greater than 99.999\%, a 'marginal detection' (MD) if it falls between 99.9\% and 99.999\%, and a 'null detection' (ND) otherwise. For a reliable  Zeeman detection, we also require no detection of signal outside the spectral line nor in the $N$ profile. 

The LSD procedure has been investigated by Kochukhov et al. (2010), using synthetic spectra. These authors concluded that as far as Stokes $V$ and $B_l$ are concerned, the method works properly for magnetic field strengths up to 1 kG. LSD efficiency has also been compared to principal component analysis (PCA) and simple line addition (SLA) on observations obtained with Narval. Even in easy cases, none of these alternatives gives better results than LSD (Paletou, 2012). In the case of Arcturus, Sennhauser and Berdyugina (2011) presented a possible Zeeman detection of the magnetic field using 3 independent observations constructed from one single Stokes $V$ sequence each, and applying the Zeeman component decomposition method (ZCD, Sennhauser and Berdyugina, 2010). For the 2 observations which are in common with this work we find a similar result, i.e. null detection on 24 August 2008 and marginal detection on 06 December 2008, with consistent longitudinal magnetic field measurements (see \S~4.2.1). However since we also obtained a MD on the null $N$ profile, we discarded the observation of 06 December 2008. The ZCD uncertainties on $B_l$ appear to be twice as small as ours derived in the LSD context. With respect to  LSD, ZCD may therefore provide smaller error bars on $B_l$, which would have to be confirmed. Nevertheless, this would be valuable in the case of very weak magnetic fields of simple topology. However, a weakness of the method is that it yields only a measurement of $B_l$, but no Stokes $V$ profile. This does not enable magnetic mapping using a method such as ZDI, which is a substantial loss of information, since studying the Stokes $V$ profile can enable the detection and study of complex fields even if $B_l$ is near or equal to zero. Ultimately, we consider that the LSD method is the most efficient method available in routine use for studying the rather weak magnetic fields of evolved stars.

\subsubsection{Magnetic field detection of giants of our sample}

Using the LSD procedure for observations of red giants provides S/N gain factors as great as 30. Since the stars are rather bright, the initial single Stokes $V$ spectra can themselves have high S/N, of the order of 1000 per 2.6 kms$^{-1}$ spectral bin, which corresponds to errors smaller than 1 G in $B_l$. Averaging series of 8-16 Stokes $V$ spectra and using LSD yields measurements of circular polarization levels as small as 10$^{-5}$ and errors on $B_l$ smaller than 0.2 G. Ultimately, all but one star ($\beta$ Boo) of our 'Active Giants' subsample were detected, in general with only one Stokes $V$ series. None of the 7 thermohaline deviants was detected: they were observed generally with single Stokes $V$ series, and we can exclude the presence of surface magnetic fields stronger than a few G in these stars. From the 'CFHT \& miscellaneous' subsample, 5 bright stars for which we had no previous evidence for activity were detected: apart for $\epsilon$ Tau, we averaged 8 to 32  Stokes $V$ series of the stars to obtained the detection, and therefore measured $B_l$ values weaker than 0.5 G. In the end, 29 giants of our sample have been detected in this work (listed in the last column of Table 1, the activity measurements are in Table 3, the journal of observations and individual measurements are in Tables 6, 7, 8).

\subsection {The characteristic strength of the magnetic field}

\subsubsection {Measurement of the averaged longitudinal magnetic field $B_l$}

From the mean LSD Stokes profiles we computed the surface-averaged longitudinal magnetic field  $B_l$ in G, using the first-order moment method (Rees \& Semel 1979), adapted to LSD profiles (Donati et al. 1997, Wade et al. 2000). These measurements of $B_l$ are presented in Tables 6, 7, 8 with their 1$\sigma$ error, in G. These errors are computed from photon statistical error bars propagated through the reduction of the polarization spectra and the computation of the LSD profiles, as described by Wade et al. (2000). 

We stress that our detection criterion of magnetic field is based upon detection of Zeeman Stokes V  features and not upon the significance of our $B_l$ measurements, as developed in \S~4.1.1. Figure 2 illustrates the power of our detection method in the case of very weak magnetic fields, presenting 3 observations of Aldebaran: weak $B_l$ measurements, even when observed with  high precision, can correspond to 'null detection'  (ND) or 'definitive detection' (DD). On 14-15 March 2010 (upper frame), we measure $B_l$ = 0.22 $\pm$ 0.08 G, but there is not significant Zeeman Stokes V feature (ND). On 05 October 2010 (middle frame), $B_l$ = -0.25 $\pm$ 0.13 G, and there is a significant Zeeman Stokes V signal with negative polarity (DD). On 16 January 2011 (lower frame), $B_l$ = 0.22 $\pm$ 0.09 G, and there is a significant Zeeman Stokes V signal with positive polarity (DD).

\begin{figure}
\centering
\includegraphics[width=10 cm,angle=0] {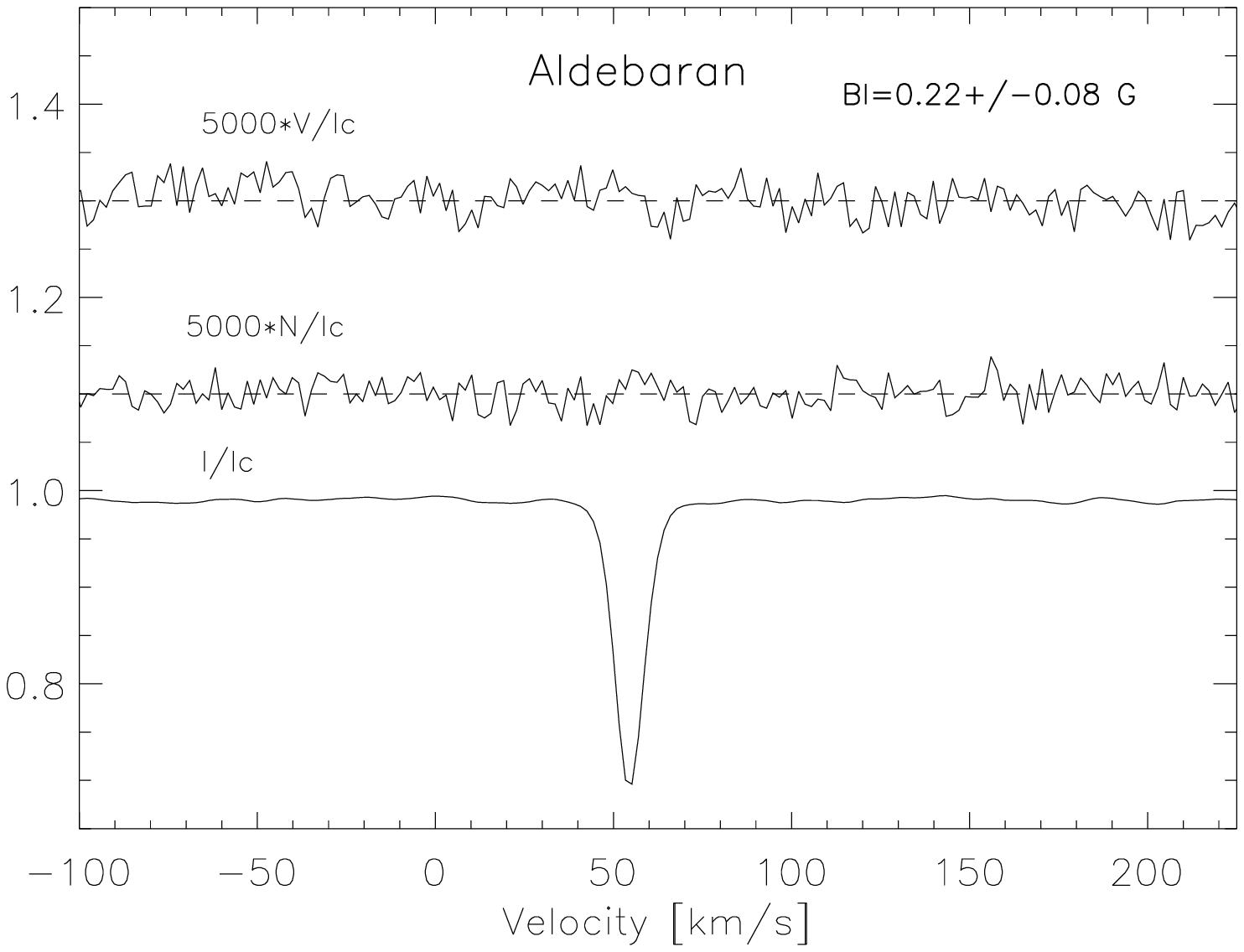}
\includegraphics[width=10 cm,angle=0] {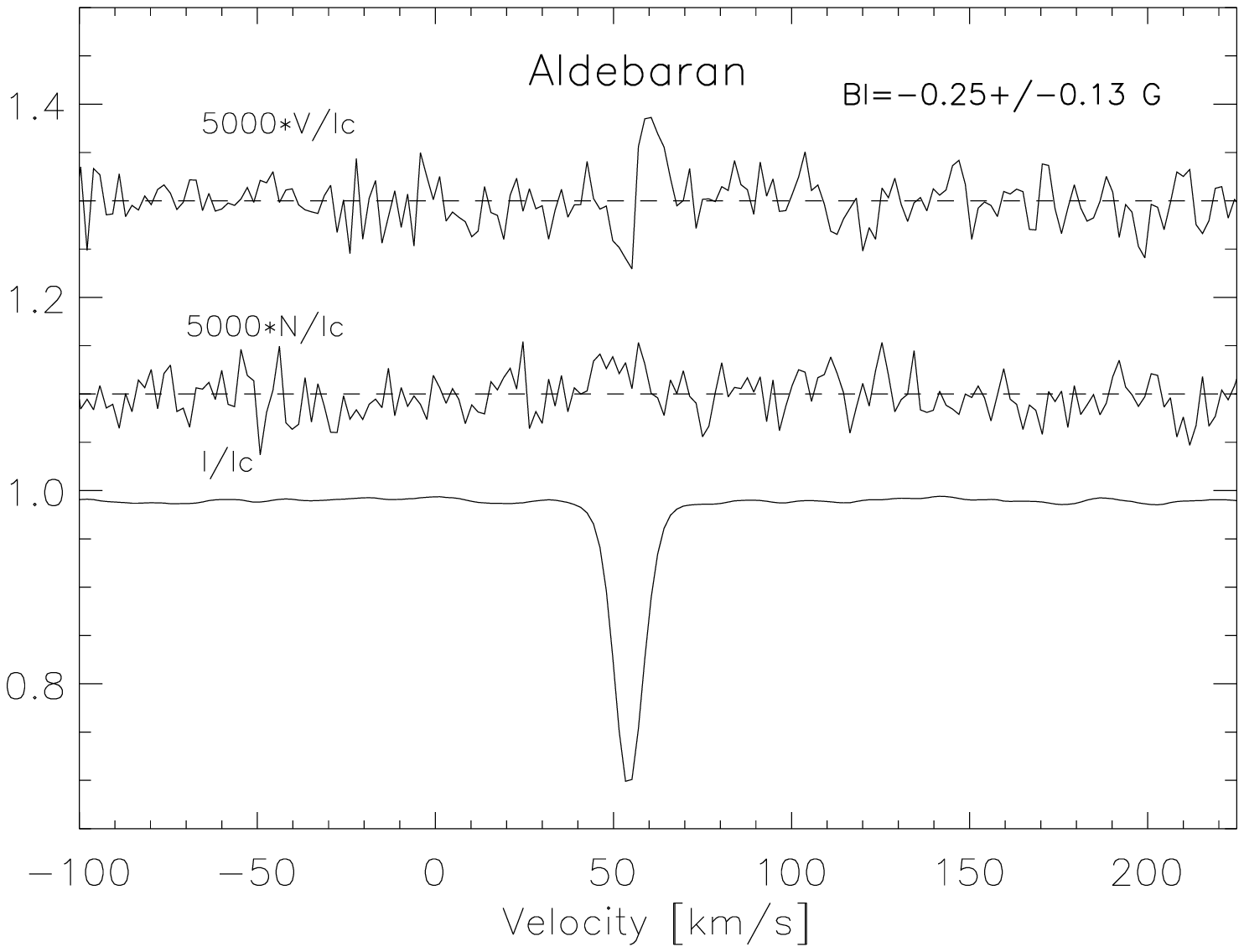} 
\includegraphics[width=10 cm,angle=0] {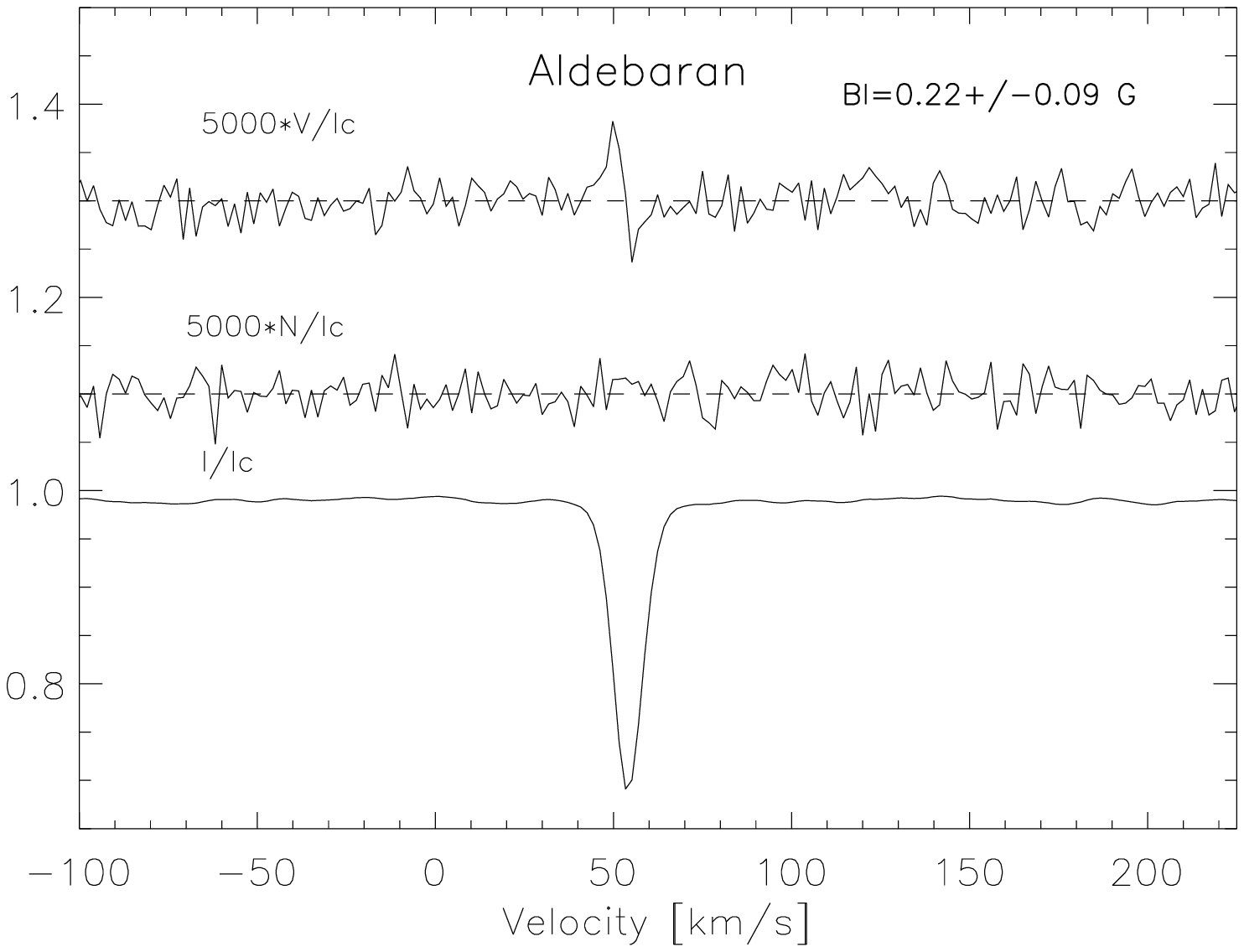}
\caption{LSD profiles of Aldebaran as observed with Narval on  14-15 March 2010 (upper panel), 05 October 2010 (middle panel) and  16 january 2011 (lower panel). For each graph, from top to bottom are Stokes $V$, null polarisation $N$, and Stokes $I$ profiles. For display purposes, the profiles are shifted vertically, and the Stokes $V$ and diagnostic $N$ profiles are expanded by a factor of 5000. The dashed lines illustrate the zero level for the Stokes $V$ and diagnostic null profiles. The Stokes $V$ profiles illustrate a non detection (ND) on 14-15 March 2010, then definite detections (DD) with a change of polarity of the magnetic field between the two later dates.}
%\label{f1}
\end{figure}  

\subsubsection {Measurement of the strength of the magnetic field}

In the Zeeman studies made with Narval and ESPaDOnS more magnetic information is included in the Stokes $V$ profile than is provided by $B_l$ measurements, and when one star is observed for more than one rotational period, it can be possible to model the large scale surface magnetic field using Zeeman Doppler imaging (ZDI, Donati et al. 2006b).  In this procedure, the surface-averaged  magnetic field $B_{\rm mean}$ is inferred from the fitted model. At the present time, we have demonstrated that ZDI is applicable to even very slowly rotating red giants (Auri\`ere et al. 2011) and have performed ZDI for 8 giants of our sample (see Table 2). Since the $v\sin i$ of most of the studied red giants is small, only the large-scale structure of the surface magnetic field can be recovered; the contribution of the small active areas of opposite polarities being cancelled out. To characterize the strength of the magnetic field detected on 29 red giants of our sample, we decided to use the observed maximum unsigned longitudinal magnetic field, $|B_l|_{\rm max}$. Table 2 compares the $|B_l|_{\rm max}$ to the $B_{\rm mean}$ for the 8 red giants of our sample for which ZDI has been performed. For the 4 stars with  $v \sin i$ $<$ 11 kms$^{-1}$, $|B_l|_{\rm max}$ and $B_{\rm mean}$ compare well. For the 3 fastest rotating giants in Table 2, $B_{\rm mean}$ is twice stronger than $|B_l|_{\rm max}$ (and even more for 31 Com). This is probably due to the fact that ZDI resolves active areas of opposite polarities which cancel their contributions when the $B_l$ is computed. 

At the end we consider that for the moderate $v \sin i$ objects of our sample, $|B_l|_{\rm max}$ is a valuable estimate of the surface magnetic field strength.

\subsubsection {Statistical distribution of the strength of the magnetic field}

Table 3 presents the activity measurements of our 29 detected giants, namely $L_{\rm x}$ from the literature, the $|B_l|_{\rm max}$ with its error in gauss, and the $S$-index at the same date. 

\begin{figure}
\centering
\includegraphics[width=9 cm,angle=0] {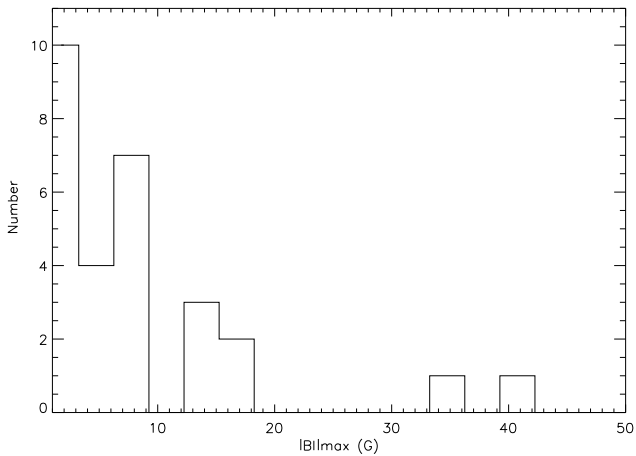}
\caption{Distribution of the strength of the magnetic field ($|B_l|_{max}$) for all  Zeeman detected stars except EK Eri (for which $|B_l|_{max}$ = 98.6 G, and lies outside the frame).}
%\label{f1}
\end{figure}

Figure 3 presents the distribution of $|B_l|_{\rm max}$ among our detected giants. Only 3 stars have $|B_l|_{\rm max}$ stronger than 20 G, 5 are between 20 G and 10 G, and 21 are weaker than 10 G. This shows that the large-scale surface magnetic fields of active single giants of our sample are not strong and that their strength distribution is dominated by those weaker than 10 G.
The incidence of this distribution is studied with the other magnetic properties of the stars in the following Sections.

\begin{table}
\caption{Magnetic strength of red giants with available magnetic field models from ZDI.}          
\label{table:2}   
\centering                         
\begin{tabular}{l l c c c c c c}     
\hline\hline
HD       &Name     &$v \sin i$ & $P_{\rm rot}$ &$|B_l|_{\rm max}$  &$B_{\rm mean}$ &Ref.\\
         &         &km$s^{-1}$& day        &  G            & G        & \\
\hline    
\hline
11812   &31 Com    & 67       & 6.8       & 9.9           & 32       &(1)         \\
33798   &V390 Aur  & 29       &  9.8      & 13            & 26       &(2)   \\
223460  &OU And    & 21.5     & 24.2      & 36            & 68       &(1) \\
%232862  &          & 20.6     & 5.34      & 60            & 128       \\
112989  &37 Com    & 11       & 111       & 6.5           & 10.8      &(3)  \\
9746    &OP And    & 8.7      &  76       & 16            & 15       &(4) \\
4128    &$\beta$ Cet  & 5.8   & 215       & 8             & 10       &(5)  \\
27536   &EK Eri    & 1        & 308.8     & 99            & 94       &(6) \\
62509   &Pollux    & 2.8      & 590       & 0.7           & 0.6      &(7) \\
\hline 
\hline                          
\end{tabular}
\tablefoot{References for $B_{\rm mean}$: (1) Borisova et al. in prep., (2) Konstantinova-Antova et al. 2012, (3) Tsvetkova et al. in prep., (4) Konstantinova-Antova, in prep., (5) Tsvetkova et al. 2013, (6) Auri\`ere et al. 2011, (7) Auri\`ere et al. 2014 and in prep. }
 \end{table}

\subsection {Measurement of the $S$-index}

In order to monitor the line activity indicators, we computed the $S$-index (defined from the Mount Wilson survey, Duncan et al. 1991) for the chromospheric Ca~{\sc ii} $H \& K$ line cores. We used two triangular bandpasses $H \& K$ with a FWHM of 0.1 nm to measure the flux in the line cores. Two 2 nm-wide rectangular bandpasses $R$ and $V$, centred on 400.107 and 390.107 nm respectively,  were used for the continuum flux in the red and blue sides of the H\&K lines. Since 1983, some red giants have been observed in the Mount Wilson survey using slits of 0.2 nm bandpasses since these stars have wider H\&K emission cores. This possibility is described by Duncan et al. (1991) and used e.g. by Choi et al. (1995). However, since the bulk of observations of red giants by the Mount Wilson survey (in particular which are in common with our survey) were observed with the smaller slit, we used it in the present work. Our procedure was calibrated independently for Narval and ESPaDOnS, using respectively 13 and 12 giant stars observed by  Duncan et al. (1991) and Young et al. (1989). Actually, as reported in Morgenthaler et al (2012), there is a spectral order overlap near the $K$ line. This overlap is different for Narval and ESPaDOnS and may introduce different normalisations of the continuum which may explain the differences observed in normalisation of the $S$-index between the two instruments. Figure 4 shows the correlation between the $S$-index from Narval (upper plot) and ESPaDOnS (lower plot) measurements and from the literature for the giant stars used for calibration (See Wright et al. 2004 and Marsden et al. 2014 for more details on the calibration procedure).

\begin{figure}
\centering
\includegraphics[width=9 cm,angle=0] {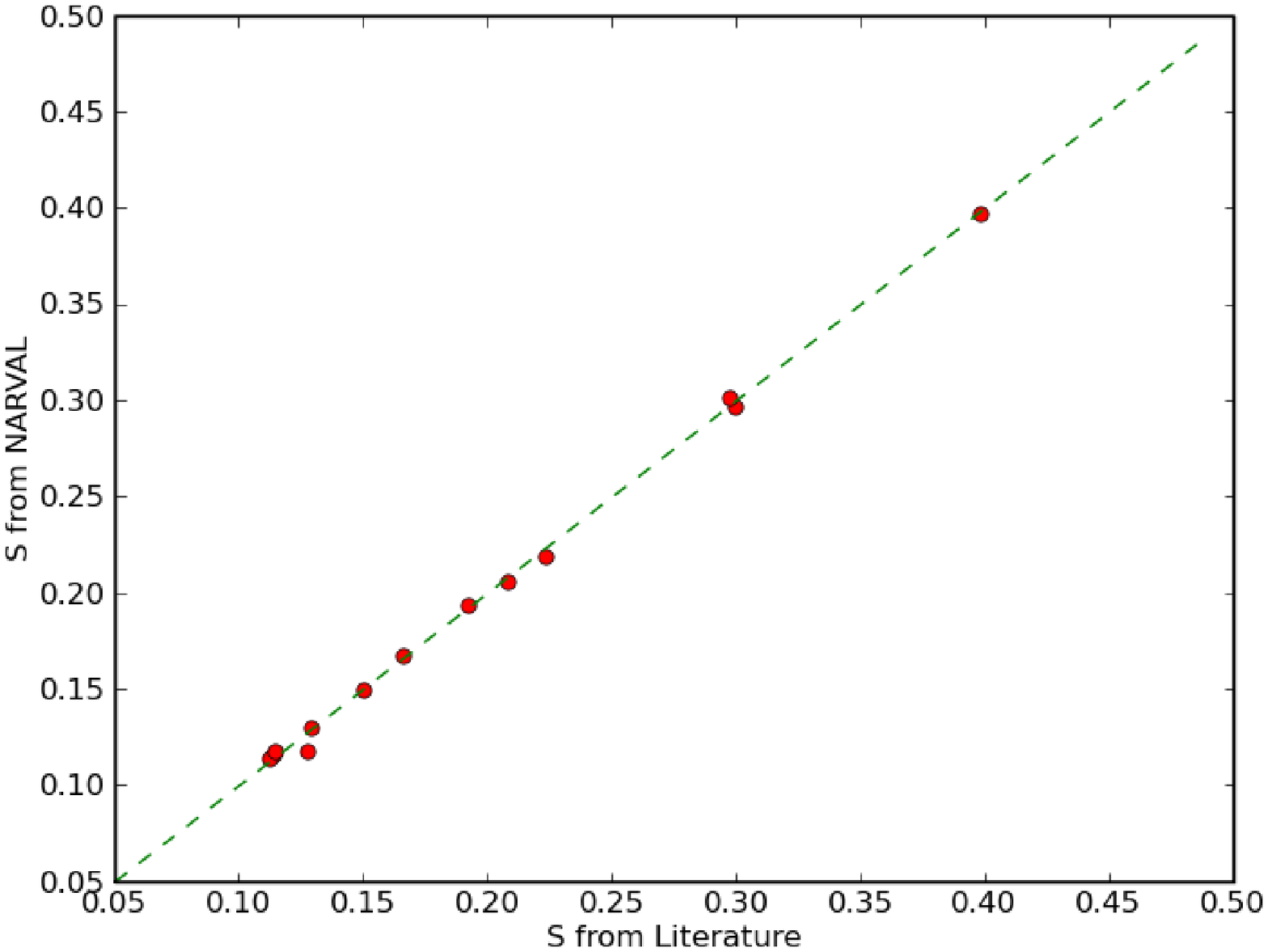}
\includegraphics[width=9 cm,angle=0] {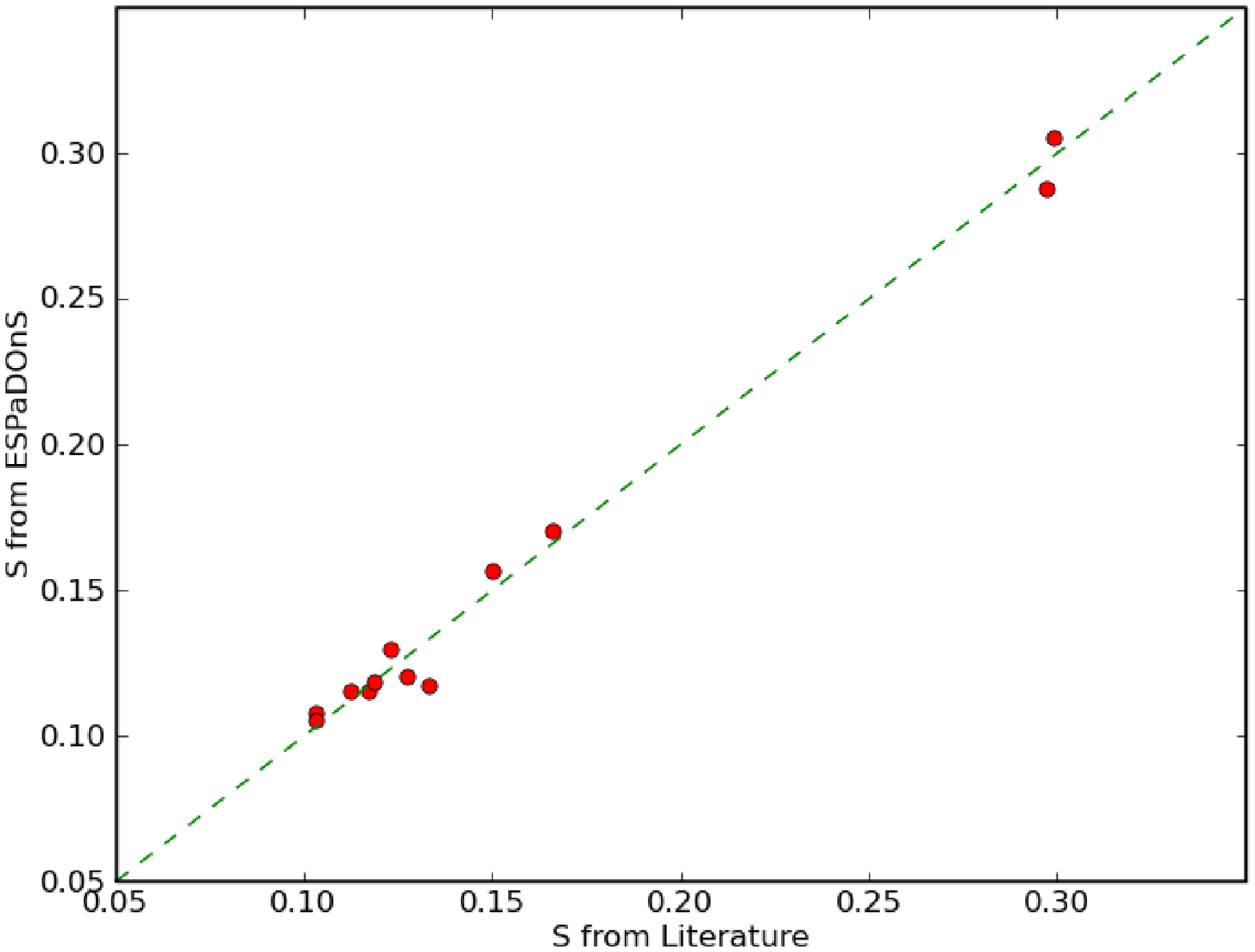}
\caption{Correlation between the $S$-index from Narval (upper) and ESPaDOnS (lower) spectral measurements and from the literature for the giant stars used for calibration.}
%\label{f1}
\end{figure}

\subsection {Measurement of the radial velocity}

 The radial velocity RV of the stars was measured from the averaged LSD Stokes $I$ profiles using a Gaussian fit. The radial velocity stability of ESPaDOnS and Narval is about 20-30 ms$^{-1}$ (Moutou et al. 2007, Auri\`ere et al. 2009) but the absolute uncertainty of individual measurements relative to the local standard of rest is about 1 km s$^{-1}$.

%\clearpage

\begin{table*}
\caption{Activity measurements of Zeeman detected red giants}          
\label{table:2}   
\centering                         
\begin{tabular}{l l c c c c c c }     
\hline\hline  

HD       &Name     &$v \sin i$ & $P_{\rm rot}$&$L_{\rm x}$                 &$|B_l|_{\rm max}$    & $\sigma$& $S$-index \\
         &         &km$s^{-1}$& day      &$10^{27}$ erg $s^{-1}$&  G        &   G       & at $|B_l|_{\rm max}$\\
\hline
\hline
\multicolumn {8} {c} {Active Giants}      \\
\hline
3229    &14 Cet    & 5        &         & 336                    &  36.2    & 1.6       & 0.239 \\
4128    &$\beta$ Cet  & 5.8   & 215     & 1585                   &  9.0     & 0.3       & 0.236 \\
9746    &OP And    & 8.7      &  76     & 25300                  & 15.7     & 0.7       & 0.798 \\
27536   &EK Eri    & 1        & 308.8   & 1000                   &  98.6    & 1.0       & 0.501 \\
28307   &77 Tau    & 4.2      & 140     & 1996                   &  3.0     & 0.5       & 0.176 \\
31993   &V1192 Ori & 33       &  25.3   & 23500                  & 14.7     & 3.2       & 0.997 \\
33798   &V390 Aur  & 29       &  9.8    & 5040                   & 15.      & 3.        & 0.681 \\
47442   &nu3 Cma   & 4.3      & 183     & 624                    &  2.2     & 0.4       & 0.170 \\
68290   &19 Pup    & 2.2 (1)  & 159     & 586                    &  4.2     & 0.4       & 0.206 \\
72146   &FI Cnc    & 17       &  28.5   & 26000                  & 17.6     & 1.7       & 1.027 \\
82210   &24 Uma    & 5.5      &         & 901                    &  3.1     & 0.7       & 0.397 \\
85444   &39 Hya    & 4.9      &         & 4690                   &  7.7     & 0.6       & 0.219 \\
111812  &31 Com    & 67       &  6.8    & 6325                   & 6.9      & 3.1       & 0.398     \\
112989  &37 Com    &11        & 111     & 5200                   & 6.5      & 0.9       & 0.368 \\
121107  &7 Boo     & 14.5     &         & 3720                   & 1.9      & 0.8       & 0.221 \\
141714  &$\delta$ CrB& 7.2    & 59      & 1456                   &  6.1     & 0.5       & 0.286 \\
145001  &$\kappa$ HerA& 9.4   &         & 2980                   & 4.6      & 0.8       & 0.296 \\
150997  &$\eta$ Her& 2.2      &         & 63                     &  6.8     & 0.5       & 0.191 \\
163993  &$\xi$ Her& 2.8       &         & 3000                   &  3.8     & 0.4       & 0.251 \\
203387  &$\iota$ Cap& 7.1     & 68      & 4482                   &  8.3     & 0.6       & 0.343 \\
205435  &$\rho$ Cyg& 3.9      &         & 1072                   &  7.3     & 0.5       & 0.268 \\
218153  &KU Peg    & 29       & 25      & 11800                  & 13.0     & 7.2       & 1.060 \\
223460  &OU And    & 21.5     & 24.2    & 8203                   & 41.4     & 1.5       & 0.515 \\
\hline
\hline
\multicolumn {8} {c} {CFHT \& miscellaneous}  \\
\hline
9270    &$\eta$ Psc   & 8.4   &         &                        &  0.4     & 0.2       & 0.133  \\
28305   &$\epsilon$ Tau&4.4   &         & 21                     &  1.3     & 0.3       & 0.116 \\ 
29139   &Aldebaran   &4.3     &         &                        &  0.25    & 0.1       & 0.235 \\
62509   &Pollux      &2.8     &590      & 5                      &  0.7     & 0.1       & 0.118 \\
81797   & Alphard    &8.5     &         &                        & 0.35     & 0.08      & 0.185          \\    
124897  &Arcturus    &4.2     &         &                        & 0.34     & 0.11      & 0.128          \\  
\hline
\hline                            
\end{tabular}
 \end{table*} 

\section {Evolutionary status, theoretical convective turnover time, and Rossby number}
\label{sect:evolutionstatus}

\subsection {Location of the sample stars in the HRD}

Figure 5 presents the positions of our sample stars in the Hertzsprung-Russell diagram. 
The values we use for the stellar effective temperatures and luminosities are those of Table 1 obtained as described in  \S~2.2. %\S~\ref{subsect:stellarproperties}.   
Filled and open symbols correspond to the detected and undetected stars, respectively. Circles correspond to stars which are in Massarotti et al. (2008) and squares correspond to other stars, as explained in \S~2.2.

In order to determine the mass and the evolutionary status of the sample stars we use the stellar evolution models with rotation and thermohaline mixing of Charbonnel \& Lagarde (2010) and Lagarde et al. (2012) completed with additional stellar masses (Charbonnel et al. in preparation;  see \S~5.2). % \S~\ref{subsect:predictionconvectionRossby}). 
For all stars we use the solar metallicity tracks shown in Fig. 5, except in the case of Arcturus for which we use the Z=0.004 tracks ([Fe/H]=-0.56). 
We distinguish stars lying in the Hertzsprung gap (HGap) before the occurrence of the first dredge-up, at base of the red giant branch (Base RGB), on the first red giant branch (RGB), in the central-helium burning phase (He burning), or on the AGB. The corresponding information is given in Table 4.
Note that it is difficult to distinguish between the base of the RGB and the central-helium burning phase for stars with masses greater than 2 $M_{\odot}$, 
(see e.g. the case of Pollux, Auri\`ere et al. 2009); in this case we indicate both possibilities in Table 4.
We have also tried to get additional information for the 12 stars with uncertain evolutionary status using the abundances of 
lithium and the carbon isotopic ratios $^{12}$C/$^{13}$C found in the literature or measured from our spectra and comparing them to the values predicted by our models for the different evolutionary phases (see Appendix C). The corresponding data and results are summarized in Table C1. %Table~\ref{table:lightelements}. 

What is striking in Fig. 5 %~~\ref{fig:hrd}
 is that most of the detected stars of our sample are in the first dredge-up phase or in the core helium-burning phase, except for four stars which are crossing the Hertzsprung gap and three which are probably on the AGB.  This zone therefore appears as a 'magnetic strip' where the most active magnetic giants are found. Among the detected stars are all our 'active giants' but one ($\beta$ Boo). On the other hand, most of our low-mass, bright RGB stars are non-detected objects. We interpret this behavior in the next sections.

%\onecolumn

\begin{figure*}
\centering

\includegraphics[width=15 cm,angle=0] {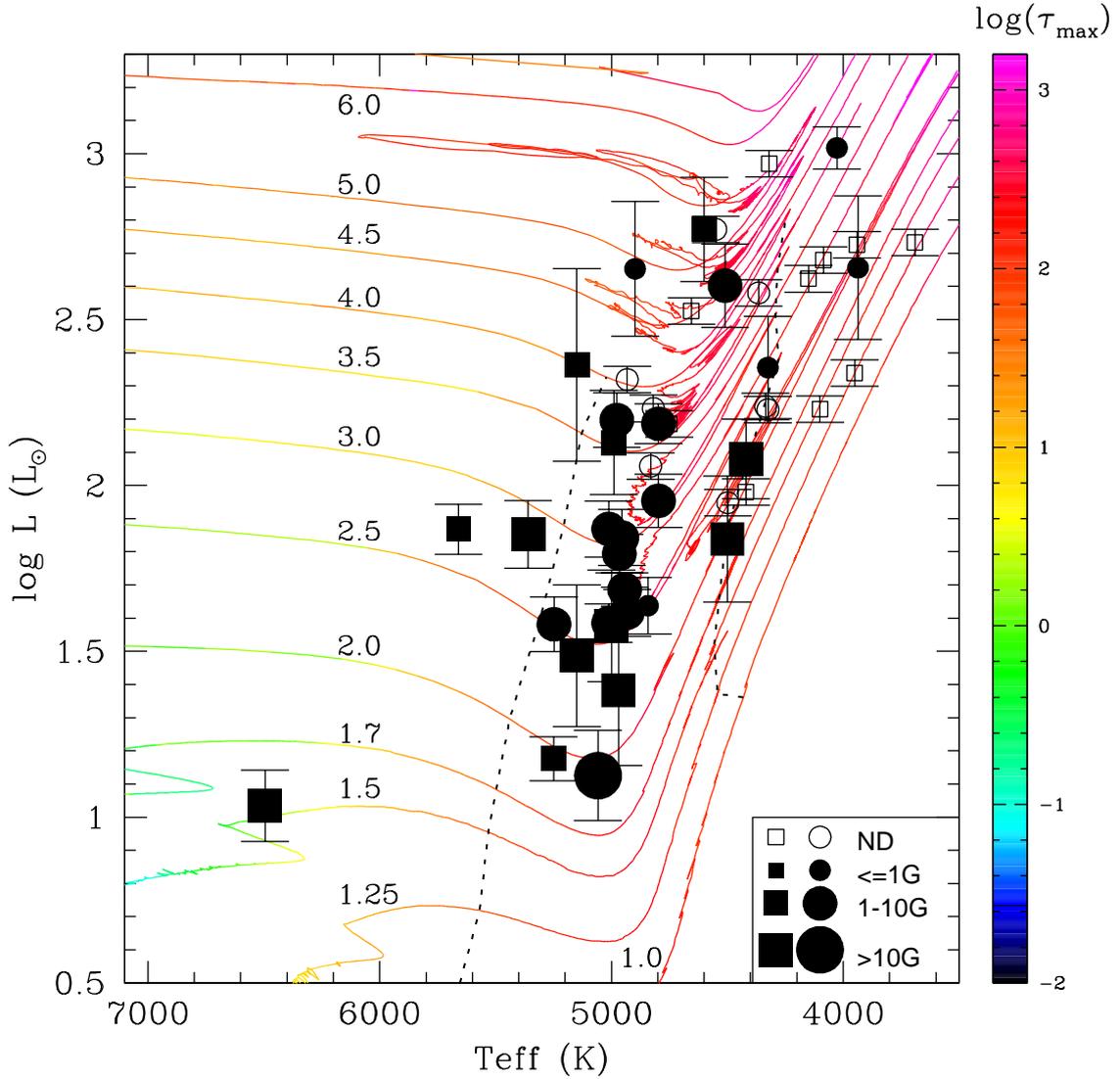}

\caption{Position of our sample stars in the Hertzsprung-Russell diagram.  Solar metallicity tracks with rotation of Charbonnel \& Lagarde (2010) and Charbonnel et al. (in prep.) are shown up to the RGB tip for the low-mass stars (below 2~M$_{\odot}$), and up to the AGB phase for the intermediate-mass stars. The initial mass of the star (in M$_{\odot}$) is indicated for each track.
The color scale indicates the value of the maximum convective turnover time within the convective envelope $\tau(max)$. 
The dotted lines delimit the boundaries of the first dredge up phase, which correspond respectively to the evolutionary points when the mass of the convective envelope encompasses 2.5\% of the total stellar mass, and when the convective envelope starts withdrawing in mass at the end of the first dredge-up. Circles correspond to stars which are in Massarotti et al. (2008) and squares correspond to other stars, as explained in the text.
}
%\label{f1}
\end{figure*}  

%\twocolumn

\begin{table*}
\caption{Theoretical quantities derived from the evolution models for the Zeeman detected giants.}         
\label{table:2}   
\centering                         
\begin{tabular}{l l c c c c c c c c c }     
\hline
\hline
HD       &Name     &   $M_*$        &Evolution &  $R_{\rm BCE}$& $\tau(H_p/2)$          & $R_{(H_p/2)}$ & $\tau(max)$      & $R_{\rm max}$   & $\tau(R_{\rm CE}/2)$& $R_{(R_{\rm CE}/2)}$ \\
         &         & $M_{\odot}$      & phase    & $R_*$   &     day                & $R_*$      & day              &$R_*$     & day                &     $R_*$ \\
\hline
\hline
\multicolumn {11} {c} {Active Giants}     \\
\hline
3229      &14 Cet  & 1.55$\pm$0.1           & HGap          & 0.936	 & 0.68	                 & 0.94           &	1.52	      & 0.94        & 0.27               & 0.97\\

4128  &$\beta$ Cet & 3.5$\pm$0.2            &He burning      & 0.504    & 121                   & 0.56          & 232                &	0.51         &  53                & 0.75 \\
    & 	   & 3.6$\pm$0.2            & Base RGB        & 0.47     & 135                   & 0.53	         & 246                & 0.48      &  59                & 0.73 \\

9746  &OP And	   & 2.0$\pm$0.5            &  RGB (Li)     & 0.044    & 25                    &  0.06         & 160                & 0.79       &  118               & 0.52  \\
     &		   & 2.0$\pm$0.5          &He burning    & 0.183    & 106                   &  0.23         & 179                & 0.45        &  80                & 0.60\\

27536 &EK Eri      & 1.9$\pm$0.3            & Base RGB  & 0.507    & 103                   & 0.56          & 234                & 0.51        & 43                 & 0.75 \\

28307 &77 Tau      & 3.$\pm$0.2             & Base RGB      & 0.553    & 95                    & 0.62          & 179                & 0.57        & 41                 & 0.79 \\

31993 &V1192 Ori   & 1.8$\pm$0.5           & RGB (Li)      & 0.054    & 32	                 & 0.07	 	 & 129	              & 0.36        & 107                & 0.53 \\  
      &            & 1.7$\pm$0.5           &He burning        & 0.244    & 99	                 & 0.32	 	 & 190	              & 0.27        & 62                 & 0.68\\  
      &            & 2.$\pm$0.5            &  RGB          & 0.059    & 37                    & 0.08          & 145                & 0.34        & 116                & 0.53\\       
		      
33798 & V390 Aur   & 2.25(*)$\pm$0.3       &  Base RGB       & 0.477	 & 119                   & 0.56          & 230                & 0.50        & 49                 & 0.77 \\
47442 & nu3 CMa	   & 4.5$\pm$0.5           &  Base RGB      & 0.415    & 170                   & 0.48          & 314                & 0.43       & 78                 & 0.71\\
      &            & 4.0$\pm$0.5           &He burning       & 0.401    & 160                   & 0.46          & 300                & 0.41       & 74                 & 0.71\\                
68290 & 19 Pup     & 2.5$\pm$0.2           & Base RGB        & 0.483    & 118                   & 0.57          & 239                & 0.51        & 49                 & 0.77\\
      &            & 2.5$\pm$0.2           & He burning        & 0.406    & 162	                 & 0.47		 & 335	              &0.41        & 67                 & 0.71\\

72146 &FI Cnc      & 2.4$\pm$0.3           &  Base RGB   & 0.608	 &  65		         & 0.68		 & 122	              & 0.63        & 28                 & 0.82 \\

82210 & 24 UMa     & 1.9$\pm$0.1           & Base RGB        & 0.625	 & 55		         & 0.67		 & 95	              & 0.63        & 24                 & 0.81 \\

85444 & 39 Hya     & 3.75(*)$\pm$0.2       & BaseRGB      & 0.639    & 59                    & 0.69           & 101                & 0.65       & 26                 & 0.83 \\

111812 & 31 Com    & 2.75(*)$\pm$0.1       & HGap        & 0.745    & 34                    & 0.75          & 58                 & 0.74        & 15                 & 0.85 \\

112989 & 37 Com    & 5.25$\pm$0.4          &He burning  & 0.977     & 91                     & 61         & 154                 & 0.57       & 38                 & 0.78  \\
       &           & 5.25$\pm$0.4          & Base RGB      & 0.527     & 124                   & 0.58          & 224                & 0.53        & 54                 & 0.76  \\

121107 & 7 Boo     & 4.0$\pm$0.7            & HGap            & 0.729	 & 34		         & 0.75 	 & 63		      & 0.72       & 14                 & 0.86 \\

141714&$\delta$ CrB& 2.5$\pm$0.1           & Base RGB            & 0.457	 & 133	                 & 0.52		 & 271	              & 0.47        & 56                 & 0.74 \\

145001&$\kappa$ HerA& 3.5$\pm$0.3          & Base RGB      & 0.654	 & 50		         & 0.71         & 85		      & 0.66       & 22                 & 0.83 \\

150997 &$\eta$ Her  & 2.5$\pm$0.3          & Base RBG         & 0.512	 & 110	                 & 0.58		 & 217	              & 0.53        & 47                 & 0.77\\
       &            & 2.7(*)$\pm$0.3       & He burning     & 0.456	 & 126	                 &  0.52         & 244	              & 0.46        & 54                 & 0.73\\

163993 &$\xi$ Her   & 3.$\pm$0.3           & Base RGB    & 0.570	 & 87		         & 0.63		 & 161 		      & 0.58         & 37                 & 0.80 \\
       &	    & 2.5$\pm$0.3	   & He burning   & 0.451	 & 119	                 & 0.52		 & 209	              & 0.47        &  51                & 0.73 \\	

203387 &$\iota$ Cap & 3.$\pm$0.2           & Base RGB       & 0.663	 & 48		         & 0.71		 & 79		      & 0.67        & 21                 & 0.84 \\

205435  &$\rho$ Cyg & 2.6$\pm$0.2          & Base RGB      & 0.449	 & 141	                 & 0.51		 & 289	              & 0.45        & 59                 & 0.73 \\

218153  &KU Peg	    & 2.5$\pm$0.2          & Base RGB     & 0.562	 & 85		         & 0.65		 & 165	              & 0.60        & 36                 & 0.82  \\

223460  &OU And	    & 2.8$\pm$0.2           & HGap             & 0.755	 & 15		         & 0.83		 & 36		      & 0.76        &  9                 &  0.88\\

%232862  &	    & (2.0)        & Base RGB        & 0.462     & 129                   & 0.52          & 252                & 0.46        &  52                & 0.73 \\

\hline
\multicolumn {11} {c} {CFHT \& miscellaneous} \\
\hline 
9270    &$\eta$ Psc & 4.9$\pm$0.5          & Base RGB    & 0.640	 & 67 	                 & 0.68 	 & 121	              & 0.644        & 28               & 0.820   \\

28305 &$\epsilon$ Tau& 3.0$\pm$0.2         & Base RGB        & 0.387	 & 168	                 & 0.45		 & 335	              & 0.39         & 73               & 0.69  \\
%        &            &   3.0$\pm$0.2       &He burning   & 0.381	 & 168	                 & 0.45		 & 292	              & 0.39         & 72               & 0.68 \\ 
	&            & 2.5$\pm$0.2	   &He burning   & 0.360	 & 146	                 & 0.43		 & 268	              & 0.37         & 67               & 0.69 \\
	    
29139  &Aldebaran    & 1.7$\pm$0.5	   & AGB	& 0.048 & 31 & 0.06 & 284 & 0.98 & 104 & 0.53 \\
       &	     & 2.0$\pm$0.5         & RGB          & 0.061      & 40   		 & 0.08		 & 143	              & 0.35         & 114              & 0.54 \\

62509  & Pollux	     & 2.5$\pm$0.3         & Base RGB     & 0.382	 & 162	                 & 0.45		 & 330	              & 0.39         & 68               & 0.68 \\
       &             & 2.5$\pm$0.3         &He burning   & 0.366      & 183	                 & 0.43		 & 391	              & 0.36         & 75               & 0.68 \\
        
81797  & Alphard     & 3.5$\pm$0.5         & AGB        & 0.064      & 59		         & 0.07		 & 768	              &0.93          & 142              & 0.49 \\

124897 & Arcturus    & 1.5$^{(a)}\pm$0.3         & AGB         & 0.095      & 50	                 & 0.12	         & 170                & 0.98         & 72             & 0.55  \\
       &             & 1.5$^{(a)}\pm$0.3         & RGB     & 0.034      & 151		         & 0.04		 & 137	              & 0.95         & 94              & 0.52 \\

\hline
\hline                           
\end{tabular}
\tablefoot{For each star we give: Stellar mass, evolution phase, radius at the base of the convective envelope, and convective turnover times at different locations within the convective envelope. $M_*$ marked (*) indicates that the values are the result of the linear interpolation between two models. For some stars we indicate two possible evolution states ('RGB(Li)' means that we use the Li abundance as an additional indicator).  The mass of Arcturus$^{(a)}$ is determined based on models computed at the metallicity of this star ([Fe/H]=-0.6; Lagarde et al. 2012).}
\end{table*}

\subsection {Predicted convection turnover time for the Zeeman detected giants; Rossby number for giants with known rotational period}
\label{subsect:predictionconvectionRossby}

We use our rotating stellar evolution models to infer the theoretical convective turnover times of our sample stars as well as the Rossby number for the stars with known surface rotational period. Before we analyze the results, we recall the main assumptions made for rotation in the model computations. 

\subsubsection{The rotating models}

The stellar models used in this study (Charbonnel \& Lagarde 2010; Charbonnel et al. in prep.) include thermohaline mixing and rotation-induced processes; we refer to Lagarde et al. (2012) for details on the input physics. 
Initial rotation velocity on the zero age main sequence is chosen at 45\% of the critical velocity at that stage for the corresponding stellar mass; this corresponds to the mean value in the observed distribution of low-mass and intermediate-mass stars in young open clusters (e.g. Huang et al. 2010). 
The evolution of the internal angular momentum profile and of the surface velocity is accounted for with the complete formalism developed  by Zahn (1992), Maeder \& Zahn (1998), and Mathis \& Zahn (2004). 
Rotation in the convective envelope is considered as solid, the rotational period at the surface of the star being that at the top of the radiative zone. 
Note that magnetic braking following the Kawaler (1988) prescription is applied for masses below or equal to 1.25~M$_{\odot}$ on the main sequence, but no magnetic braking is assumed in the following evolution phases nor for the more massive stars. 
Additional models including magnetic braking after the main sequence turnoff will be presented by Charbonnel et al. (in prep.) where predictions for the rotation periods will be compared to the observed periods of our sample stars (when available). 

\subsubsection{Theoretical turnover timescales and semi-empirical Rossby numbers}

Convective turnover timescale and Rossby number are important quantities to infer magnetic activity and dynamo regime. Figure 5 shows the variations of the maximum convective turnover time\footnote{The local convective turnover time at a given radius $r$ inside the convective envelope is defined as $\tau(r) = \alpha H_p(r)/V_c(r)$. $H_p(r)$ and $V_c(r)$ are the local convective pressure and velocity scale height. In our models, $\alpha=1.6$.} $\tau(max)$ in the convective envelope along the evolutionary tracks for different masses. 
We see that for all stellar masses, $\tau(max)$ increases when the stars move towards lower effective temperature across the Hertzsprung gap up to a maximum value at about the middle of the first dredge-up, i.e., at the base of the RGB.  
$\tau(max)$ then decreases when the stars climb along the RGB, before increasing again when the stars settle in the central He-burning phase. Although not shown here, $\tau(H_p/2)$ computed at half a pressure scale height above the base of the convective envelope, and the convective turnover time at half radius within the convective envelope $\tau(R/2)$ follow the same behavior along the evolution tracks. These results are discussed by Charbonnel et al. (in prep.). 

For each detected giant we present in Table 4 the stellar mass and evolutionary status derived as described in \S~5.1, as well as the theoretical values (as predicted by the relevant model or interpolated between tracks of different masses) for the radius at the base of the convective envelope, for the convective turnover time at different locations in the convective envelope (at the $H_p/2$ level, at the half convective envelope radius, and the maximum value within the convective envelope)  as well as the corresponding radii in $R_*$. 
Table 4 shows that $\tau(max)$ is found very near the base of the convective envelope for stars in the HGap, Base RGB and He burning phases. It is found higher in the convective envelope for stars ascending the RGB.

For giants with known rotational periods, we compute a semi-empirical Rossby number $Ro$, defined as the ratio of the observed $P_{\rm rot}$ and the maximum $\tau$ value, $\tau(max)$, within the convective envelope (Table 5). 
$Ro$ values are discussed in  \S~6.1.2%\S~\ref{subsubsection:Ro}. 

\begin{table*}
\caption{Rossby number for stars with measured $P_{\rm rot}$.}          
\label{table:2}   
\centering                         
\begin{tabular}{l l c c c c c c}     
\hline\hline  

HD       &Name     &obs $P_{\rm rot}$& $M$        & Branch   & $R$    & $\tau(max)$& $Ro$ \\
         &         &  day       & $M_{\odot}$ &          & $D_{CE}$ &    day     &       \\
\hline
\hline
\multicolumn {8} {c} {Active Giants}      \\
\hline
4128    &$\beta$ Cet&215       & 3.5        & He burning& 0.01   & 232         & 0.93   \\
9746    &OP And    &  76       & 2.         & RGB       & 0.78   & 160         & 0.47   \\
27536   &EK Eri    & 308.8     & 1.9         & Base RGB & 0.01   & 234         & 1.3  \\
28307   &77 Tau    & 140       & 3.         & Base RGB  & 0.03   & 179         & 0.78 \\
31993   &V1192 Ori &  28       & 1.8        & RGB       & 0.32   & 129         & 0.22 \\
33798   &V390 Aur  &  9.8      & 2.25       & Base RGB  & 0.04   & 230         & 0.04 \\
47442   &nu3 CMa   & 183       & 4.5        & Base RGB  & 0.02   & 314         & 0.58 \\
68290   &19 Pup    & 159       & 2.5        &He burning & 0.07   & 335           & 0.47 \\
72146   &FI Cnc    &  28.5     & 2.4        & Base RGB  & 0.06   & 122           & 0.23 \\
111812  &31 Com    &  6.8      & 2.75       & HGap      & 0.02     & 58            & 0.12 \\
112989  &37 Com    &  111      & 5.25       &He burning & 0.     & 132           & 0.84 \\
141714  &$\delta$ CrB&59       & 2.5        & HGap      & 0.01   & 271           & 0.22 \\
203387  &$\iota$ Cap& 68       & 3.         & Base RGB  & 0.02   & 79            & 0.86 \\
218153  &KU Peg    &  25       & 2.5        & Base BRG  & 0.08   & 165           & 0.15 \\
223460  &OU And    &  24.2     &3.          & HGap      & 0.01   & 36            & 0.68 \\
%232862  &          &  5.34     & 2.         & Base RGB  & 0.     & 252           & 0.02  \\
\hline
\hline
\multicolumn {8} {c}  {CFHT \& miscellaneous}   \\
\hline
62509   &Pollux     & 590      & 2.5        &Base RGB   &  0.01  & 330           & 1.78    \\
\hline
\hline                          
\end{tabular}
\tablefoot{The radius $R$ (6$^{\rm th}$ column) where $\tau(max)$ is measured, is counted above the base of the convective envelope (CE) and is in units of the depth of the CE.}
 \end{table*}

\section {Analysis of the data of the detected giants}

\subsection {A relation between the strength of the magnetic field and the stellar rotation}

\subsubsection {The strength of the magnetic field with respect to the rotational period}

As discussed in \S~2.3, $P_{\rm rot}$ has been determined for 16 stars of our sample (apart from Pollux, they are all in the 'Active Giants' subsample). Figure 6 shows the variations of the strength of the magnetic field $|B_l|_{\rm max}$ (as defined in \S~4.2.2) as a function of the observed rotational period in a log/log scale for this subsample. The straight line is the least squares regression, excluding EK Eri which is known to be overactive with respect to its rotational period (e.g. Auri\`ere et al. 2008), and our three faster rotators for which $|B_l|_{\rm max}$  is significantly smaller than $B_{\rm mean}$ (Table 2 and see next Section).
The overall fit is good, with a regression index of -0.83. Ten stars are very close the regression line. Therefore, Fig. 6  shows clearly that there is a rather tight relation between magnetic field strength and rotational period in the range of 7 - 200 days. This indicates that the majority of the stars classified as 'Active Giants' obey to the same $|B_l|_{\rm max}$-$P_{\rm rot}$ relation, that the strength of their magnetic fields depends on rotation, and that the origin of their magnetic field should be the same. We also identify several outliers: these are discussed in the next Section.

\begin{figure}
\centering

\includegraphics[width=9 cm,angle=0] {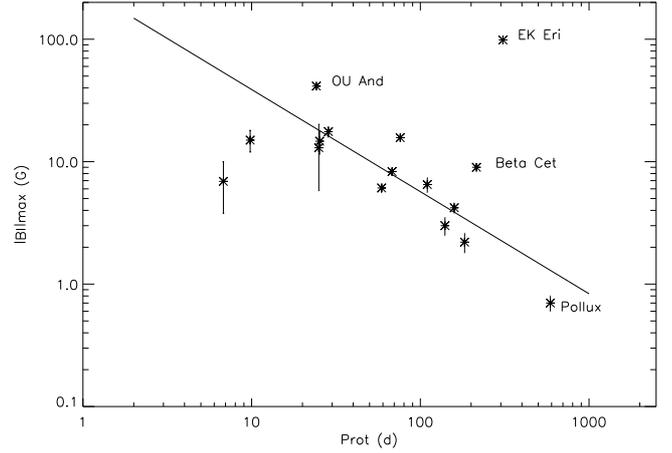}

\caption{Correlation of the strength of the magnetic field ($|B_l|_{\rm max}$ in G) with the rotational period (in days). The positions of Pollux and of 3 possible descendants of magnetic Ap stars are shown.}
%\label{f1}
\end{figure}  

\subsubsection {The strength of the magnetic field with respect to the Rossby number}

To better understand the dynamo regime which likely causes the relation between the magnetic field strength and the rotation, we plot in Fig. 7 $|B_l|_{\rm max}$ as a function of the semi-empirical Rossby number $Ro$ (see \S~5.2.2) for the same 16 giants with known $P_{\rm rot}$. We use $\tau(max)$ as the most representative quantity of the dynamo that might operate at different depths within the convective envelope of giant stars. 
In these conditions, the Rossby number spans mainly between 0.04 and 1 and we obtain a satisfactory correlation (regression index of -0.68 in logarithmic coordinates) excluding the same 4 giants as in  \S~6.1.1. 
This indicates that an $\alpha - \omega$ type dynamo probably operates in these evolved stars with $P_{\rm rot}$ shorter than 200 day, as predicted by Durney \& Latour (1978). However, due to its high $Ro$ of 1.8, an $\alpha - \omega$ type dynamo appears unlikely for Pollux (Auri\`ere et al. 2014a and in prep.).

 Some stars deviate from the relationship and  deserve special comments. EK Eri appears completely out of the plot. This illustrates its status as the archetype of giants descended from magnetic Ap stars ($P_{\rm rot}$ = 308.8 d, $|B_l|_{\rm max}$= 98.6 G, Auri\`ere et al. 2012). OU And appears also in a similar situation with a very strong $|B_l|_{\rm max}$ of 36 G for a $P_{\rm rot}$ of 24.2 d. It is worth comparing OU And with 31 Com, since they lie close from each other within the Hertzsprung gap and have similar mass (about 2.8 $M_{\odot}$) but very different rotation periods.  With its small $P_{\rm rot}$ and small $Ro$, 31 Com illustrates a fast rotator. OU And has a period 3 times longer and an Rossby number 6 times greater, but a stronger $|B_l|_{\rm max}$ and in general is as magnetically active as 31 Com. The Ap star descendant hypothesis to explain its high activity is therefore very likely (Borisova et al. in prep.). $\beta$ Cet, with a period of 215 d, Rossby number of 0.93 and $|B_l|_{\rm max}$ of 9 G, is well off the relation. This star is also consistent with the Ap star descendant hypothesis, as proposed by Tsvetkova et al. (2013). 

Looking to both Fig. 6 and Fig. 7, one may suspect that some saturation of the dynamo occurs for our two faster rotators ($P_{\rm rot} <$ 20d or $Ro <$ 0.1). In this case, we should consider saturated and non-saturated activity-rotation relations as used for X-ray studies by e.g. Wright et al. (2011). Actually, since we excluded our fastest rotators from the regression, the straight lines presented in Fig. 6 and Fig. 7 represent the unsaturated regime. We can then compare our results to that obtained recently by Vidotto et al. (2014) concerning main sequence stars. These authors find relations similar to ours between $B_{\rm mean}$, which is inferred by ZDI as described in Sect. 4.2.2, and rotation. We found in Sect. 4.2.2 that  $B_{\rm mean}$ is similar to our $|B_l|_{\rm max}$ for stars with $v \sin i$ $\leq$ 11 km$s^{-1}$  which corresponds to our slowest rotators. The rotators studied by Vidotto et al. 2014 have $P_{\rm rot}$ smaller than 43 days, i.e. they are in general faster rotators than our giants (see Fig. 1). On the other hand, since the convective envelope of our giants is fairly deep and the maximum convective turnover time $\tau(max)$ high, our $Ro$ range has a significant overlap with theirs. Ultimately, the  logarithmic slopes (or coefficients of the power law in linear units) are found to about 1.6 times steeper (larger) for the dwarfs than for the giants. This may due to the fact that $P_{\rm rot}$ is generally shorter for the dwarfs, and their magnetic fields are much stronger, which corresponds to a different regime, and implying that our $|B_l|_{\rm max}$ cannot be directly compared to their $B_{\rm mean}$.

\begin{figure}
\centering

\includegraphics[width=9 cm,angle=0] {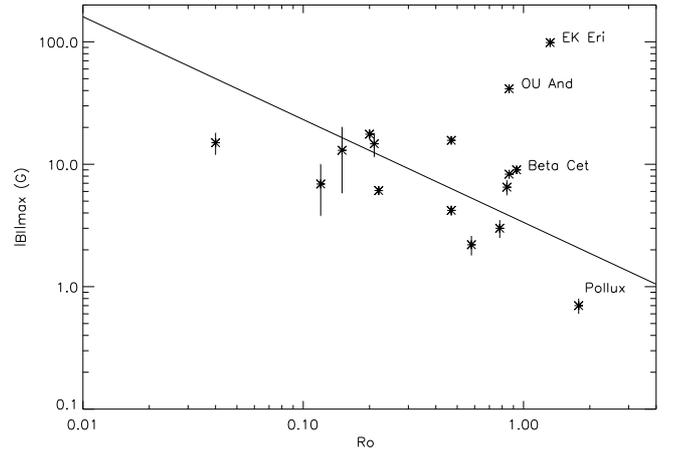}

\caption{Correlations of the strength of the magnetic field ($|B_l|_{\rm max}$ in G) with the Rossby number. The positions of Pollux and of 3 possible descendants of magnetic Ap stars are shown.}
%\label{f1}
\end{figure}

\subsection {The S-index (Ca~{\sc ii} $H \& K$ emission) as a measure of magnetic activity}

The measurement of the flux of the chromospheric Ca~{\sc ii} $H \& K$ emission is a classical proxy of the magnetic activity of cool stars (e.g. Strassmeier et al. 1990, Young et al. 1989, Pasquini et al. 2000). We choose here to use the $S$-index defined by the Mount Wilson $H K$ survey (Duncan et al.1991) as described in \S~4.3. Figure 8 presents the distribution of the observed $S$-index for the 48 stars of our sample: the upper plot shows, for the detected stars, the $S$-index at the date of $|B_l|_{\rm max}$; the lower plot shows, for the non detected stars, the maximum value observed for the $S$-index. Figure 8  shows that some rather strong values are measured for the most active stars, but that about half of the whole sample (26 of 48 giants) consists of stars with $S$-index smaller than 0.2. It also shows that even stars with detected magnetic field can have a very weak $S$-index. The relation with the basal chromospheric flux will be discussed in \S~6.2.3. Figure 8 also shows that almost all giants with $S$-index greater than 0.2 are detected, a result similar to that obtained by Marsden et al. (2014) for dwarf stars, but with $S$-index greater than 0.3. Actually, Schr\"oder et al. (2012) show that although exhibiting the same basal flux, giants show a systematic offset (towards weaker values) in $S$-index as compared to main sequence stars, because their photospheric spectral properties are gravity-sensitive, which may explain the result reported above.

\begin{figure}
\centering
\includegraphics[width=9 cm,angle=0] {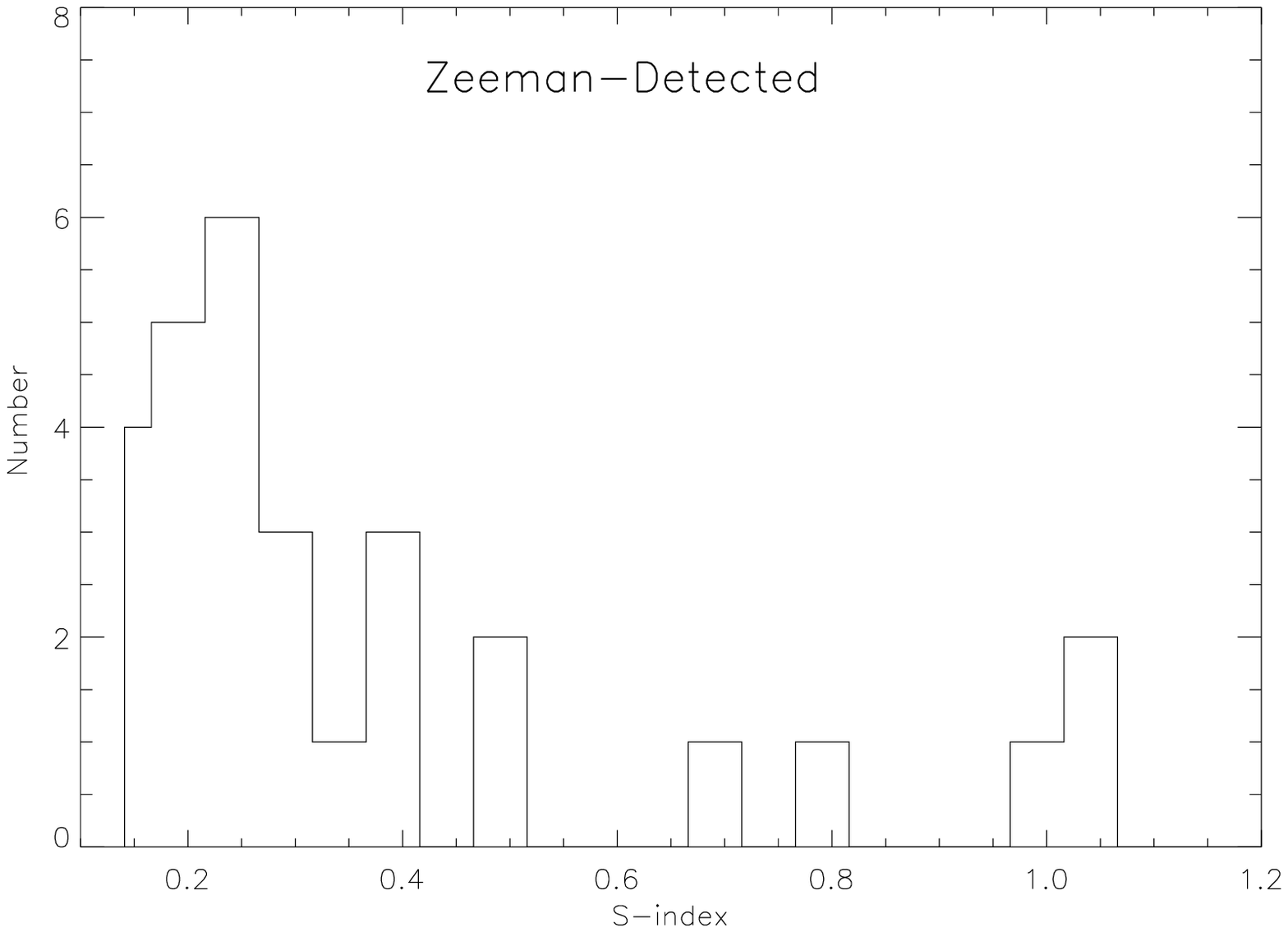}
\includegraphics[width=9 cm,angle=0] {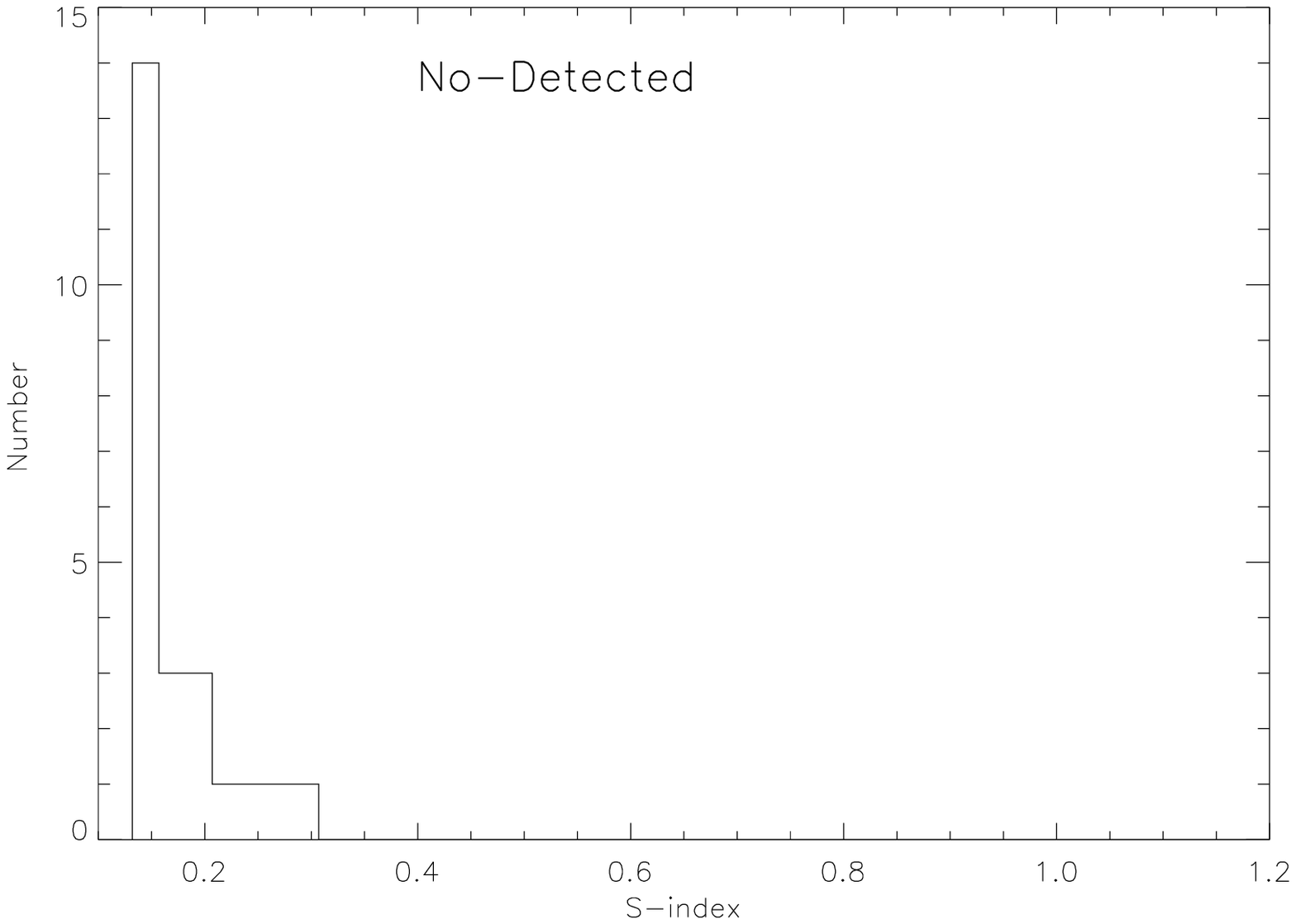}
\caption{Distribution of the S-index: upper plot, detected giants; lower plot, non-detected giants.}
%\label{f1}
\end{figure}

\subsubsection {Variations of the $S$-index with the rotational period}

Figure 9 shows a significant correlation (index of -0.90)  between $S$-index measured at $|B_l|_{\rm max}$ and the rotational period. As for Fig. 6 and correlation of $|B_l|_{\rm max}$ vs. $P_{\rm rot}$, EK Eri and the 3 faster rotators of our sample (31 Com, V390 Aur and OU And) were not considered for the linear regression. A similar relationship between $S$-index and $P_{\rm rot}$ for single and binary stars (both dwarfs and RS CVn stars) was found and used by Young et al. (1989) to predict the rotational period of active red giants. However, even if these predicted periods can be used further (e.g. Gondoin 2005b), they can lead to erroneous rotational periods as in the case of $\beta$ Ceti (80 d predicted and 215 d measured, Tsvetkova et al., 2013).

\begin{figure}
\centering

\includegraphics[width=9 cm,angle=0] {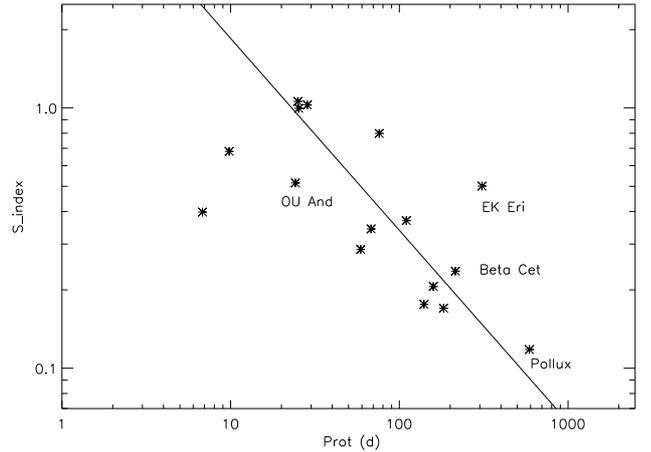}

\caption{Correlations of the S-index with the rotational period (in days). The positions of Pollux and of 3 possible descendants of magnetic Ap stars are shown.}
%\label{f1}
\end{figure}

\subsubsection {Correlation between the S-index and the strength of the magnetic field}

Figure 10 shows for the first time for single giants the correlation of the $S$-index with the strength of the magnetic field. The $S$-index is the only activity indicator used in this paper which can be determined for each of our 29 detected giants. 
For about twenty giants, there is a good correlation between $S$-index and $|B_l|_{\rm max}$. These stars include all the stars with determined  $P_{\rm rot}$ which already fulfilled the relation between $|B_l|_{\rm max}$ and $P_{\rm rot}$. As discussed in  \S~6.1, the observed magnetic field of these giants is concluded to be dynamo-driven. In this case several scales of magnetic field can exist, all contributing to the heating of the chromosphere, while the small scales which are with opposite polarities can cancel each other, reducing the observed averaged longitudinal field. However, giants with $|B_l|_{\rm max}$ greater than 20 G deviate from this relationship, as well as those with $|B_l|_{\rm max}$ weaker than 1 G. 

The outliers with strong magnetic fields are 3 stars that we consider to be possible descendants of magnetic Ap stars (EK Eri, OU And and 14 Cet; see \S~7.2.2).
 In the case of Ap star descendant fossil fields, the large scale dipole dominates, resulting in a simple 
topology and smaller contribution of the small scale elements to the chromospheric heating and $S$-index. This leads to comparatively higher $B_\ell$ measurements for a given $S$-index than for a dynamo-driven field.
Figure 10 shows that the $S$-index saturates at values below about 1, while $|B_l|_{\rm max}$ does not saturate. 

Among the stars with weak $|B_l|_{\rm max}$, Aldebaran, which has a weak, intermittently detectable magnetic field, has an $S$-index as large as those of stars with $|B_l|_{\rm max}$ of a few G.  On the other hand, Pollux, which has one of the smallest measured $S$-indices, has a consistently detectable magnetic field at the sub-G level. A Zeeman Doppler imaging study of Pollux (Auri\`ere et al. 2014a and in prep.) shows that its magnetic topology is dominated by a magnetic dipole, i.e. a large scale structure. One may alternatively suggest that the intermittently detectable magnetic field of Aldebaran is composed of small scale magnetic regions of opposite polarities which all contribute to the $S$-index but cancel each other's contributions to $B_l$. These small scale magnetic regions could also exist in the other 5 giants with sub-G $|B_l|_{\rm max}$ to explain why they have a stronger $S$-index than expected by the relationship drawn in Fig. 10.

\begin{figure}
\centering

\includegraphics[width=9 cm,angle=0] {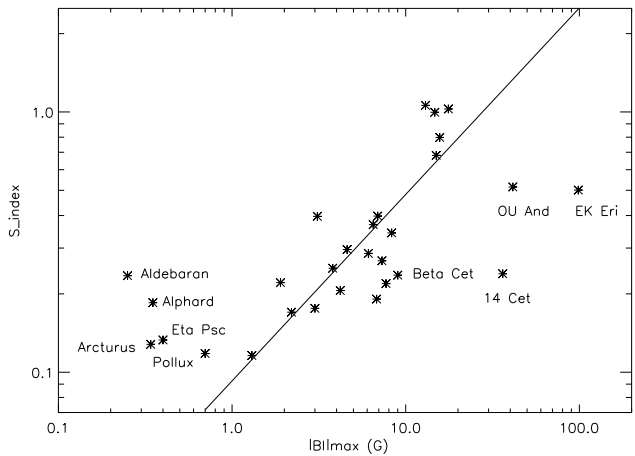} 

\caption{Variations of the $S$-index with the strength of the magnetic field ($|B_\ell|_{\rm max}$ in G). The positions of 4 possible descendants of magnetic Ap stars are shown, as well as those of 5 giants with sub-G $|B_l|_{\rm max}$.}
%\label{f1}
\end{figure}  

\subsubsection {The basal chromospheric flux and the minimum activity level}

The smallest measured $S$-indices ($<$ 0.2) in our sample of giants correspond to the basal chromospheric flux observed on red giants (Hall 2008, Schr\"oder et al. 2012). For all the stars measured in this study except 14 Ceti, an emission core was found for Ca~{\sc ii} $H \& K$ lines. Figure 11 illustrates the case of $\mu$ Peg, which is among the stars with the smallest $S$-index measured in this study. Two giants with $S$-index about the smallest observed in this study were detected (Pollux and $\epsilon$ Tau). Their $S$-index of about 0.12 could correspond to the basal chromospheric flux as described by Schr\"oder et al. (2012), as well as their magnetic strength of 0.7 and 1.3 G respectively, if it is of magnetic origin only. Obviously further observations to verify the temporal behaviour of these cases are of much interest, as these may help to answer the long-standing question of the nature of the basal flux energy source and the role (if any) of magnetic field, and whether it is created by a local dynamo (V\"ogler \& Sch\"ussler 2007) or by a global dynamo.

\begin{figure}
\centering

\includegraphics[width=6 cm,angle=-90] {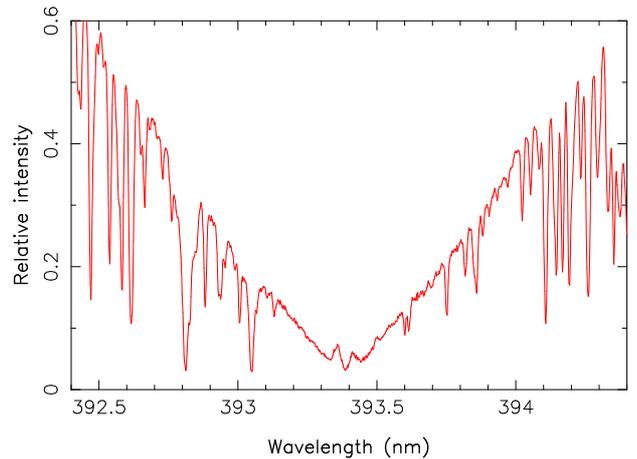} 

\caption{Spectrum of $\mu$ Peg on 20 September 2008 showing the emission core of the Ca~{\sc ii} $K$ line.}
%\label{f1}
\end{figure}  

\subsection {X-ray luminosity, in relation to $P_{\rm rot}$ and $|B_l|_{\rm max}$}

All the 'Active Giants' of Table 1 and $\nu$ Hya are X-ray emitters and are included in the ROSAT All-Sky Survey catalogues of H\"unsch et al. (1998a,b) and Voges et al. (1999). In addition, 2 stars of the sample were X-ray detected by long exposure, pointed ROSAT observations (Pollux: H\"unsch et al. 1996, Schr\"oder et al. 1998; $\epsilon$ Tau, Collura et al. 1993).  We first study the properties of the 25 giants which are detected in both X-rays and magnetic field. 

Figure 12 shows the histogram of the X-ray luminosities $L_{\rm x}$ for the Zeeman detected giants, which is peaked at high values, with a small tail towards weak values. 

Figure 13 shows the variations of $L_{\rm x}$ with respect to the rotational period for the 16 X-ray detected giants of our sample with determined $P_{\rm rot}$. These giants have been all magnetically detected in our work. Figure 13 shows a general trend with $L_{\rm x}$ decreasing with increasing $P_{\rm rot}$, with a correlation index of -0.71. Pollux is the main outlier of this plot. Gondoin (1999, 2005b) performed a study of X-ray emission and rotation for active red giants. He remarked that in general active red giants are more luminous in X-rays than expected by the classical Pallavicini et al. (1981) relation: $L_{\rm x} \approx   (v \sin i)^2  10^{27}$ erg\, s$^{-1}$. Actually, Pollux, with $v \sin i$ of about 1 km\,s$^{-1}$, has the $L_{\rm x}$ as expected from this relation, and is an outlier with weak $L_{\rm x}$ in Fig. 13. Gondoin (2005b) has found that for his sample of intermediate mass G giants that the X-ray surface flux decreases linearly (in log scale) with the rotational period, which is also consistent with the trend of our Fig. 13. However, the rotational periods employed in that study come mainly from Young et al. (1989) and are predicted, rather than measured, periods. Therefore Gondoin's work is biased towards a relation between X-ray emission and chromospheric emission (Ca~{\sc ii} $H \& K$).

In the present work, for the first time, the X-ray luminosity can be compared to a measurement of the magnetic field strength. Figure 14 shows the variations of $L_{\rm x}$ with $|B_l|_{\rm max}$. The straight line is the linear least-squares regression obtained, in log-log scale, using all the detected X-ray stars apart from the 3 stars with $|B_l|_{\rm max}>20$~G (3 Ap star descendant candidates). The main outliers of this plot are the 3 strongest magnetic stars and $\eta$ Her.

All the stars detected in the present work were known as X-ray emitters but 4: Aldebaran, Alphard, Arcturus and $\eta$ Psc.  Aldebaran could not be detected even with Chandra (Ayres et al., 2003), but a tentative detection of Arcturus was obtained ($L_{\rm x}$  about 1.5 $10^{25}$ erg $s^{-1}$, Ayres et al. 2003). Aldebaran and Alphard are the coolest detected giants of our sample.  Ayres et al. (2003) have shown that several reasons may explain this null result, and the X-ray dividing line was revised by H\"unsch \& Schr\"oder (1996).

As to the 3 X-ray detected giants of our sample which are not Zeeman detected ($\beta$ Boo, $\nu$ Hya and $\mu$ Peg), the geometry of the field at the time of the spectropolarimetric observations may explain the present null result.

\begin{figure}
\centering

\includegraphics[width=9 cm,angle=0] {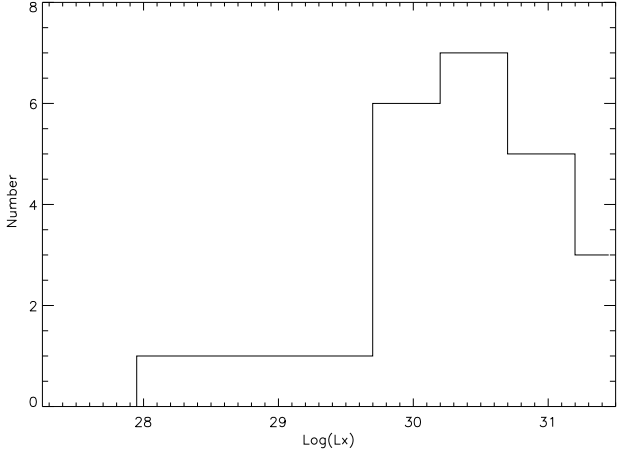}
\caption{Distribution of the X-ray luminosity $\log L_{\rm x}$ (in units of $10^{27}$ erg\,s$^{-1}$).}
%\label{f1}
\end{figure}  

\begin{figure}
\centering

\includegraphics[width=9 cm,angle=0] {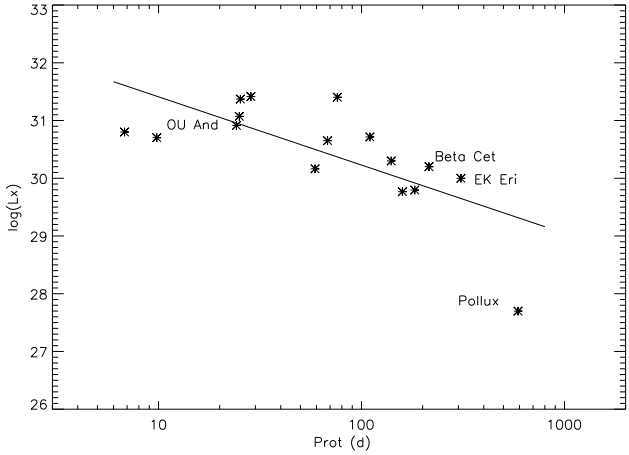}

\caption{Correlations of the X-ray luminosity $\log L_{\rm x}$ (in units of $10^{27}$ erg\,s$^{-1}$) with the rotational period (in days). The positions of Pollux and of 3 possible descendants of magnetic Ap stars are shown.}
%\label{f1}
\end{figure}

\begin{figure}
\centering

\includegraphics[width=9 cm,angle=0] {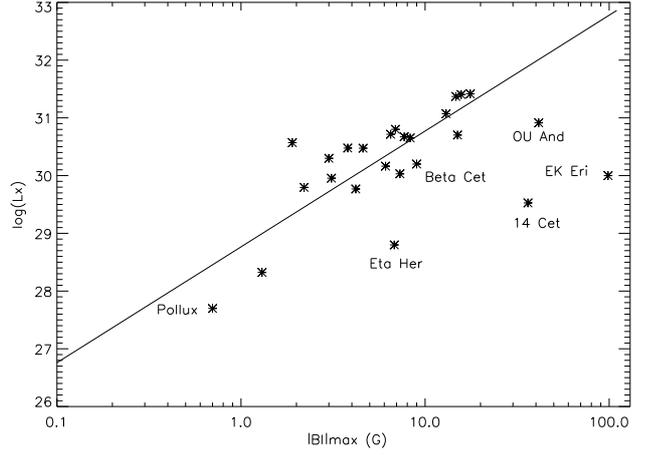}

\caption{Correlations of the X-ray luminosity $\log L_{\rm x}$ (in units of $10^{27}$ erg\, s$^{-1}$) with the strength of the magnetic field ($|B_l|_{\rm max}$ in G). The positions of Pollux, $\eta$ Her and of 4 possible descendants of magnetic Ap stars are shown.}
%\label{f1}
\end{figure}

\section {Main results and discussion}

 The most important new results from this work come from the direct detection of magnetic fields in 29 red giants. Among the 48 stars of our sample are 24 giants already known to present indirect signs of activity. Zeeman effect is detected in 23 of them, demonstrating conclusively that the indirect activity indicators are associated with magnetic fields. Among the Zeeman detected sample, 16 stars have a determined rotational period, $P_{\rm rot}$. We show in \S~6.1 that a relation exists between the magnetic strength and rotation for a majority of them. We discuss different possible origins of magnetic fields in the following subsections. Rotationally induced (dynamo) fields are discussed in Sect 7.1. Three of the outliers are identified as probable descendants of magnetic Ap stars. These cases and the incidence of such stars are discussed in \S~7.2. Finally, the ESPaDOnS/CFHT snapshot program of bright giants led to the detection of weak magnetic fields in 3 objects. A follow up program with Narval was also performed, adding 3 more detections; these results are discussed in \S~7.3. Some properties of the Zeeman detected giants are also given in Appendix A and Appendix B.

\subsection {Stars with rotational periods shorter than 200 days}
\subsubsection{The solar-type dynamo regime}

As discussed in \S~6 we find a good correlation between the magnetic strength and the observed rotational period for a majority of the sample stars with $P_{\rm rot}$ between 7 and 200 days. 
The correlation remains when the semi-empirical Rossby number (defined as $Ro= P_{\rm rot}/\tau(max)$, with observed $P_{\rm rot}$ and theoretical $\tau(max)$; see \S5.2.2) is considered. 
This suggests that the magnetic activity of these stars has the same dynamo origin. 
Since for these stars the semi-empirical Rossby number is smaller than 1, an $\alpha - \omega$ type dynamo could be the origin of their magnetic activity (Durney \& Latour, 1978).

\subsubsection{Evolution considerations}

All our detected stars appear to be located within or very close to the magnetic strip identified in the Hertzsprung-Russell diagram (\S~5.1). 
Most of them are undergoing the first dredge-up (including the three stars that are crossing the Hertzsprung gap and lie just at the beginning of the first dredge-up episode), two are in the core He-burning phase, while the very weak-field stars Alphard, Aldebaran, and Arcturus are probable AGB stars (as indicated in Sect. 5.1 and Table 4, some ambiguity remains concerning the evolutionary status of a few stars with masses greater than 2~$M_{\odot}$).
This is in good agreement with our theoretical stellar models, which predict that the convective turnover time within the stellar convective envelope is highest precisely during these evolution phases (see Fig.~5; we refer to Charbonnel et al. in prep. for more details) reaching 100 to 300 days. As shown in Table 5 for the stars with surface rotational periods from our sample, when $P_{\rm rot}$ is smaller than 200 days, we obtain Rossby numbers smaller than 1 and a dynamo-driven magnetism is expected.
We note that a relation between surface rotation and magnetic field strength may also exist in the case of active M giants (Konstantinova-Antova et al. 2013). 
 Our first results on a magnitude-limited subsample of single cool giants in the Solar vicinity (70\% of the sample observed) revealed evidences for a second 'magnetic strip' corresponding to the bright part of the RGB and to the AGB (Konstantinova-Antova et al. 2014). Analyzing additional observations will allow us to test the existence of this second 'magnetic strip' (Konstantinova-Antova et al. in prep.).

As explained in \S~5.2.1, the rotating models we use are computed assuming an initial rotation velocity on the zero age main sequence that corresponds to the mean value observed for low- and intermediate-mass stars belonging to young open clusters (see Charbonnel \& Lagarde 2010 for details), and no magnetic braking is applied after the turnoff. Therefore and although the general observed behavior of surface rotation as a function of evolution is well reproduced by such models (e.g. Tsvetkova et al. 2013), a detailed comparison between the theoretical and observed rotation periods for individual stars cannot be performed at this stage. Such an investigation will be presented in a forthcoming paper where additional models will be computed for different initial rotation velocities and taking into account magnetic braking.

 In particular, the small number of very active red giants that exhibit high rotation velocities are of interest as already stated in previous studies (e.g. Balanchandran et al. 2000). Indeed, these objects cannot be explained even by the rotators with initial velocity near the critical velocity (e.g. V390 Aur, Konstantinova-Antova et al., 2012). 
Some transfer of angular momentum from the stellar core towards the envelope may be required, as indicated independently by Kepler asteroseismic data for red giants with slowly rotating cores (e.g. Mosser et al. 2012).

\subsection {The descendants of magnetic Ap stars}

\subsubsection {The incidence of the descendants of magnetic Ap/Bp stars}

From studies of magnitude-limited samples, magnetic Ap/Bp stars (Ap stars in this paper) were found to represent about 7\% of the early A and late B type main sequence stars (Wolff 1968, Johnson 2004). Studying a sample of about 3300 main sequence stars of intermediate mass located within 100 pc of the sun, Power et al. (2007) derived the mass dependance of the incidence of the Ap stars: from 0\% at 1.5 $M_{\odot}$ and 1\% at 2 $M_{\odot}$ to about 10\% at 2.5 $M_{\odot}$, the bulk incidence being about 2\%. In the same volume, Masserotti et al (2008) selected 761 evolved stars drawn from the Hipparcos catalogue. Looking to their Hertzsprung-Russell diagram (their Fig. 12) we infer that less than half of their stars are more massive than 1.5 $M_{\odot}$ which represents the lower limit for the mass of Ap stars (Power et al. 2007). If the mass distribution of the giant stars was the same as on the main sequence, about 6 stars of the sample of Masserotti et al. (2008) could be descendants of Ap stars, i.e. inside the 100 pc neighbouring sphere. 

Ap stars host surface magnetic fields which are essentially dipolar (e.g. Landstreet, 1992), with dipole strength greater than about 300 G (Auri\`ere et al. 2007) and a distribution for nearby stars peaking at about 2500 G (Power et al. 2008). One may hypothesize that when an Ap star leaves the main sequence, the magnetic flux is conserved (Stepie\'n, 1993) and therefore the dipole strength decreases as $1/R^2$. Using the models of Charbonnel and Lagarde (2010), we find that for a 3 $M_{\odot}$ star at the base of the RGB the dilution of the surface magnetic field will have reached a factor of about 30, and dipole strength would be expected in the range of a few G to a few tens of G. The dilution will strongly increase along the RGB; at the position of Alphard it will be about 1000, i.e. the dipole strength of the descendant of the most strongly magnetic Ap stars would be only a few G. However, geometrical effects may considerably weaken the field as diagnosed using the longitudinal magnetic field $B_l$.

From these numbers, we expect a few Ap star descendants to exist in the 100 pc sphere, as well as among our active giants sample (some of which are outside this sphere).

\subsubsection {The giants identified as possible descendants of magnetic Ap-stars}

We have proposed 4 stars as being Ap star descendants, which are shown to be outliers on some of our plots in Sect. 6 and whose properties are presented by Auri\`ere et al. (2014b). We list them again below.  EK Eri, identified as such an object by Stepie\'n  (1993) was confirmed by our Zeeman studies (Auri\`ere et al. 2008, 2011) to be the archetype of this class. We then proposed 14 Cet, an outstanding F5IV magnetic star as being an Ap star descendant entering the Hertzsprung gap (Auri\`ere et al. 2012). OU And was identified as such an object in the present work (\S~6.1.2) and by Borisova et al. (in prep.). $\beta$ Cet, a star at the He-burning phase, is a possible candidate for being an Ap star descendant (Tsvetkova et al. 2013). All these candidates are in the mass range where Ap stars are the most frequent (Power et al. 2007) except for 14 Cet: with its mass of $1.55 \pm 0.1$ $M_{\odot}$, it is in the lowest mass part of the range. However, Auri\`ere et al. (2012) show that a dynamo alternative scenario is very unlikely for 14 Cet.

Fig. 6 and Fig. 7 show a rather tight relation between $|B_l|_{\rm max}$ and rotational period or Rossby number. EK Eri and OU And,  for which the rotational periods are known, are outliers on these plots (as well as $\beta$ Ceti, more marginally). The stars 14 Ceti, EK Eri and OU And are also outliers on the plot of $S$-index vs. $|B_l|_{\rm max}$ (Fig. 10), showing that in these stars, the large-scale surface magnetic structure dominates the smaller scales structures. 
For the 3 proposed Ap star descendant giants for which ZDI studies exist (EK Eri, OU And, $\beta$ Ceti) a simple magnetic topology (inclined dipole) is observed, as well as some long term stability of the periodic variations of the magnetic field and activity indicators. For all four stars, the hypothesis of magnetic flux conservation as the radius increases provides reasonable values of the magnetic strength for possible Ap star progenitors. The present investigation shows that an outstanding magnetic strength with respect to other stars of the same class is the most efficient way to detect a candidate being an Ap star descendant.

\subsubsection {Are there as-yet unidentified descendants of magnetic Ap stars in our sample?}

The four active giants identified as Ap star descendants are distributed along the HGap/Base RGB, and at the He burning phase (or base of RGB), which correspond to evolution phases with moderate increases of the stellar radius, and hence a moderate magnetic dipole strength dilution. So far, the interaction of a strong, preexisting magnetic field with energetic convection has been simulated mainly in the solar case. For example, Cattaneo et al. (2003) have predicted that the interaction would evolve from magnetoconvection to pure dynamo regime. Strugarek et al. (2011a,b) show that for a star with one solar mass and a convective envelope representing 30\% of the radius, a fossil dipolar magnetic field deeply buried in the radiative zone will permeate the convective envelope and will be present at the surface of the star. The extension of these simulations to a deeper convective envelope and to more massive stars would be worthwhile. Featherstone et al. (2009) have studied the effects of fossil magnetic fields on convective core dynamos in A-type stars. Their simulations result in a more laminar but stronger dynamo state. Our study of EK Eri (Auri\`ere et al. 2012) suggests that this star undergoes an interaction between a remnant strong dipolar field and deep convection that produces a certain level of field variability.
The 12 stars with known $P_{\rm rot}$ not identified as possible Ap star descendants do not appear as strongly magnetic outliers in any of the relationships presented in \S~6 : we may therefore consider that all potential Ap star descendants with  known $P_{\rm rot}$ have been identified in our sample. In addition, there are no obvious outliers in the $S$-index-$|B_l|_{\rm max}$ or $L_{\rm x}$-$|B_l|_{\rm max}$ relations which have other properties suggesting that they could be Ap star descendants. We therefore suggest that if unidentified Ap star descendants exist in our sample, they correspond to weakly magnetic Ap stars or unfavourable geometries, and/or to evolution phases with large radii and possible large dilution of possible surface fossil fields.

\subsubsection {The thermohaline deviants}

Charbonnel \& Zahn (2007b) suggested that the descendants of Ap stars hosting a strong 
fossil magnetic field should escape 
the thermohaline mixing that occurs at the bump of the red-giant branch 
in $\sim 95 \%$ of low-mass stars (Charbonnel \& Zahn 2007a). 
These 'thermohaline deviants'  could be recognized from abundance anomalies in Li and $^{12}C/^{13}C$ 
with respect to the non-magnetic stars. We selected 7 stars in our sample as THDs, as described in \S~2.1. Their properties are presented Table 1, and the journal of observations is given in Table 7. These stars are more evolved than the bump, are located outside the 100 pc sphere around the sun, and did not present any hint of magnetic activity. None of them is detected during our survey at the level of $B_l$ about 1 G.
This survey therefore does not support the hypothesis of Charbonnel \& Zahn (2007b). However, surface magnetic field of less than 1 G or, more importantly, a fossil magnetic field confined beneath the stellar surface, could have escaped to our survey.

\subsection {Weak magnetic activity}

The snapshot survey of bright giants with ESPaDOnS led to the discovery of very weak magnetic fields ($\le 1$ G) in Pollux, Aldebaran and $\epsilon$ Tau. A follow up with Narval added 3 more Zeeman detections, in Alphard, Arcturus and $\eta$ Psc. Apart from Arcturus (Sennhauser \& Berdyugina 2011), these stars were not considered as magnetic giants before this study. All these weakly magnetic giants (apart from Aldebaran) have very small $S$-index, almost corresponding to the basal chromospheric flux (Schr\"oder et al. 2012). Pollux and $\epsilon$ Tau were detected by pointed observations of ROSAT (H\"unsch et al. 1996, Collura et al. 1993, respectively) at the level of a few $10^{27}$ erg\, s$^{-1}$. %However, $\mu$ Peg detected by ROSAT at 1.1$10^{27}$ erg $s^{-1}$, not known so far as presenting stable RV variations, was not Zeeman-detected in our survey. 
On the other hand, no detection in X-rays was obtained for $\eta$ Psc (ROSAT) which is at the base of RGB, nor Aldebaran (ROSAT and Chandra) and Alphard (ROSAT) which are cooler and ascending the RGB. Arcturus, which is of smaller mass, was not detected in X-rays by ROSAT, but Ayres et al. (2003) obtained a tentative detection with Chandra ( $L_{\rm x}$  about 1.5 $10^{25}$ erg\,s$^{-1}$).
 Only Pollux has an established rotational period, of about 590 d (Auri\`ere et al. 2014a, and in prep.), which is found to be consistent with the period of the radial velocities (Hatzes et al. 2006). Pollux is not a strong outlier in the plots presented in this work, apart from Fig. 13 in which its $L_{\rm x}$ is smaller than expected with respect to its $P_{\rm rot}$.

$\epsilon$ Tau, Aldebaran and $\eta$ Psc are  known to present stable RV variations with periods of respectively 595 days (Sato et al. 2007), 630 days (Hatzes 2008) and 629 days (Hekker et al. 2008). These variations are suspected to be due to a hosted planet for the two former giants, and to pulsations for the latter. In  Appendix A we present observations of $\epsilon$ Tau and Aldebaran spanning  several years. For these 2 stars (as well as for 77 Tau which is also study there) we do not see any clear correlations of $B_l$ with RV or $S$-index. 

The knowledge of the $P_{\rm rot}$ of these stars would allow us to infer the type of the dynamo occurring in them: we would see if all the giants magnetic field strengths follow the same law with respect to rotation. Long-term stability of this rotation is also an important property, as well as the presence (or not) of differential rotation. We obtained all of this information in the case of Pollux, which may be a rare object or more simply the archetype of a class of weakly magnetic G K giants (Auri\`ere et al. 2014a and in prep.).

 Therefore, a survey of a sample of giants not biased towards activity (as in the present study) will be useful to obtain statistics on weak activity on all the HRD branches. Such a project is currently in progress (Konstantinova-Antova et al. 2014). Two of the main results of this new project (to be confirmed when the sample will be completed) are that about 50\% of the giants are magnetic at the level of Pollux or above, and the discovery of numerous magnetic stars located on the upper part of the RGB and on the AGB, hence defining a second 'magnetic strip'.

\section {Conclusions}

We have conducted a spectropolarimetric survey of 48 single red giant stars that included known active giants, thermohaline deviants, as well as bright giants. The results of this work are the following:

- 1) Magnetic fields are unambiguously detected in 29 stars of our sample via the Zeeman effect. This has enlightened our understanding of magnetism along the RGB:

  * The majority of the detected stars are located during the first dredge-up phase and at the core He-burning phase. A few other detected stars are in the Hertzsprung gap or ascending the RGB or the AGB.

  * For the 16 detected stars with a known rotational period, we find robust correlations between the magnetic strength and rotation, namely exponential relations with rotational period ($P_{\rm rot}$) and semi-empirical Rossby number. 

  * The models of Charbonnel et al. (2010, and in prep.) show that during the evolution along the RGB, the convective turnover time $\tau$ is maximum during the first dredge-up phase, when the stars are at the foot of the RGB, and at the core He burning phase. The semi-empirical Rossby numbers for our detected stars with $P_{\rm rot}$ determined from observations are found in the range 0.04-1, which indicate that an $\alpha-\omega$ dynamo could be at work there. 

  * These results reveal a 'magnetic strip' on the RGB (corresponding to the first dredge up + central He-burning phase) where activity is predicted to occur more frequently by the evolutionary models that we used and which is actually observed in the present investigation.

  * We identified 4 stars for which the magnetic field is measured to be outstandingly strong with respect to their rotation or evolutionary status as being probable Ap star descendants.

  * Apart from the 24 giants already known to show evidence of activity, 5 giants not previously known to be magnetic (Pollux, Aldebaran, Alphard, $\epsilon$ Tau, $\eta$ Psc), and Arcturus, were detected. Their surface magnetic field is measured to be equal or weaker than 1~G. 
The only star with a determined period and a sub-G magnetic field strength is Pollux: it may be the archetype of a class of weakly magnetic G K giants.

- 2) For all the stars and all observations, the chromospheric (Mount Wilson survey) $S$-index was measured. For the giants with detected magnetic fields, this proxy is well correlated with $P_{\rm rot}$. The correlation with our measurements of the maximum longitudinal magnetic field strength ($|B_l|_{\rm max}$) is also good apart from those stars with the strongest fields (which we identify as possible Ap star descendants) and the stars with sub-G fields. The weakest values of the $S$-index, which correspond to the basal chromospheric flux (Schr\"oder et al., 2012), are observed both in magnetically detected and non-detected giants. This may suggest that if the magnetic field is responsible for a part of the basal chromospheric flux (Schr\"oder et al., 2012), it is near our detection limit, and that we may have detected it in some giants.

- 3) Twenty-eight stars of our sample were detected in X-rays. All but $\mu$ Peg and $\nu$ Hya were also Zeeman detected. A correlation between X-ray luminosity $L_{\rm x}$ and $P_{\rm rot}$ is inferred.  Pollux is the greatest outlier from the relation.

In the future, a new unbiased sample will be needed to quantitatively evaluate the incidence and systematics of magnetic activity among G and K single giant stars. Such a work is already in progress (Konstantinova-Antova et al. 2014).

\begin{acknowledgements}
       We thank  the TBL and CFHT teams for providing service observing. The observations obtained with Narval in 2008 were supported by the OPTICON trans-national access program. We acknowledge the use of PolarBase at: http://polarbase.irap.omp.eu/. We thank Richard I. Anderson for careful reading of the manuscript and constructive comments. R.K.-A. acknowledges partial support by the Bulgarian-French mobility contract DRILA 01/3 funded by the Bulgarian NSF. GAW and JDL acknowledge support from the Natural Sciences and Engineering Research Council of Canada (NSERC). T.D. acknowledges financial support from the Swiss National Science Foundation (FNS) and from ESF-Eurogenesis, and from the UE Programe (FP7/2007-2013) under grant agr. n$^o$ 267251 'Astronomy Fellowships in Italy' (ASTROFit).  N.A.D. acknowledges the support of Saint Petersburg State University for research grant 6,38.18.2014 and PCI/MCTI (Brazil) grant under the Project 302350/2013-6. WW acknowledges support from the Austrian Science Fond (P17890). MA and CC thank the French National Program of Stellar Physics (PNPS) of INSU CNRS.
\end{acknowledgements}

\clearpage

\onecolumn
\begin{longtable}{llllcccccc}
\caption{Journal of observations of the active giants.} \\
\hline
\hline
HD      &Name      &Instrum.& Date      &HJD        &Detect. & $B_l$ & $\sigma$ &  RV      & $S$-index \\
        &          &        &           &(2450000+) &        &  G       &  G       &km $s^{-1}$&        \\
\hline
\hline
\endfirsthead
\caption{continued.}\\
\hline\hline
HD      &Name      &Instrum.& Date      &HJD        &Detect. & $B_l$ & $\sigma$ &  RV      & $S$-index \\
        &          &        &           &(2450000+) &        &  G       &  G       &km $s^{-1}$&        \\
\hline
\hline
\endhead
\hline
\hline
\endfoot
\hline
3229    &14 Cet    & N      &Auri\`ere et al. &2012 & DD     \\
4128    &$\beta$ Cet  &E    &29 Sep. 07 &4373.97    & DD     &   3.83   &  0.54    & 13.349    & 0.243\\   
        &          & E      &30 Sep. 07 &4374.97    & DD     &   4.22   &  0.56    & 13.332    & 0.243\\   
        &          & E      &01 Oct. 07(2)&4375.99  & DD     &   4.62   &  0.39    & 13.356    & 0.242\\   
        &          & E      &02 Oct. 07(2)&4376.98  & DD     &   4.59   &  0.35    & 13.352    & 0.243\\   
        &          & N      &31 Dec. 07(2)&4466.22  & DD     &   7.01   &  0.30    & 13.347    & 0.286\\   
        &          & E      &21 Aug. 08(4)&4701.09  & DD     &   9.01   &  0.31    & 13.375    & 0.236\\ 
        &          & N      &14 Sep. 08 &4724.51    & DD     &   4.93   &  0.55    & 13.333    & 0.217\\   
        &          & N      &19 Sep. 08 &4729.59    & DD     &   3.11   &  0.56    & 13.336    & 0.211\\   
        &          & N      &24 Sep. 08 &4734.471   & DD     &   1.59   &  0.46    & 13.374    & 0.217\\   
        &          & N      &30 Sep. 08 &4740.466   & DD     &   0.59   &  0.45    & 13.422    & 0.210\\   
        &          & E      &15 Oct. 08(3)&4755.99  & nd     &  -0.79   &  0.66    & 13.389    & 0.201\\   
        &          & E      &16 Oct. 08(2)&4756.92  & MD     &  -1.44   &  0.51    & 13.389    & 0.221\\ 
        &          & E      &28 Sep. 09 &5102.93    & DD     &   1.13   &  0.34    & 13.327    & 0.246\\   
        &          & E      &02 Oct. 09 &5106.94    & DD     &   0.25   &  0.43    & 13.305    & 0.239\\   
 %       &          & E      &06 Oct. 09 &           & ND     &  -3.25   &  2.56    & 13.324    & 0.192\\   
        &          & E \& N &Tsvetkova et al. &2013 & DD     & \\ 
9746    &OP And    & N      &15 Sep. 08 & 4725.58   & DD     &  -15.75   & 0.69     & -42.396   & 0.798\\
        &          & N      &20 Sep. 08 & 4730.46   & DD     &   -7.27   & 1.21     & -42.380   & 0.743 \\
        &          & N      &25 Sep. 08 & 4735.54   & DD     &   +1.85   & 0.92     & -42.359   & 0.834\\
        &          & N      &27 Sep. 08 & 4737.53   & DD     &   +5.84   & 0.98     & -42.345   & 0.800\\
        &          & N      &29 Sep. 08 & 4739.55   & DD     &   +6.68   & 0.75     & -42.377   & 0.799\\
        &          & N      &21 Dec. 08 & 4822.32   & DD     &   +5.86   & 0.80      &-42.395   & 0.750\\
        &          & N&Konstantinova-Antova&in prep.& DD     \\
27536   &EK Eri       &N    &Auri\`ere et al.& 2008 & DD      \\
        &             &N    &Auri\`ere et al.& 2011 & DD      \\
28307   &77 Tau       &  N  &31 Dec. 07    &4466.25 & nd      &   1.00  &    0.99  &  39.608   &  0.170 \\
        &             &  N  &01 Jan. 08    &4467.25 & DD      &   0.86  &    0.54  &  39.613   &  0.170 \\
        &             &  N  &22 Jan. 08    &4488.43 & DD      &   1.42  &    0.41  &  39.646   &  0.169 \\ 
        &             &  N  &20 Sep. 08    &4730.58 & DD      &   3.01  &    0.50  &  40.253   &  0.176 \\ 
        &             &  N  &30 Sep. 08    &4740.53 & DD      &   2.46  &    0.33  &  40.365   &  0.171   \\
        &             &  N  &25 Feb. 09    &4888.33 & DD      &   2.19  &    0.31  &  40.643   &  0.161   \\
        &             &  E  &05 Sep. 09    &5080.15 & DD      &   0.05  &    0.34  &  41.291   &  0.174   \\
        &             &  E  &28 Sep. 09    &5102.96 & nd      &   1.20  &    0.38  &  41.406   &  0.145   \\
        &             &  E  &08 Mar. 10    &5263.71 & DD      &   1.12  &    0.38  &  41.802   &  0.171   \\
        &             &  N  &14 Mar. 10    &5270.30 & DD      &   1.98  &    0.31  &  41.789   &  0.183   \\
        &             &  N  &16 Apr. 10    &5303.31 & ND      &  -0.32  &    0.48  &  42.046   &  0.208  \\
        &             &  E  &17 Jul. 10    &5396.13 & DD      &   1.18  &    0.37  &  42.420   &  0.178   \\
        &             &  N  &18 Sep. 10    &5458.66 & DD      &  -1.44  &    0.36  &  42.500   &  0.172  \\
        &             &  E  &17 Oct. 10 (2)&5487.86 & DD      &   1.64  &    0.31  &  42.765   &  0.190  \\
        &             &  N  &18 Oct. 10    &5488.65 & DD      &   1.10  &    0.33  &  42.739   &  0.172  \\
        &             &  E  &18 Oct. 10    &5489.15 & DD      &   1.19  &    0.30  &  42.727   &  0.194   \\
        &             &  N  &12 Nov. 10    &5513.57 & nd      &   0.55  &    0.36  &  42.882   &  0.167  \\
        &             &  E  &21 Nov. 10    &5522.85 & nd      &  -0.45  &    0.30  &  42.762   &  0.175  \\
        &             &  E  &27 Nov. 10    &5528.86 & nd      &  -0.14  &    0.47  &  42.733   &  0.175  \\
        &             &  N  &13 Dec. 10    &5544.48 & nd      &   0.76  &    0.51  &  42.946   &  0.174  \\
        &             &  N  &14 Jan. 11    &5576.27 & MD      &  -0.51  &    0.32  &  43.006   &    0.164 \\
31993   &V1192 Ori & N      &14 Sep. 08 & 4724.63   & DD     &   +10.52  & 4.15    &  14.752    & 0.934 \\
        &          & N      &19 Sep. 08 & 4729.70   & DD     &   -1.59   & 4.43    &  16.301    & 0.785\\
        &          & N      &25 Sep. 08 & 4735.62   & DD     &   -14.67  & 3.19    &  14.695    & 0.997 \\
        &          & N      &28 Sep. 08 & 4738.62   & DD     &   -11.09  & 2.95    &  14.887    & 0.933\\
        &          & N      &30 Sep. 08 & 4740.60   & DD     &   -10.18  & 2.96    &  14.969    & 0.946\\
33798   &V390 Aur  & N &Konstantinova-Antova& 2008  & DD     &           &         &            &          \\
        &          & N &Konstantinova-Antova&2012   & DD     &           &         &            &          \\
47442   &nu3 Cma      &  N  &27 Sep. 08    &4737.69 & DD      &  -1.24  &    0.39  &  -0.899   &  0.167 \\   
        &             &  E  &18 Oct. 10    &5489.03 & DD      &  -2.25  &    0.45  &  -0.850   &  0.170 \\   
        &             &  E  &15 Nov. 10    &5517.16 & nd      &  -1.01  &    0.62  &  -0.916   &  0.167 \\  
        &             &  E  &28 Nov. 10    &5530.01 & MD      &  -1.15  &    0.34  &  -0.964   &  0.147 \\  
68290   &19 Pup       &  N  &25 Feb. 09    &4888.41 & DD      &  -4.18  &    0.44  &  35.774   &  0.206  \\ 
72146   &FI Cnc    & N      &02 Apr. 08 & 4559.40   & DD     &   -16.13  & 2.18    &  -1.637    & 1.068 \\
        &          & N      &04 Apr. 08 & 4561.40   & DD     &   -17.57  & 1.74    &  -0.978    & 1.027 \\
        &          & N      &15 Apr. 08 & 4572.46   & DD     &   +3.80   & 4.36    &  -1.695    & 1.372\\
82210   &24 Uma       &  N  &01 Jan. 08    &4467.71 & DD      &  -3.05  &    2.58  & -26.944   &  0.408 \\  
        &             &  N  &02 Jan. 08    &4468.75 & DD      &   1.52  &    2.62  & -26.888   &  0.404 \\  
        &             &  N  &02 Apr. 08    &4559.45 & DD      &  -3.09  &    0.66  & -26.848   &  0.397 \\  
85444   &39 Hya       &  N  &02 Apr. 08    &4559.43 & DD      &  -7.75  &    0.58  & -14.016   &  0.219 \\   
111812  &31 Com    & N      &01 Mar.12 (2)&5988.57  & DD     &   +6.28   & 2.86    &  -0.360    & 0.392 \\
        &          & N      &10 Mar.12 (2)&5997.65  & DD     &   +6.95   & 3.08    &  -0.610    & 0.398 \\ 
        &          & N&Borisova et al. &in prep.    & DD     & \\
112989  &37 Com  & N\&E & Tsvtekova et al.& in prep.& DD     &           &         &            &          \\
121107  &7 Boo     & N      &05 Apr. 08 & 4562.50   & MD     &   +1.91   & 0.84    &  -11.501   & 0.221 \\
        &          & N      &15 Apr. 08 & 4572.50   & nd     &   +0.65   & 1.55    &  -11.444   & 0.234\\
        &          & N      &24 Feb. 09 & 4887.61   & nd     &   +0.02   & 0.95    &  -11.419   & 0.233 \\
133208  &$\beta$ Boo  &  N  &06 Feb. 08    &4503.77 & nd      &   0.30  &    0.33  & -18.254   &  0.115 \\   
        &             &  N  &02 Apr. 08    &4559.52 & nd      &   0.02  &    0.32  & -18.237   &  0.114 \\  
        &             &  E  &23 Aug. 08    &4702.73 & nd      &  -0.02  &    0.53  & -18.105   &  0.115 \\   
%        &             &  E  &17 Dec. 08    &        & nd      &  -0.29  &    0.54  & -18.091   &  0.112 \\   
        &             &  N  &25 Feb. 09 (4)&4888.62 & nd      &  -0.04  &    0.16  & -18.256   &  0.115 \\  
141714  &$\delta$ CrB &  N  &30 Dec. 07    &4465.77 & DD      &  -1.17  &    0.98  & -20.114   &  0.305 \\  
        &             &  N  &31 Dec. 07    &4466.77 & DD      &   1.99  &    0.59  & -20.105   &  0.307\\  
        &             &  N  &16 Sep. 08    &4726.32 & DD      &  -4.69  &    0.54  & -20.081   &  0.297\\  
        &             &  N  &24 Feb. 09    &4887.70 & DD      &  -2.00  &    0.61  & -20.148   &  0.286\\  
        &             &  E  &26 Jan. 10    &5223.11 & DD      &  -6.09  &    0.53  & -20.024   &  0.286\\  
        &             &  E  &02 Feb. 10    &5230.17 & DD      &  -3.1   &    0.61  & -20.019   &  0.317\\  
        &             &  E  &27 Feb. 10    &5255.01 & DD      &  -2.17  &    0.47  & -20.065   &  0.321\\  
        &             &  E  &08 Mar. 10    &5263.98 & DD      &   4.68  &    0.97  & -20.162   &  0.310\\  
        &             &  E  &03 Jun. 10    &5351.76 & DD      &  -2.41  &    0.47  & -20.124   &  0.305\\  
        &             &  E  &12 Jun. 11    &5725.79 & DD      &  -5.4   &    0.84  & -20.068   &  0.303\\
145001  &$\kappa$ HerA& N   &04 Apr. 08 & 4561.55   & DD     &   -3.35   & 0.68    &  -10.021   & 0.290 \\
        &          & N      &14 Sep. 08 & 4724.35   & DD     &   -4.61   & 0.81    &  -10.067   & 0.296\\
        &          & N      &17 Sep. 08 & 4727.33   & DD     &   -3.64   & 0.92    &  -10.074   & 0.285 \\
        &          & N      &21 Sep. 08 & 4731.28   & DD     &   -3.02   & 0.93    &  -10.076   & 0.275\\
        &          & N      &25 Sep. 08 & 4735.30   & DD     &   -3.37   & 0.72    &  -10.076   & 0.277\\
        &          & N      &30 Sep. 08 & 4740.29   & DD     &   -2.69   & 1.07    &  -10.003   & 0.282 \\
        &          & E      &17 Oct. 08 & 4757.70   & DD     &   +4.48   & 0.81    &   -9.890   & 0.283 \\
        &          & N      &24 Feb. 09 & 4887.72   & DD     &   -3.62   & 0.87    &  -10.041   & 0.305\\
        &          & E      &26 Jan. 10 & 5223.12   & DD     &   -1.69   & 0.61    &   -9.886   & 0.307 \\
        &          & E      &01 Feb. 10 & 5229.12   & DD     &   -0.94   & 0.52    &   -9.965   & 0.297 \\
        &          & E      &27 Feb. 10 & 5255.03   & DD     &   -1.53   & 0.53    &   -9.888   & 0.292 \\
        &          & E      &01 Mar. 10 & 5257.17   & DD     &   -1.19   & 0.70    &   -9.938   & 0.290 \\
        &          & E      &05 Mar. 10 & 5261.04   & DD     &   -1.45   & 0.59    &   -9.924   & 0.290  \\
        &          & E      &08 Mar. 10 & 5264.00   & DD     &   -3.18   & 1.31    &   -9.973   & 0.307  \\
        &          & E      &03 Jun. 10 & 5351.77   & DD     &   -1.62   & 0.54    &  -10.021   & 0.327  \\
        &          & E      &21 Jun. 10 & 5369.75   & DD     &   -1.71   & 0.52    &   -9.973   & 0.328  \\
        &          & E      &16 Oct. 10 & 5486.72   & DD     &   +0.37   & 0.80    &   -9.844   & 0.324  \\
150997  &$\eta$ Her   &  N  &17 Sep. 08    &4727.35 & DD      &  -5.33  &    0.54  &   8.611   &  0.193 \\  
        &             &  N  &20 Sep. 08    &4730.32 & DD      &  -6.81  &    0.54  &   8.588   &  0.191 \\  
163993  &$\xi$ Her    &  N  &17 Sep. 08    &4727.37 & DD      &  -2.63  &    1.00  &  -1.582   &  0.242 \\   
        &             &  N  &30 Sep. 08    &4740.31 & DD      &   3.77  &    0.37  &  -1.521   &  0.251 \\  
203387  &$\iota$ Cap  &  N  &20 Sep. 08 (2)&4730.36 & DD      &   0.28  &    0.67  &  12.444   &  0.313 \\  
        &             &  N  &29 Sep. 08    &4739.46 & DD      &  -3.45  &    0.51  &  12.394   &  0.302 \\  
        &             &  E  &02 Oct. 09    &5106.83 & DD      &  -7.39  &    0.75  &  12.386   &  0.317 \\  
        &             &  E  &05 Oct. 09    &5109.87 & DD      &  -7.55  &    0.46  &  12.450   &  0.311 \\  
        &             &  E  &16 Nov. 10    &5517.68 & DD      &  -1.87  &    0.40  &  12.417   &  0.288 \\  
        &             &  E  &20 Nov. 10    &5521.74 & DD      &  -3.96  &    0.53  &  12.467   &  0.289 \\   
        &             &  E  &16 Dec. 10    &5547.72 & DD      &  -7.33  &    0.45  &  12.505   &  0.296 \\ 
        &             &  E  &16 Jun. 11    &5730.03 & DD      &  -8.33  &    0.56  &  12.454   &  0.343 \\ 
205435  &$\rho$ Cyg   &  E  &20 Aug. 08    &4700.03 & DD      &   5.06  &    0.67  &   6.914   &  0.278 \\   
        &             &  E  &21 Aug. 08    &4701.06 & DD      &   5.06  &    0.61  &   6.934   &  0.270\\   
        &             &  N  &19 Sep. 08    &4729.47 & DD      &   5.73  &    0.51  &   6.920   &  0.306 \\  
        &             &  N  &25 Sep. 08    &4735.40 & DD      &   6.51  &    0.44  &   6.924   &  0.310 \\  
        &             &  N  &20 Dec. 08    &4821.23 & DD      &   7.28  &    0.52  &   6.948   &  0.268 \\  
218153  &KU Peg    & N      &14 Sep. 08 & 4724.45   & DD     &   -0.53   & 4.44    &  -79.797   & 1.117 \\
        &          & N      &19 Sep. 08 & 4729.51   & DD     &   +13.05  & 7.24    &  -79.797   & 1.060 \\
        &          & N      &24 Sep. 08 & 4734.55   & DD     &   +10.06  & 5.10    &  -79.797   & 1.083 \\
        &          & N      &29 Sep. 08 & 4739.49   & DD     &    -3.81  & 4.87    &  -79.797   & 0.979  \\
223460  &OU And    & N      &14 Sep. 08 & 4724.46   & DD     &   -28.02  & 5.47    &   0.       & 0.492 \\
        &          & N      &16 Sep. 08(4)& 4726.52 & DD     &   -24.75  & 3.16    &   -0.5     & 0.467 \\
        &          & N      &19 Sep. 08 & 4729.57   & DD     &   -10.28  & 1.71    &   -1.1     & 0.498  \\
        &          & N      &21 Sep. 08 & 4731.41   & DD     &   +5.56   & 2.69    &   -1.3     & 0.515  \\
        &          & N      &25 Sep. 08 & 4735.43   & DD     &   +30.02  & 1.68    &   -2.1     & 0.517  \\
        &          & N      &29 Sep. 08 & 4739.52   & DD     &   +40.86  & 1.50    &   -1.5   & 0.515  \\
        &          & N& Borisova et al. &in prep.   & DD     &           &  \\ 
\hline                                  
%\end{tabular}
% \end{table*}                          
\end{longtable}

\clearpage

\begin{table*}
\caption{Journal of observations of the thermohaline deviants}          
\label{table:1}   
\centering                         
\begin{tabular}{l l l l c c c c c c}     
\hline\hline  

HD         &Name    &Instrum.& Date      &HJD        &Detect. & $B_l$ & $\sigma$ &  RV      & $S$-index \\
           &        &        &           &(2450000+) &        &  G       &  G       &km $s^{-1}$&        \\
\hline
\hline
50885      &        &    N   & 03apr08   & 4560.41   &  nd   & +0.34    & 0.38    & -17.765 &  0.136 \\
           &        &    N   & 28sep08   & 4738.70   &  nd   & -0.70    & 0.41    & -17.654 &  0.138 \\
           &        & N      & 30sep08   & 4740.69   &  nd   & +0.20    & 0.41    & -17.590 &  0.132 \\ 
           &        & E      & 18oct08   & 4759.06   &  nd   & -0.41    & 0.53    & -17.561 &  0.123 \\
95689   & $\alpha$ UMa & N   & 05feb08(2)& 4502.73   &  nd   & -0.06    & 0.34    & -11.303 &  0.135 \\
        &           & N      & 03apr08   & 4560.54   & nd    & -0.42    & 0.45    & -11.425 &  0.136 \\
        &           & E      & 19oct08   & 4760.14   & nd    & +0.17    & 0.45    & -11.370 &  0.134 \\
150580 &            & N      & 02apr08   & 4559.58   & nd    & +0.37    & 0.57    & -68.586 &  0.145 \\
       &            & N      & 25sep08   & 4735.33   & nd    & -0.05    & 0.53    & -68.563 &  0.134 \\
178208 &            & N      & 03apr08   & 4560.61   & nd    & +0.26    & 0.75    &  +4.963 &  0.128 \\
       &            & N      & 14sep08   & 4724.39   & nd    & +0.29    & 0.58    &  -4.206 &  0.126 \\
       &            & N      & 30sep08   & 4740.38   & nd    & +0.39    & 0.55    &  -4.328 &  0.129 \\
       &            & E      & 17oct08   & 4757.73   & nd    & +0.06    & 0.41    &  -4.396 &  0.126 \\
       &            & E      & 19oct08   & 4759.81   & nd    & +0.12    & 0.45    &  -4.419 &  0.123 \\
186619 &            & E      & 29jul08   & 4678.02   & nd    & -0.11    & 0.85    & -46.023 &  0.208 \\
       &            & E      & 19aug08   & 4699.03   & nd    & +0.36    & 0.47    & -46.027 &  0.241 \\
       &            & N      & 14sep08   & 4724.42   & nd    & +1.34    & 0.61    & -45.908 &  0.242 \\
       &            & N      & 29sep08 (4)&4739.42   & nd    & -0.23    & 0.32    & -45.997 &  0.247 \\
       &            & N      & 21may09 (8)&4973.64   & nd    & +0.03    & 0.55    & -45.973 &  0.272 \\
199101 &            & E      & 27jul08   & 4676.08   & nd    & -0.27    & 0.52    & -12.107 &  0.237 \\
       &            & E      & 19aug08   & 4699.05   & nd    & -0.74    & 0.46    & -12.083 &  0.216 \\
       &            & E      & 20aug08   & 4700.07   & nd    & -0.07    & 0.51    & -12.025 &  0.223 \\
       &            & N      & 19sep08   & 4729.44   & nd    & -0.20    & 0.50    & -11.925 &  0.221 \\
       &            & N      & 30sep08   & 4740.42   & nd    & -1.02    & 0.34    & -11.996 &  0.219 \\
218452 &   4 And    & E      & 20aug08   & 4700.04   & nd    & -0.69    & 0.43    & -11.562 &  0.141 \\
       &            & E      & 21aug08 (2)&4701.07   & nd    & -0.20    & 0.28    & -11.577 &  0.147 \\
       &            & N      & 16sep08   & 4726.42   & nd    & -0.39    & 0.31    & -11.619 &  0.139 \\
       &            & N      & 26sep08   & 4736.54   & nd    & -0.36    & 0.31    & -11.662 &  0.150 \\
       &            & E      & 17oct08   & 4757.75   & nd    & -0.59    & 0.33    & -11.416 &  0.157 \\
\hline
\hline                            
\end{tabular}
 \end{table*} 

\clearpage

\onecolumn
\begin{longtable}{l l l l c c c c c c}
\caption{Journal of observations of the CFHT snapshot and miscellaneous giants} \\
\hline
\hline
HD      &Name      &Instrum.& Date      &HJD        &Detect. & $B_l$ & $\sigma$ &  RV      & $S$-index \\
        &          &        &           &(2450000+) &        &  G       &  G       &km $s^{-1}$&        \\
\hline
\hline
\endfirsthead
\caption{continued.}\\
\hline\hline
HD      &Name      &Instrum.& Date      &HJD        &Detect. & $B_l$ & $\sigma$ &  RV      & $S$-index \\
        &          &        &           &(2450000+) &        &  G       &  G       &km $s^{-1}$&        \\
\hline
\hline
\endhead
\hline
\hline
\endfoot
9270    &$\eta$ Psc   &N    &15 Sep. 08 (4)&4725.55 & DD      &  -0.45  &    0.20  &  13.843   &  0.133  \\   
        &             &N    &20 Sep. 08    &4730.32 & nd      &  -0.33  &    0.85  &  13.787   &  0.190 \\ 
        &             &N    &21 Sep. 08 (4)&4731.48 & nd      &  -0.28  &    0.27  &  13.795   &  0.132  \\ 
        &             &N    &24 Sep. 08 (3)&4734.53 & nd      &   0.06  &    0.35  &  13.786   &  0.135  \\ 
9927    &$\upsilon$ Per     & E      & 20aug08   &4700.05    & nd     &  -0.56   &  0.53    &  16.417   &  0.116 \\    
12929   & $\alpha$ Ari      & E      & 29sep07   &4373.99    & nd     &  -0.17   &  0.57    & -14.362   &  0.119\\    
        &                   & E      & 30sep07   &4374.99    & nd     &   0.87   &  0.58    & -14.359   &  0.122 \\    
        &                   & E      & 01oct07   &4375.99    & nd     &   -0.47  &  0.53    & -14.362   &  0.120 \\    
        &                   & E      & 26dec07   &4461.83    & nd     &    0.68  &  0.52    & -14.429   &  0.123 \\    
        &                   & E      & 20aug08   &4700.06    & nd     &   -0.32  &  0.46    & -14.280   &  0.115 \\    
        &                   & E      & 17oct08   &4757.94    & nd     &    0.20  &  0.40    & -14.425   &  0.118 \\   
        &                   & E      & 06dec08   &4806.76    & nd     &    0.56  &  0.38    & -14.423   &  0.118 \\   
28305   &$\epsilon$ Tau     & E      & 22aug08   &4702.02    & MD     &    1.4   &  0.58    &  38.464   &  0.122 \\   
        &                   & E      & 17oct08   &4757.94    & nd     &    0.68  &  0.45    &  38.436   &  0.134 \\   
        &                   & E      & 17dec08 (3)&4817.80   & DD     &   -1.34  &  0.27    &  38.423   &  0.116 \\  
        &                   & N      & 25feb09   &4888.35    & DD     &    0.44  &  0.30    &  38.451   &  0.134 \\   
        &                   & E      & 02oct09   &5107.01    & nd     &    0.61  &  0.34    &  38.586   &  0.121 \\   
        &                   & E      & 07oct09   &5112.01    & nd     &    0.59  &  0.48    &  38.627   &  0.097 \\   
        &                   & E      & 08mar10   &5263.72    & nd     &    0.97  &  0.38    &  38.465   &  0.123 \\  
        &                   & E      & 19jul10   &5398.12    & nd     &   -0.52  &  0.35    &  38.498   &  0.116 \\  
        &                   & E      & 18oct10   &5489.16    & MD     &    -1.00 &  0.32    &  38.493   &  0.121 \\   
        &                   & E      & 15nov10 (2)&5517.02   & DD     &  -0.59   &  0.23    &  38.603   &  0.127 \\   
        &                   & E      & 21nov10   &5522.86    & nd     &   0.02   &  0.34    &  38.565   &  0.115 \\  
29139   & Aldebaran         & E      &26 Sep. 07 & 4371.14   & nd     & -1.08    &  0.74    &  54.425   &  0.209 \\
        &                   & E      &30 Sep. 07 & 4374.98   & nd     & -0.77    &  0.60    &  54.356   &  0.222 \\
        &                   & E      &21 Aug. 08 & 4701.13   & nd     & -0.69    &  0.37    &  54.350   &  0.222 \\
        &                   & E      &17 Oct. 08 & 4757.93   & nd     & -1.04    &  0.38    &  54.112   &  0.219 \\
        &                   & E      &15 Dec. 08 (8)&4815.82 &7 nd \& 1 MD& 0.13  & 0.16    &  54.087   &  0.235 \\ 
        &                   & N      &20-21 Dec. 08 (16)&4821.84& DD  &  0.18     & 0.10    &  54.127   &  0.239 \\   
        &                   & N      &02 Oct. 09 (16)& 5107.56 & MD   & -0.19     & 0.07    &  54.595   &  0.236 \\
        &                   & N      &26 Oct. 09 (16)& 5131.61 & nd   & -0.13     & 0.14    &  54.498   &  0.249 \\
        &                   & N      &14-15 Mar. 10 (32)&5270.83& nd  &  0.22    &  0.08    &  54.493   &  0.233 \\
        &                   & N      &21 Sep. 10 (16)&5461.66& nd     &  0.28    &  0.12    &  54.186   &  0.225 \\
        &                   & N      &05 Oct. 10 (16)&5475.64& DD     & -0.25    &  0.13    &  54.293   &  0.235 \\
        &                   & N      &20 Oct. 10 (16)&5490.67& nd     & -0.31    &  0.20    &  54.173   &  0.235 \\
        &                   & N      &16 Jan. 11 (16)&5578.27& DD     &  0.22    &  0.09    &  54.096   &  0.224 \\
        &                   & N      &18 Mar  11 (16)&5639.32& nd     &  0.03    &  0.11    &  54.254   &  0.236 \\            
32887   &$\epsilon$ Lep     & E      &17 Oct. 08     &4758.15& nd     &   -0.46  &  0.49    &   1.380   &  0.180 \\    
        &                   & E      &16 Dec. 08 (4) &4816.81& nd     &    0.25  &  0.32    &   1.353   &  0.178 \\    
62509   & Pollux     & E \& N & Auri\`ere et al.& 2009       & DD     &       \\
        &            & E \& N & Auri\`ere et al.& in prep.   & DD     &       \\
76294   &$\zeta$ Hya        & E      &17 Oct. 08     &4758.16& nd     &   -0.05  &  0.61    &  23.065   &  0.108  \\
        &                   & E      &08 Dec. 08 (8) &4809.17& nd     &   -0.24  &  0.26    &  22.987   &  0.104 \\   
81797   & Alphard           & E      &29 Dec. 07 (2) &4465.12& nd     &   -0.40  &  0.65    &  -4.401   &  0.109  \\  
        &                   & N      &21-22 Jan. 12 (32)&5949.14& MD  &  0.07    &  0.08    &  -4.448   &  0.173 \\
        &                   & N      &26 Mar. 12 (16)&6013.43& DD     &  0.35    &  0.08    &  -4.370   &  0.185 \\
89484   &$\gamma$ Leo A     & E      &29 Dec. 07 (3) &4465.17& nd     &    0.29  &  0.38    & -35.848   &  0.130  \\  
        &                   & E      &18 Oct. 08     &4759.15& nd     &   -0.66  &  0.54    & -35.849   &  0.130 \\   
        &                   & E      &06 Dec. 08 (4) &4807.15& nd     &   -0.22  &  0.28    & -35.829   &  0.130 \\ 
        &                   & N      &24 Mar. 12 (16)&6011.47& nd     &   -0.13  &  0.13    & -36.145   &  0.129 \\ 
        &                   & N      &10 Dec. 12 (15)&6272.73& nd     &    0.25  &  0.28    & -36.310   &  0.132 \\
93813   &$\nu$ Hya          & E      &19 Oct. 08     &4760.15& nd     &    0.32  &  0.50    &  -0.332   &  0.116 \\    
        &                   & E      &06 Dec. 08 (4) &4807.17& nd     &    0.07  &  0.25    &  -0.440   &  0.121 \\    
105707  &$\epsilon$ Crv     & E      &17 Dec. 08 (2) &4818.13& nd     &    0.38  &  0.35    &   5.288   &  0.122 \\    
124897  & Arcturus          & E      &23 Aug. 08     &4702.72& nd     &    0.27  &  0.45    &   -4.982  &  0.123 \\
%        &                   & E     & 06 Dec. 08     &4807.17& nd     &    0.54  &  0.46    &   -4.842  &  0.128 \\
        &                   & N     &22-23 Jan. 12 (32)&5950.19& MD   &   0.34   &  0.11    &  -5.014   &  0.123 \\
        &                   & N     & 25 Mar. 12 (16)&6012.56&MD      &  -0.05   &  0.11    &  -5.110   &  0.121 \\ 
        &                   & N     & 23 Jun. 12 (16)&6102.39& nd     &   0.07   &  0.11    &  -5.054   &  0.117 \\
129989  &$\epsilon$ Boo A   & E     & 23 Aug. 08     &4702.72& nd     &   -0.23  &  0.58    &  -16.131  &  0.129 \\   
        &                   & E     & 17 Dec. 08     &4818.14& nd     &    0.98  &  0.63    &  -15.998  &  0.119 \\   
131873  &$\beta$ UMi        & E     & 23 Aug. 08     &4702.73& nd     &   -0.01  &  0.40    &   16.712  &  0.191 \\   
        &                   & E     & 17 Dec. 08     &4818.14& nd     &   -0.28  &  0.36    &   16.955  &  0.196 \\   
163917  &$\nu$ Oph          & E     & 23 Aug. 08     &4702.74& nd     &   -0.98  &  0.64    &   12.777  &  0.107 \\    
        &                   & E     & 17 Oct. 08     &4757.69& nd     &    1.02  &  0.58    &   12.703  &  0.105 \\    
216131  &$\mu$ Peg          &  N    & 19 Sep. 08     &4729.49& nd     &    0.47  &  0.54    &   13.865  &  0.117 \\  
        &             &  N  &20 Sep. 08    &4730.39 & nd      &   0.00  &    0.50  &  13.836   &  0.118 \\  
        &             &  N  &21 Sep. 08 (4)&4731.37 & nd      &  -0.43  &    0.21  &  13.879   &  0.116 \\  
        &             &  N  &21 Dec. 08 (4)&4822.25 & nd      &   0.23  &    0.19  &  13.756   &  0.117 \\  
        &             & N&05-06 Sep. 12 (8)&6176.95 & nd      &  -0.04  &    0.13  &  13.861   &  0.117 \\ 
\hline                                  
%\end{tabular}
% \end{table*}                          
\end{longtable}

\clearpage

\begin{appendix}
\twocolumn

\section {Complementary results for 77 Tau, $\epsilon$ Tau and Aldebaran which have been followed up during several seasons}

\subsection{77 Tau, HD 28307}

77 Tau, also known as $\theta^1$ Tau is one giant of the Hyades cluster. It is a close binary with a period of 16.3 y (Torres et al. 1997). The star presents some hints of activity, namely variations of Ca~{\sc ii} $H \& K$ emission, inducing periodic variations of its $S$-index (Choi et al. 1995, P = 140 d), and a rather strong emission in X-rays at the 10 $^{30}$ erg s$^{-1}$ level (Gondoin 1999). Since the $v\sin i $ of 77 Tau is only 4.2 km\,s$^{-1}$ , 77 Tau appeared as a possible slow rotator and we included it in the Zeeman survey from Pic du Midi and CFHT. We observed it during 3 years, on 20 dates,  from 31 December 2007 to 14 January 2011 with both Narval and ESPaDOnS (see Table 6). Figure A1 plots the variations of RV, $S$-index and $B_l$ during the 3 years. The variations of RV are consistent with the binary status (Torres et al. 1997). On the other hand, the $S$-index measurements do not show clear variations consistent with the period of 140 d inferred by Choi et al. (1995). The magnetic field at the surface of 77 Tau is detected for more than 60\% of the observations (Table 6). However, even if $B_l$ varies significantly and changes its sign, we were unable to determine a period. At the end, although we were unable to confirm it as the genuine rotational period, we use the period of 140 d as the rotational period of 77 Tau for our investigation in \S~6. Recently, Beck et al. (2014) performed a high-precision spectroscopic multisite campaign including 77 Tau as a target. They infer long term variations with a period of 165 days which supports the order of magnitude of the rotational period proposed from variations of the $S$-index.

\begin{figure}
\centering
\includegraphics[width=9 cm,angle=0] {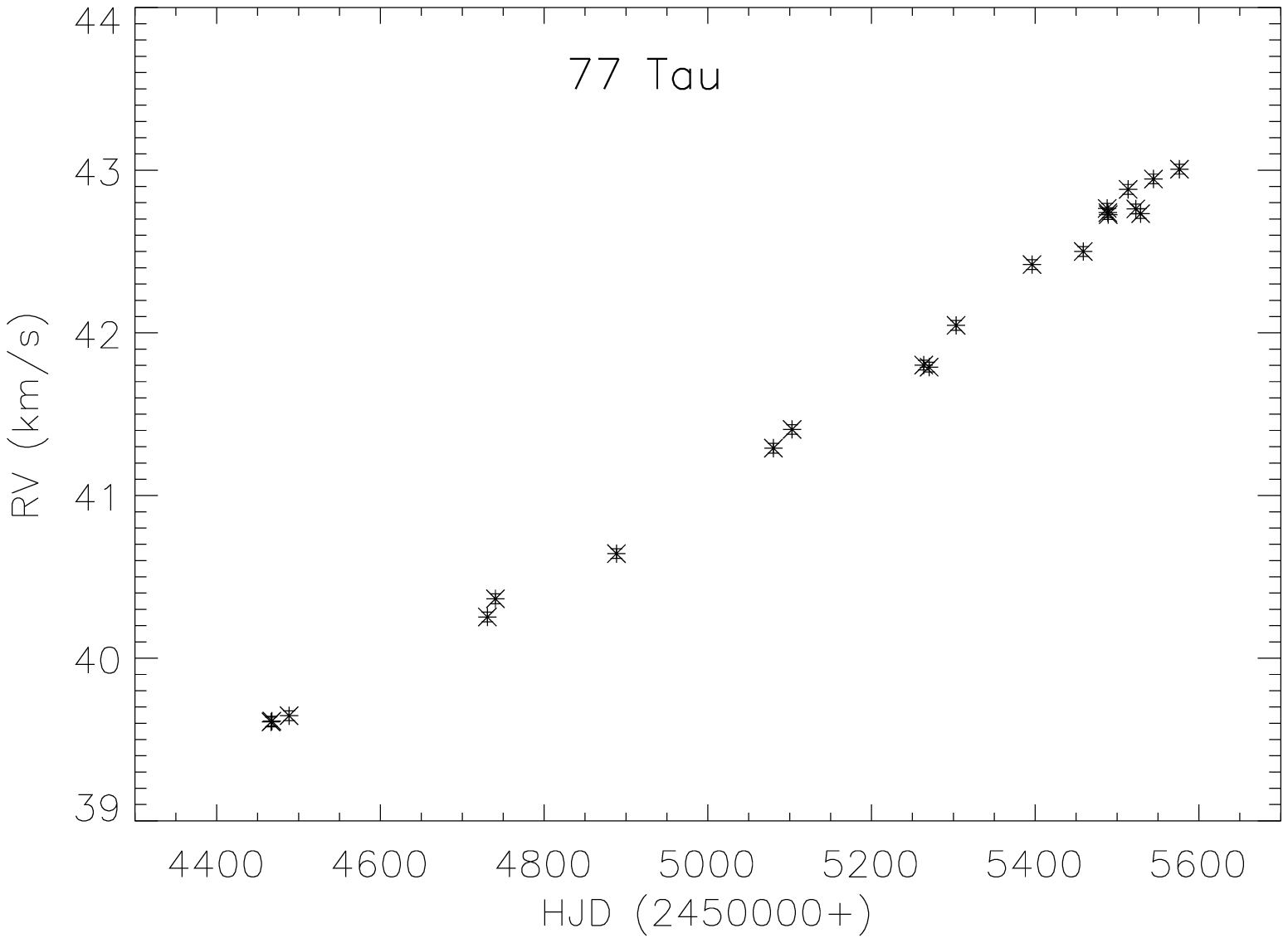} 
\includegraphics[width=9 cm,angle=0] {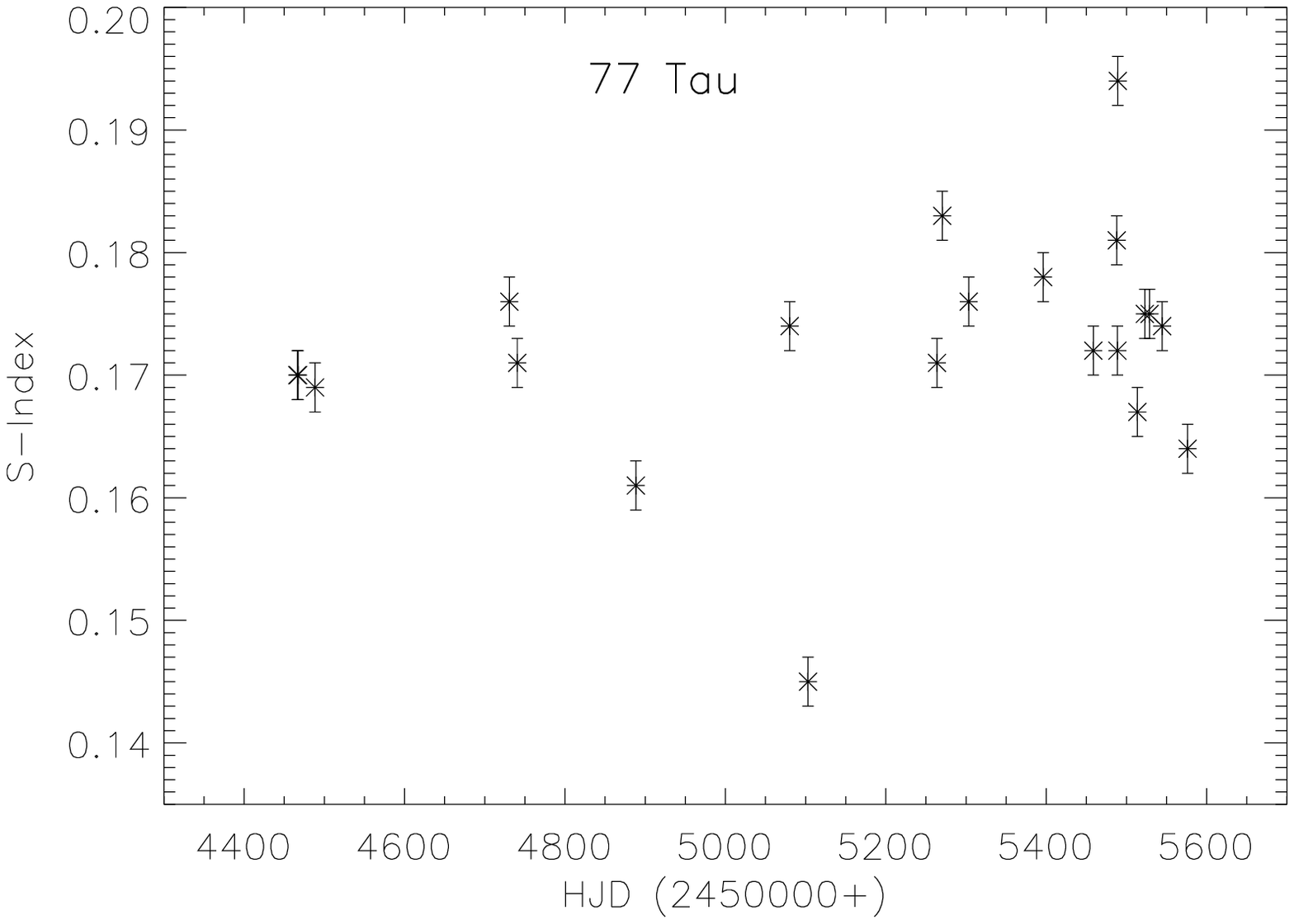}
\includegraphics[width=9 cm,angle=0] {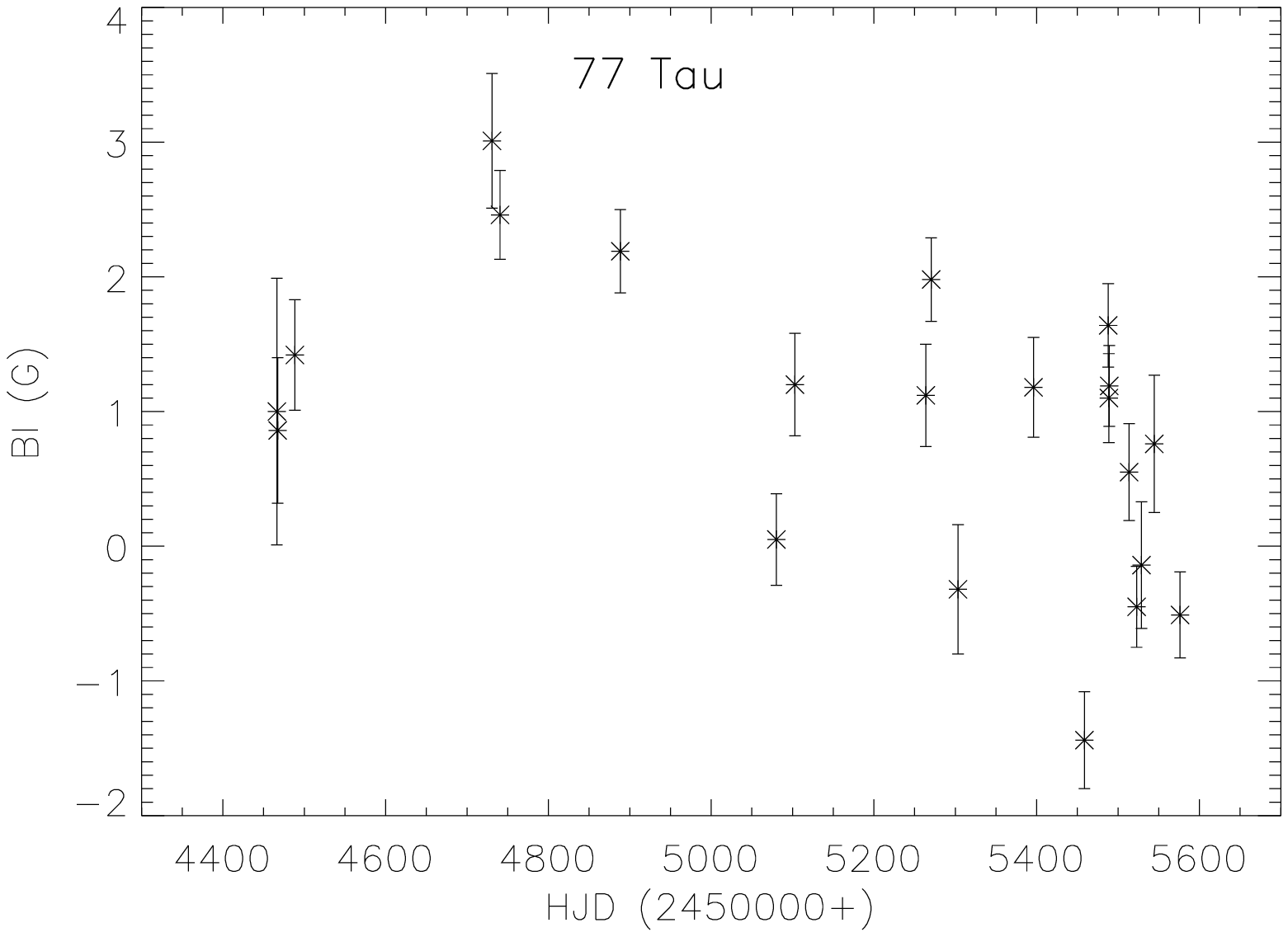}
\caption{Variations of the radial velocity (RV, upper graph), $S$-index (middle graph) and longitudinal magnetic field ($B_l$, lower graph) of 77 Tau between 31 December 2007 and 14 January 2011, as observed with Narval and ESPaDOnS. Error bars of 30 m/s and 0.05 are illustrated for RV and $S$-index respectively. As to the $B_l$ plot, error bars are those of Table 6.}
%\label{f1}
\end{figure} 

\subsubsection{$\epsilon$ Tau, HD 28305}

$\epsilon$ Tau is a giant star of the Hyades known to be single (Mason et al. 2009). A weak X-ray emission of about $10^{28}$ ergs\, s$^{-1}$ was detected with ROSAT (Collura et al. 1993). Sato et al. (2007) detected variations of its radial velocity with a semiamplitude of 95.9 m\,s$^{-1}$ and a period of 594.9 d that they interpret as due to an hosted planet. $\epsilon$ Tau was included in the survey with ESPaDOnS, then with Narval, and was observed 11 times between 22 August 2008 and 21 November 2010.  A magnetic field is detected 5 times: its longitudinal component reverses its polarity and reaches 1~G. Table 8 and Fig. A2 present our results. Figure A2 shows variations of RV in which the amplitude and timescale of variations are consistent with the detection of Sato et al. (2007). While some of our magnetic field measurements are significant, we cannot confirm or exclude a correlation between $B_l$ and RV variations. The variation of the $S$-index is even more scattered. To investigate the possibility that in $\epsilon$ Tau the RV variations are due to magnetic spots and not to a planet, as in Pollux (Auri\`ere et al. 2014a and in preparation), would require a spectropolarimetric follow-up study of the star with both higher precision and longer time base.

\begin{figure}
\centering
\includegraphics[width=9 cm,angle=0] {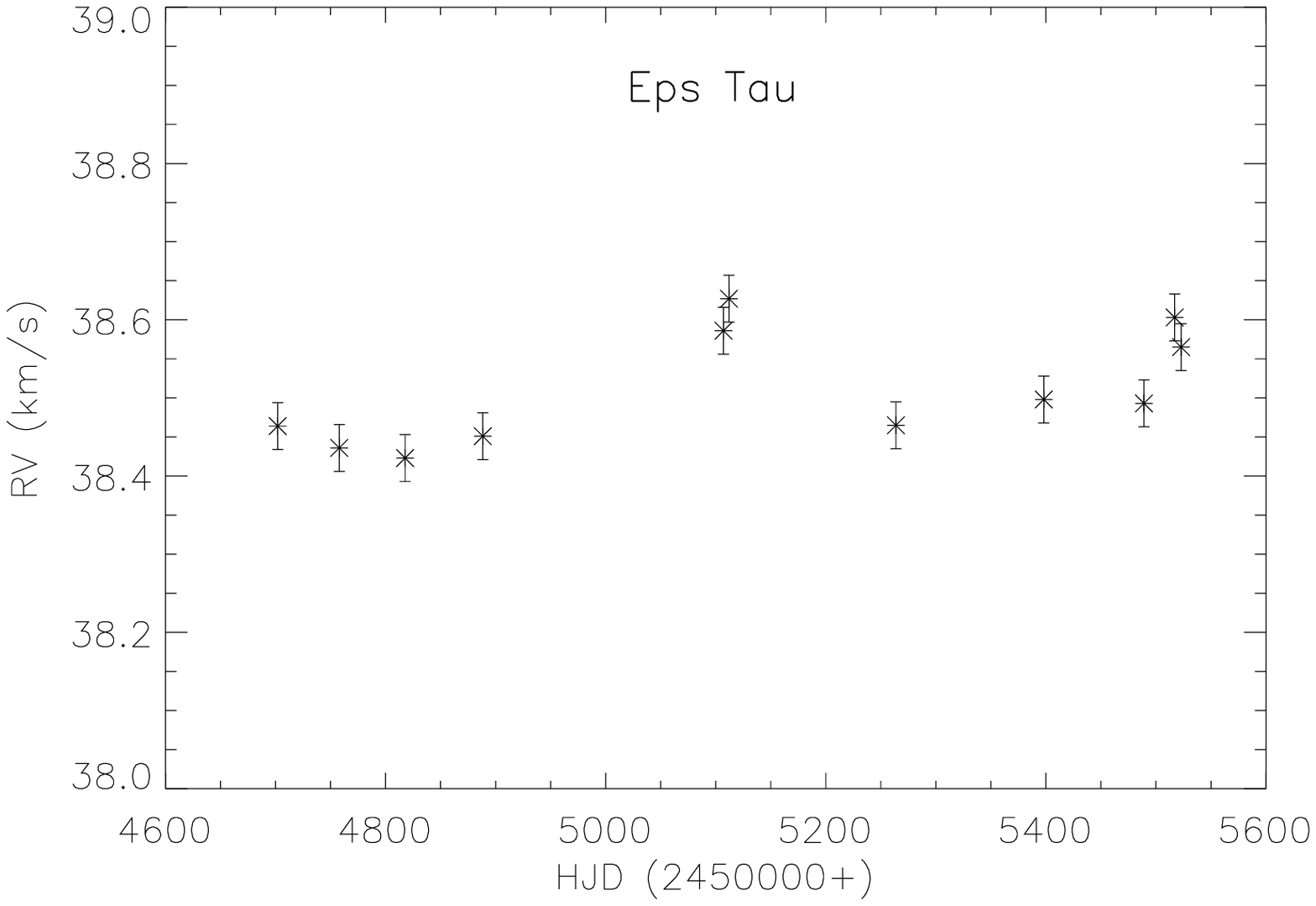}
\includegraphics[width=9 cm,angle=0] {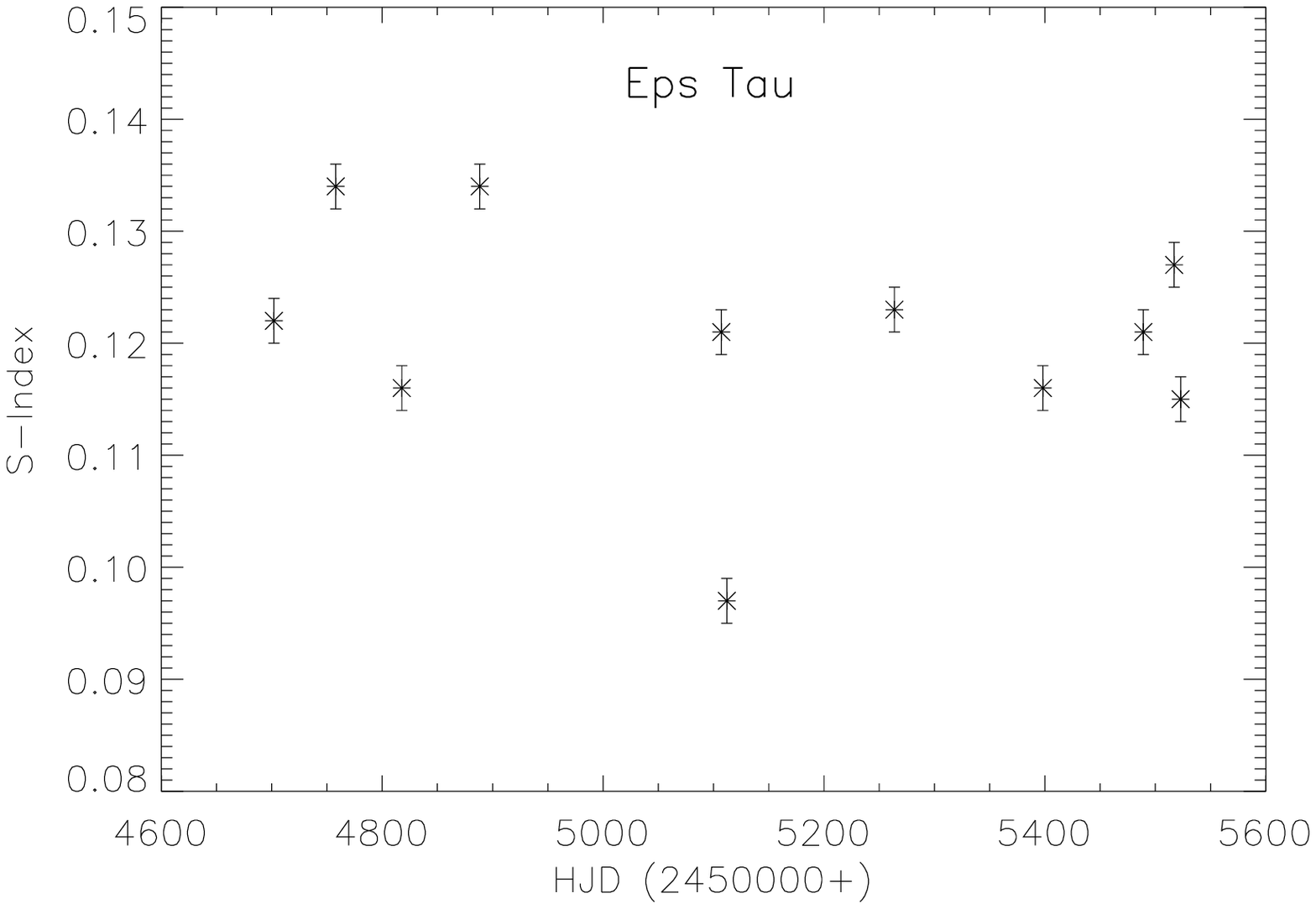}
\includegraphics[width=9 cm,angle=0] {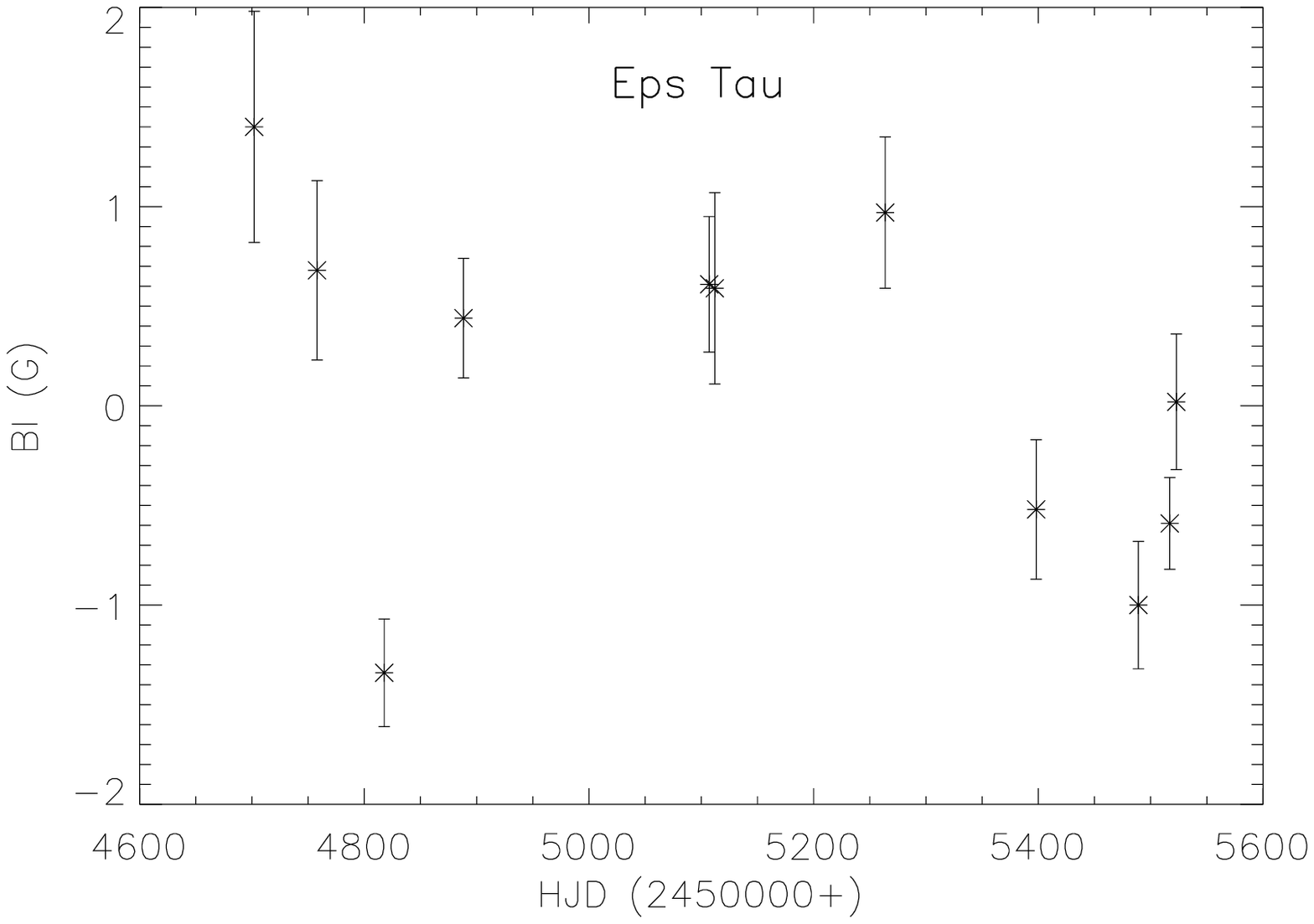}
\caption{Variations of the radial velocity (RV, upper graph), $S$-index (middle graph) and longitudinal magnetic field ($B_l$, lower graph) of $\epsilon$ Tau between 22 August 2008 and 21 November 2010, as observed with ESPaDOnS and Narval. Error bars of 30m/s and 0.02 are illustrated for RV and $S$-index respectively. As to the $B_\ell$ plot, error bars are those of Table 8.}
\end{figure}

\subsubsection{Aldebaran, $\alpha$ Tau, HD 29139}

Aldebaran was observed on 5 dates with ESPaDOnS, then on 9 dates with Narval, spanning between 26 September 2007 and 18 March 2011 (see Table 8). The magnetic field is detected only 4 times. The $B_l$ is weak and reverses its sign, being comprised between -0.25 and +0.22 G during our observations (see Fig. 2). Aldebaran has been identified a long time ago as one of the red giants presenting a long-period RV variation (Hatzes \& Cochran 1993). Hatzes (2008) revisited all the radial velocity observations and their periodogram analysis 'yielded a strong peak corresponding to a period of 630 days'. The RV measurements show considerable scatter about Hatzes's RV curve. Hatzes (2008) analysis concludes that the RV variations may be explained by an hosted planet orbiting the star with the 630 day period and radial oscillations with a period of 5.8 days. Hatzes (2008) also suggested that the rotational period of Aldebaran could be of 892 days, corresponding to periodic variations of the equivalent width of the Balmer H${\alpha}$ line. Figure A3, upper graph, shows the variations of RV during our investigation. An error of 30 m/s is illustrated. Our plots, which span about 3.5 years, show a total amplitude of variations reaching about 600 m/s as expected and is consistent with a period of about 600-800 days. The $S$-index of Aldebaran is stronger than 0.2 which may indicate a chromospheric flux higher than the basal flux. Its variations, while scattered, mimic those of RV. On the other hand, we considered that only the four dates corresponding to significant Zeeman detections gave significant $B_l$ measurements. These measurements are plotted with their error bars in the lower graph of Fig. A3.  As illustrated in Fig. 2 and in Fig. A3, we measure weak values of $B_l$ which reverses its sign. No Stokes $V$ Zeeman signal was visible in the mean LSD Stokes $V$ profiles on the other dates.   
H\"unsch et al. (1996) give $L_{\rm x}<0.7\ 10^{27}$~erg\,s$^{-1}$ as an upper limit for the X-ray luminosity. Nor were Ayres et al. (2003) able to detect the X-ray emission of Aldebaran with Chandra ($L_{\rm x}<7\, 10^{25}$ erg\,s$^{-1}$).

\begin{figure}
\centering
\includegraphics[width=9 cm,angle=0] {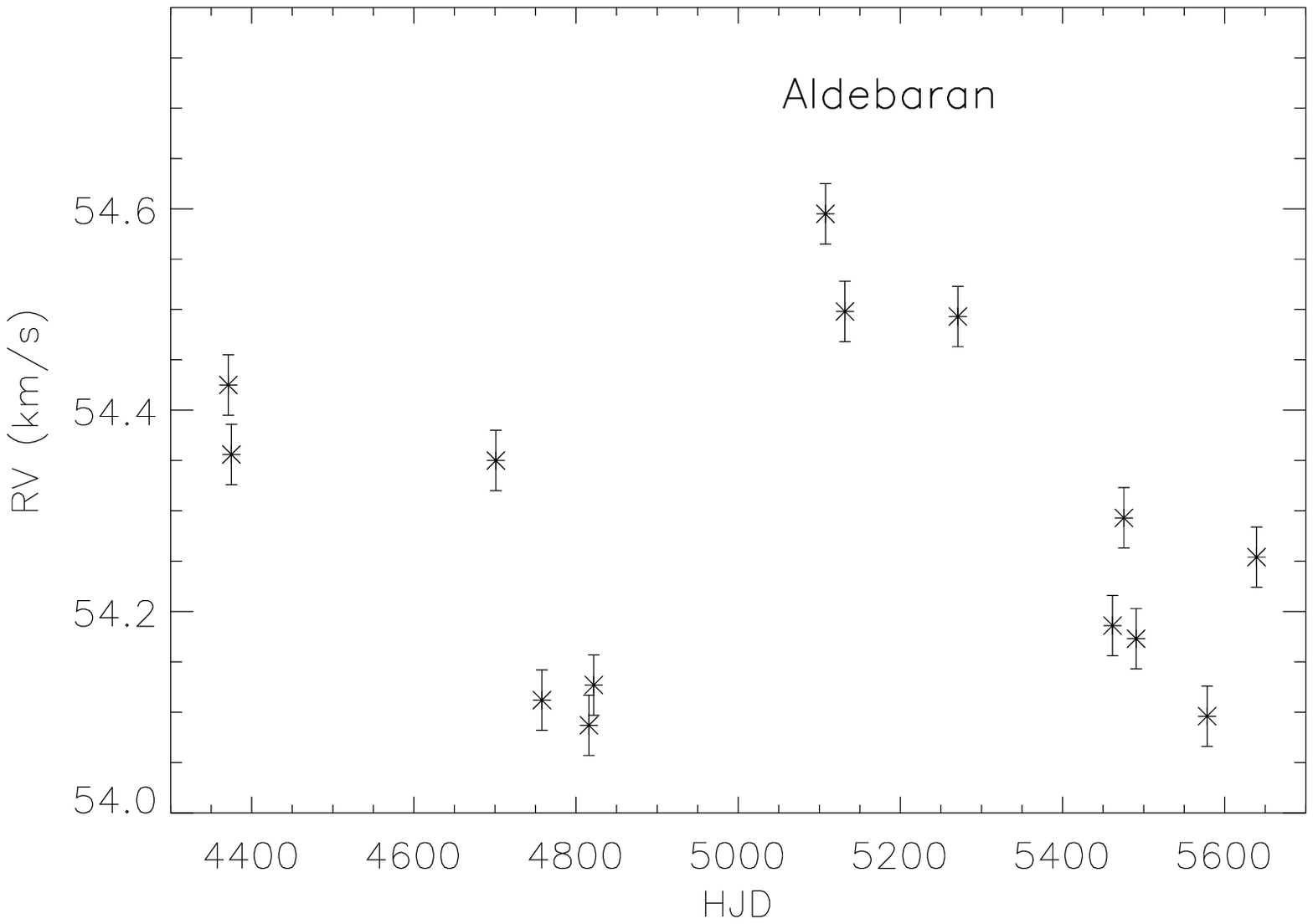} 
\includegraphics[width=9 cm,angle=0] {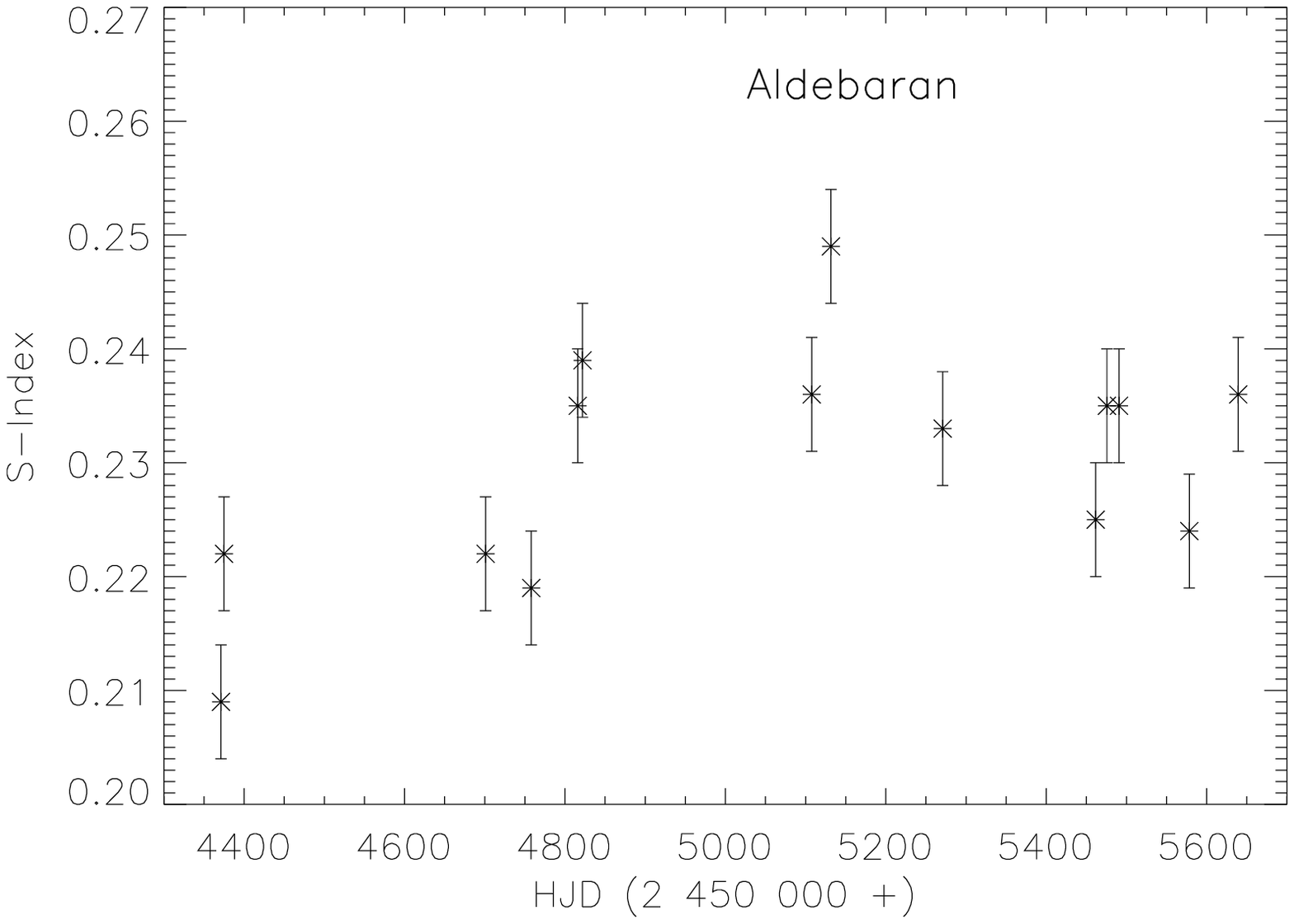}
\includegraphics[width=9 cm,angle=0] {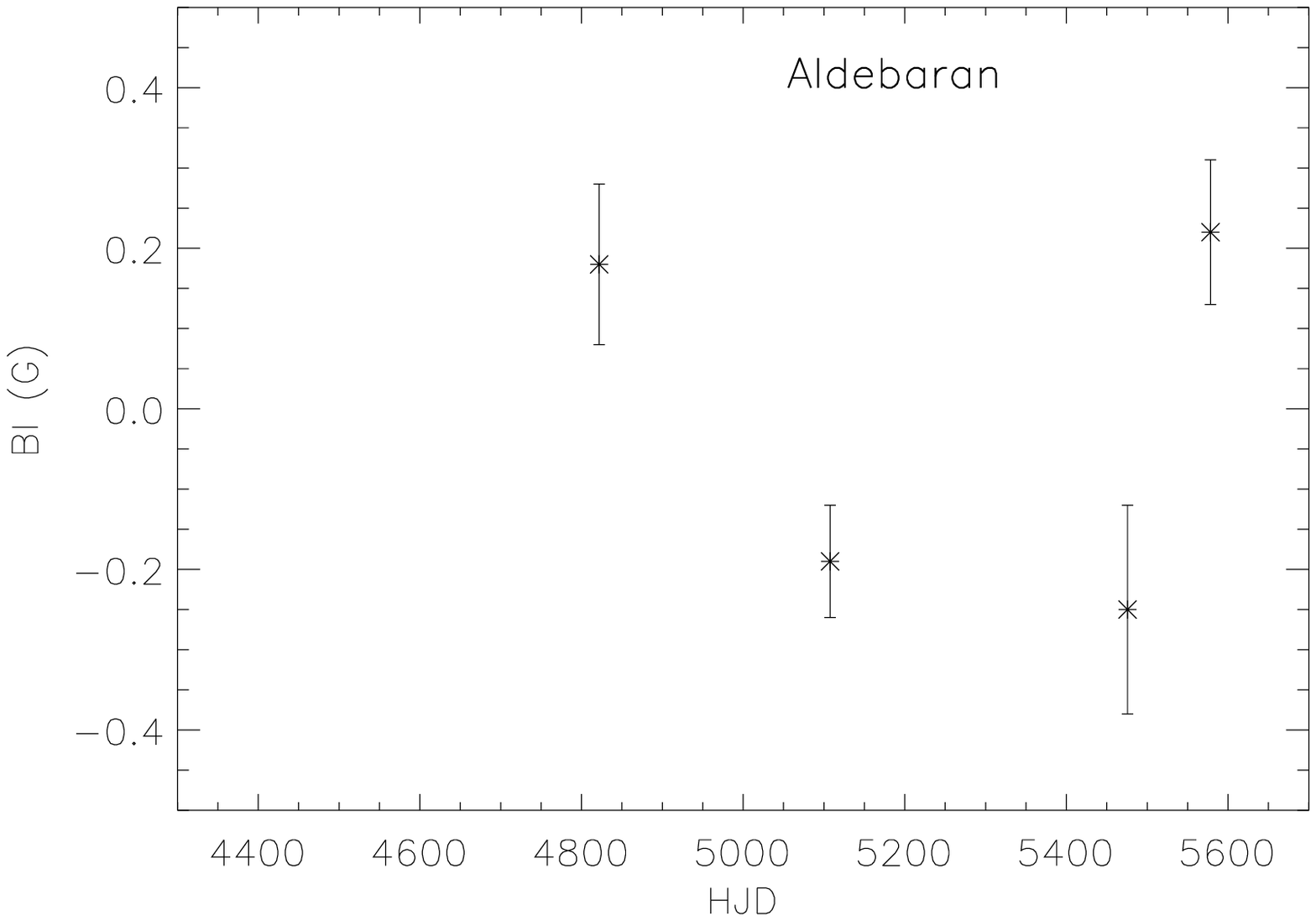}
\caption{Variations of the radial velocity (RV, upper graph), $S$-index (middle graph) and longitudinal magnetic field ($B_l$, lower graph) of Aldebaran between 26 September 2007 and 18 March 2011, as observed with ESPaDOnS and Narval. Error bars of 30m/s and 0.05 are illustrated for RV and $S$-index respectively. As to the $B_l$ plot, only the observations corresponding to Zeeman detections are shown (with their error bars). }
%\label{f1}
\end{figure}  

\end{appendix}

\begin{appendix}
\twocolumn
\section{Stellar parameters of the stars which are not in Massarotti et al. (2008) and complementary informations for all Zeeman detected stars}

 The stars are ordered as in Table 1. The different subsamples are described in \S~2.
The journal of observations, including measurements of  $B_l$, RV and $S$-index is presented, subsample by subsample, star by star, in Tables 6, 7 and 8. We give here the references for atmospheric parameters and $v \sin i$ given in Table 1 and which are not from Massarotti et al. (2008). In addition, for each detected star, we give references of related work with Narval/ESPaDOnS  magnetic observations or more detailed results than in the plain part of the paper. For these giants, the used spectra are in general already public and available in the ESPaDOnS/Narval stellar spectra data-base PolarBase (Petit et al. 2014): access at http://polarbase.irap.omp.eu/. For the strongest fields, when the Zeeman detection was obtained with only one Stokes $V$ spectrum, the LSD profile showing the detection may be viewed online on PolarBase. Otherwise, when an average of several Stokes $V$ spectra was necessary to yield the Zeeman detection, we present here one mean LSD Stokes $V$ profile corresponding to a detection.

\subsection{Active giants}

\subsubsection{14 Ceti, HD 3229}

The magnetic field of 14 Ceti is detected and studied by Auri\`ere et al. (2012). These authors present the star as a candidate for being an Ap star descendant. The $T_{\rm eff}$, $\log g$, metallicity and upper limit for $v \sin i$ used in the present work and in that of  Auri\`ere et al. (2012), are from Van Eck et al. (in prep.).

\subsubsection{$\beta$ Ceti, HD 4128}
The magnetic field of  $\beta$ Ceti was detected for the first time with ESPaDOnS then Narval in 2007. Zeeman Doppler imaging (ZDI) and the rotational period are presented by Tsvetkova et al. (2012, 2013). These authors present this giant as possibly burning He in its core and being an Ap star descendant candidate.

\subsubsection{OP And, HD 9746}

For OP And, we use here the atmospheric parameters and $v \sin i$ retained by Balachandran et al. (2000). Its magnetic field has been detected on each observation. Borissova et al. (2012) studied the activity of the star in the period 1979-2010, using Narval data for the last seasons. Konstantinova-Antova et al. (in prep.) obtained a ZDI map of OP And whose $B_{\rm mean}$ is given in Table 2.

\subsubsection{EK Eri, HD27536}
Auri\`ere et al. (2008, 2011) have detected the magnetic field of EK Eri and obtained a ZDI map. This giant is confirmed as the archetype of the Ap star descendant candidates (see \S~7.2.2 and Auri\`ere et al. 2014b). EK Eri is in Massarotti et al. (2008) who do not give a value for metallicity. In Table 1 we give the metallicity from Dall et al. (2010).

\subsubsection{V1192 Ori, HD 31993}

We use here the atmospheric parameters and  $v \sin i$ retained by Fekel and Balachandran (1993). V1192 Ori was detected in X-rays by ROSAT (Voges et al. 1999). The magnetic field is detected in this work for each observation (Table 6). 

\subsubsection{V390 Aur, HD 33798}
V390 Aur is a wide binary but synchronization plays no role in its fast rotation. The magnetic field of V390 Aur has been detected and studied by Konstantinova-Antova et al. (2008). A ZDI map was presented by Konstantinova-Antova et al. (2012). The atmospheric parameters and $v \sin i$ of the star were redetermined by Konstantinova-Antova et al. (2012) and used here.

\subsubsection{$\nu$3 CMa, HD 47442}

$\nu$3 CMa is a well known K0 giant with moderate magnetic activity. Its chromospheric activity induces variations of the $S$-index with a period of 183 d (Choi et al. 1995) that we use as $P_{\rm rot}$. It also appears in the ROSAT catalog of X-ray emitting giants of H\"unsch et al. (1998a). Parameters for $\nu$3 CMa are given in several papers as catalogued by Soubiran et al. (2010). We give $T_{\rm eff}$ and $\log g$ from McWilliams (1990) and $v \sin i$ from Hekker \& Mel\'endez (2007). We detected its magnetic field, which is visible in plots from PolarBase.

\subsubsection{19 Pup, HD 68290}

19 Pup was X-ray detected by ROSAT and appears in the catalog for X-ray emitting giants of H\"unsch et al. (1998a). Its $S$-index shows periodic variation and we use the 159 d period of Choi et al. (1995) as its rotational period. We observed it once with Narval and detected its magnetic field (see on PolarBase). 

\subsubsection{FI Cnc, HD 72146}

We use for FI Cnc the $T_{\rm eff}$ inferred by Wright et al. (2003) from its spectral type. This value is somewhat hotter that the value given by Strassmeier et al. (2000). FI Cnc is a well studied active giant for which the rotational period is known from photometric studies (see \S~2.3). We have detected its magnetic field on each of our observations.

\subsubsection{24 UMa, HD 82210}

24 UMa, also known as DK UMa is a star crossing the Hertzsprung gap known to be active in X-rays and ultraviolet (Ayres et al. 2007). We observed it and detected its magnetic field easily with Narval. The atmospheric parameters that we present in Table 1 are that given by Leborgne et al. (2003). The $v \sin i$ is that of de Medeiros and Mayor (1999).

\subsubsection{39 Hya, HD 85444}

39 Hya is an active star already observed in X-rays (e.g. Gondoin 1999) and for which the $S$-index was measured (Young et al. 1989). We have observed it once and detected its magnetic field.

\subsubsection{31 Com, HD 111812}

31 Com is well known as a very active star crossing the Hertzsprung gap (Ayres et al. 2007). This star was studied by Strassmeier et al. (2010) from whom we take the numbers presented in Table 1. 31 Com was observed with Narval in March 2012 in two different rotational phases and its magnetic field was detected.
Then Borisova et al. (2014 and in prep.) observed again 31 Com with Narval and could get a ZDI map.

\subsubsection{37 Com, HD 112989}

The activity of 37 Com and its evolutionary status were studied by De Medeiros et al. (1999). Its magnetic field is detected in this work. We present in Table 1 the value of $T_{\rm eff}$ from Wright et al. (2003) which is consistent with the one choosen by De Medeiros et al. (1999). Gravity and metallicity come from Soubiran et al. (2010). The given $v \sin i$ is that of De Medeiros and Mayor (1999).
A photometric period of  $P_{\rm rot}$ = 70 d was inferred by photometry (Henry et al.  2000, Strassmeier et al. 1996), but a ZDI study suggested $P_{\rm rot}$ = 110 d (Tsvetkova et al. 2014, and in prep.). The very low $^{12}$C/$^{13}$C ratio (see appendix C) and comparison to the predictions of Charbonnel and Lagarde (2010) evolutionary models used in this work show that 37 Com is in the core Helium-burning phase.

\subsubsection{7 Boo, HD 121107}

7 Boo is a weakly active giant (Konstantinova-Antova 2001) though with a moderate $v \sin i$ (de Medeiros and Mayor 1999) and a  rather strong X-ray luminosity (Gondoin, 1999).  We use here the $T_{\rm eff}$ inferred by Wright et al. (2003) from its spectral type and the metallicity from Franchini et al. (2004). From our investigation we infer that 7 Boo is a 4 $M_{\odot}$ giant at the end of the Hertzsprung gap. We observed 7 Boo 3 times and detected its magnetic field once (marginal detection from the LSD statistics on 05 April 2008, see the LSD profiles in PolarBase).

\subsubsection{$\beta$ Boo, HD 133208}

An X-ray flare from $\beta$ Boo was observed by ROSAT on 8 August 1993 (H\"unsch \& Reimers, 1995). However this star is known to be a low-activity giant, as confirmed by the very small $S$-index that we measured on our 4 observations with Narval and ESPaDOnS. $\beta$ Boo was not Zeeman-detected on any of our 4 observations. From our study we can infer that $B_l$ was smaller than 1 G during our observations. The given $v \sin i$ is that of de Medeiros and Mayor (1999).

\subsubsection{$\delta$ CrB, HD141714}

$\delta$ CrB is a moderately active giant with a 59 d period determined both photometrically ( Fernie 1987, 1999) and by variations of the chromospheric Ca~{\sc ii} $H \& K$ lines (Choi et al. 1995). It was studied in X-rays (Gondoin, 2005a). We have detected the magnetic field of $\delta$ CrB on each of our observations with Narval or ESPaDOnS.

\subsubsection{$\kappa$ HerA, HD 145001}

$\kappa$ HerA is a giant active in X-rays (e.g. Gondoin 1999). Its chromospheric properties were studied by Konstantinova-Antova (2001). We detected its magnetic field in each of our Narval or ESPaDOnS observations. For $\kappa$ HerA, we use here the atmospheric parameters provided by McWilliam (1990), and the $v \sin i$ of de Medeiros and Mayor (1999).

\subsubsection{$\eta$ Her, HD 150997}

$\eta$ Her was detected in X-rays by ROSAT (H\"unsch et al. 1998a).  We detected its magnetic field in each of our Narval observations.

\subsubsection{$\xi$ Her, HD163993}

$\xi$ Her was detected in X-rays by ROSAT (H\"unsch et al. 1998a).  We detected its magnetic field in each of our Narval observations.

\subsubsection{$\iota$ Cap, HD203387}

$\iota$ Cap is an active giant which has been X-ray-detected by ROSAT (H\"unsch et al. 1998a). Its rotational period of 68 d is determined both by variations of the chromospheric Ca~{\sc ii} $H \& K$ lines (Choi et al. 1995) and photometrically (Henry et al. 1995). We detected its magnetic field in each of our Narval or ESPaDOnS observations.

\subsubsection{$\rho$ Cyg, HD 205435}

$\rho$ Cyg was detected in X-rays by ROSAT (H\"unsch et al. 1998a).  We detected its magnetic field in each of our Narval or ESPaDOnS observations.

\subsubsection{KU Peg, HD 218153}

For KU Peg, we use the fundamental parameters and $v \sin i$ provided by L\`ebre et al. (2009), who detected for the first time, with Narval, its surface magnetic field.
KU Peg was detected in X-rays by ROSAT (Voges et al. 1999). Using a count rate C=0.173 ct s$^{-1}$, the Hipparcos distance and the relation of Jorissen et al. (1996) we get $L_{\rm x}$ = 1.1 10$^{31}$ erg s$^{-1}$.

\subsubsection{OU And, HD 223460}

OU And is a well known active red giant, with small rotation period for a giant (24.2 d, Strassmeier et al. 1999) and very active in X-rays (Gondoin et al. 2003, Ayres et al. 2007).  In the literature is given a $T_{\rm eff}$ value inferred for OU And by Wright et al. (2003) from its spectral type. Gondoin (2003, 2005b) uses a cooler temperature but we could not find how it was measured. We then made our own mesurements of  $T_{\rm eff}$ and $\log g$ presented in Table 1 and used in this work. We present in Table 1 the $v \sin i$ from de Medeiros and Mayor (1999). We detected its magnetic field in this study and showed in \S~6  that this giant is very likely an Ap-star-descendant. Then Borisova et al. (in prep.) made new Narval observations and got a ZDI image of the surface magnetic field.

\subsection{Thermohaline deviants (THD)}

\subsubsection{$\alpha$ UMa, HD 95689}

$\alpha$ UMa is one of the 'deviant' of Charbonnel \& do Nascimento (1998). We used here in Table 1 the parameters of Houdashelt et al. (2000). The star was not detected on any of our 3 observations.

\subsubsection{HD 50885, HD 150580 \&HD 178208}

HD 50885,HD 150580 \&HD 178208 were selected as possible 'Thermohaline deviants' (Charbonnel \& Zahn 2007a,b) from their Li abundance (Charbonnel \& Jasniewicz in prep.). We use the $T_{\rm eff}$ given by  Wright et al. (2003). The stars were not detected on any of our observations.

\subsubsection{HD 186619}

HD 186619  was selected as a possible 'Thermohaline deviant' (Charbonnel \& Zahn 2007a,b) from its Li abundance (Charbonnel \& Jasniewicz in prep.). We use the $T_{\rm eff}$ and $\log g$ given by  Bord\'e et al. (2002). The star was not detected on any of our 5 observations.

\subsubsection{HD 199101 \& 4 And (HD 218452)}

HD 199101 \& 4 And were selected as possible 'Thermohaline deviants' (Charbonnel \& Zahn 2007a,b) from their Li abundance (Charbonnel \& Jasniewicz in preparation). We use the atmospheric parameters given by Cayrel de Strobel et al. (2001). The stars were not detected on any of our observations.

\subsection{CFHT sample and miscellaneous}

\subsubsection{$\eta$ Psc, HD 9270}

This star was taken from Tarasova (2002) selection of possible magnetic stars. It was not detected by ROSAT, but we detected its magnetic field once (definitive detection DD with the LSD statistics on 15 September 2008) as shown in Table 8 and Fig. B1. Following the observations on Pollux (Auri\`ere et al. 2009, 2014a) and Arcturus (Sennhauser \& Berdyugina 2011) this is one of the 3 new detection of sub-G magnetic field detected at the surface of G-K giants presented in this paper.

\begin{figure}
\centering
\includegraphics[width=9 cm,angle=0] {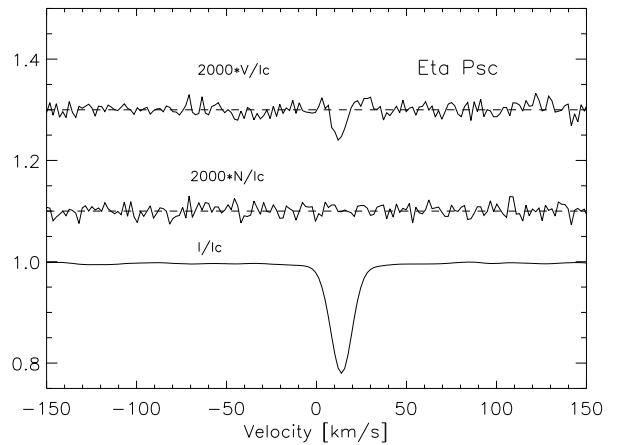} 
\caption{Mean LSD profiles of  $\eta$ Psc as observed  with  Narval on 15 September 2008. From top to bottom are Stokes $V$, null polarisation $N$, and Stokes $I$ profiles. For display purposes , the profiles are shifted vertically, and the Stokes $V$ and diagnostic $N$ profiles are expanded by a factor of 2000. The dashed lines illustrate the zero level for the Stokes $V$ and diagnostic null profiles.}
%\label{f1}
\end{figure}

%\subsubsection{$\upsilon$ Per, HD 9927}

%in Masserotti et al. (2008) nd

%\subsubsection{$\alpha$ Ari, HD 12929}

%in Masserotti et al. (2008) nd

\subsubsection{$\epsilon$ Lep, HD 32887}

$T_{\rm eff}$, $\log g$, $v \sin i$ for $\epsilon$ Lep are given in Table 1 from Hekker and Mel\`endez (2007). 
The star was not detected on any of our 2 observations.

\subsubsection{Pollux, HD 62509}

The surface magnetic field of Pollux was detected by Auri\`ere et al. (2009). Then new Narval's observations enabled to get the $P_{\rm rot}$ of Pollux and a ZDI map (Auri\`ere et al., 2014a and in preparation). These authors also suggest that the presence of the surface magnetic field may be sufficient to explain the observed periodic $RV$ variations (Hatzes et al. 2006), and that the hypothesis of an hosted planet may be unnecessary. Weak activity among giants is discussed in \S~7.3 and in Konstantinova-Antova et al. (2014 and in preparation). Pollux maybe the archetype of a class of weakly magnetic G K giants. 3D MHD simulations of the convective envelope of Pollux are in progress which will enable to understand the development and the action of the dynamo there (Palacios \& Brun 2014).

\subsubsection{Alphard, $\alpha$ Hya, HD 81797}

This star has been first observed with ESPaDOnS without detection on December 2007. We then made deep observations with Narval, adding up to 32 spectra, to approach the detection limit of the spectropolarimeter in January and March 2012 (see Table 8). We then got Zeeman-detections (MD and DD) of a very weak magnetic field ($B_l$ = 0.3 $\pm$ 0.08 G). Fig. B.2 shows our Zeeman-detection of Alphard on 26 March 2012.
The atmospheric parameters of Alphard presented in Table 1 come from Massarotti et al. (2008). However, Gray (2013) discussed their high $v \sin i$ value of 8.5 km $s^{-1}$ derived using the cross-correlation technique. We then made our own estimate using the macroturbulent velocity of Gray (2003) of 4.9 km $s^{-1}$ and the spectral synthesis method. We found $v \sin i$=2.3 km $s^{-1}$, which is consistent with Gray (2013) result and is presented in Table 1.

\begin{figure}
\centering
\includegraphics[width=9 cm,angle=0] {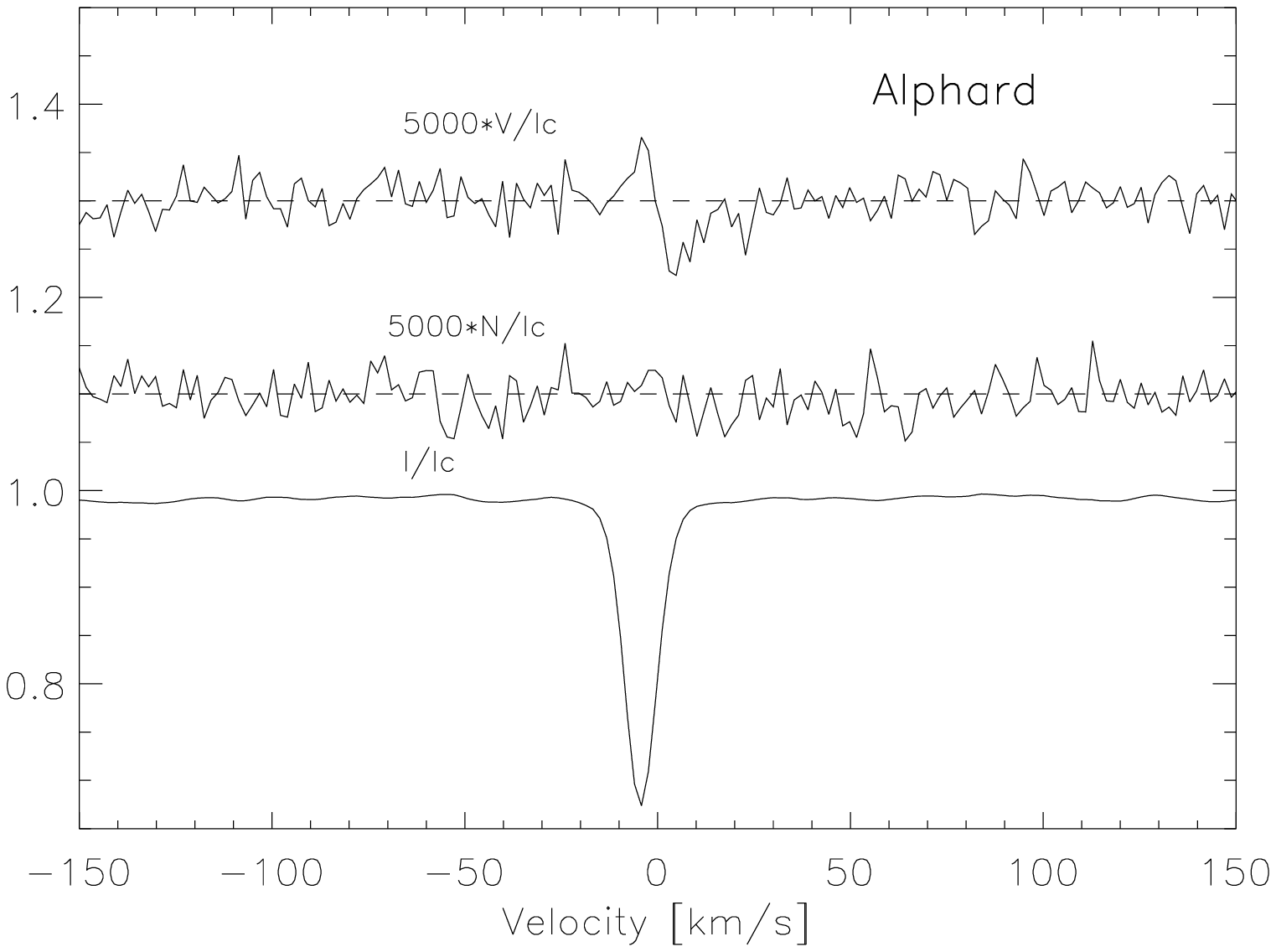} 
\caption{ Mean LSD profiles of  Alphard as observed  with  Narval on 26 March 2012. Presentation of the plots is the same as for Fig. B1. }
%\label{f1}
\end{figure}

\subsubsection{$\gamma$ Leo A, HD 89484}
$\gamma$ Leo is a double star. The secondary is about 2 magnitudes fainter than the primary and distant of 4.6 arcsec (L\'epine \& Bongiorno 2007). All the observations made with ESPaDOnS and reported in Table 8 concern $\gamma$ Leo A. Gondoin (1999) gives $L_{\rm x}$ $<$ 28 $10^{27}$ erg $s^{-1}$. Han et al. (2010) report the detection of periodic RV variations ($P$ = 429 d) which they explain by a planetary companion. We could not detect a magnetic field on none of our ESPaDOnS or Narval observation.

\subsubsection{$\nu$ Hya, HD 93813}

This star has been detected in X-rays (H\"unsch et al. 1998b, Gondoin et al. 1999; $L_{\rm x}$ = 71 $10^{27}$ erg $s^{-1}$ ). However its $S$-index is rather small (0.116, 0.121 for our two observations) and it was not detected on our two single Stokes V spectrum observations obtained with ESPaDOnS.

\subsubsection{$\epsilon$ Crv, HD105707}

$T_{\rm eff}$, $\log g$, [Fe/H], $v \sin i$ of Table 1 are given from Carney et al. (2008). $\epsilon$ Crv was not detected on our single ESPaDOnS's observation.

\subsubsection{Arcturus, $\alpha$ Boo, HD 124897}

Arcturus was considered as a non active low-mass giant, not detected in X-rays with ROSAT observations (H\"unsch et al. (1996), $L_{\rm x}$ $<$ 0.05 $10^{27}$ erg $s^{-1}$). However, Ayres et al. (2003) obtained a tentative X-ray detection of Arcturus with Chandra ( $L_{\rm x}$  about 1.5 $10^{25}$ erg $s^{-1}$). Then Sennhauser \& Berdyugina (2011) announced the discovery of the surface magnetic field, using the ZCD method (see \S~4.1.1). We then did a deep Zeeman investigation in 4 nights of 2012 first semester and got detections on January and March (MD from LSD statistics, see Table 8). Figure B3 shows our detection of Arcturus on 25 March 2012.

\begin{figure}
\centering
\includegraphics[width=9 cm,angle=0] {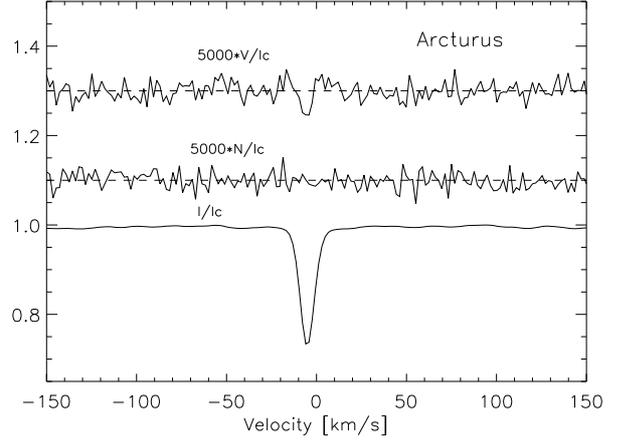} 
\caption{Mean LSD profiles of  Arcturus as observed  with  Narval on 25 March 2012. Presentation of the plots is the same as for Fig. B1.}
%\label{f1}
\end{figure}

%\subsubsection{$\epsilon$ Boo A, HD 129989}

%Gondoin (1999) gives $L_x$ $<$ 51 $10^{27}$ erg $s^{-1}$.

%in Masserotti et al. (2008) nd

\subsubsection{$\beta$ UMi, HD 131873}

$T_{\rm eff}$, $\log g$, [Fe/H] for $\beta$ UMi given in Table 1 are taken from Decin et al. (2003). For $v \sin i$, we take the value from de Medeiros and Mayor (2009). H\"unsch et al. (1996) give $L_{\rm x}$ $<$ 0.4 $10^{27}$ erg $s^{-1}$. The star was not detected on any of our 2 observations.
 
%\subsubsection{$\nu$ Oph, HD 163917}

%in Masserotti et al. (2008) nd

\subsubsection{$\mu$ Peg, HD 216131}

This star was taken from Tarasova (2002) selection of possible magnetic stars. It is (with $\beta$ Boo and $\nu$ Hya) one of the three stars which is detected in X-rays (Schr\"oder et al. 1998, ROSAT, $L_{\rm x}$ = 1.1 $10^{27}$ erg $s^{-1}$ ) but which is not Zeeman-detected in the present work. 

\end{appendix}

\clearpage

\begin{appendix}
\twocolumn
\section{CNO abundances and $^{12}$C/$^{13}$C 
ratios for 12 stars with ambiguous evolutionary status}

\subsection{Method}
The evolution status for 12 of our Zeeman-detected stars could not be unambiguously determined using only their position on the HRD; 
these stars could indeed be either at the base of the red giant branch (base RGB), or in the central helium-burning phase (He burning). 
In order to have additional constraints for these objects, we compared the observed values of the $^{12}$C/$^{13}$C ratio and of the CNO abundances with the predictions of the rotating models of Charbonnel \& Lagarde (2010) used in this study.
The strongest constraint comes from the $^{12}$C/$^{13}$C, when plotting this quantity versus stellar mass in Fig.17 of Charbonnel \& Lagarde (2010).
For nine of them we found the useful data in the litterature. For nu3\,CMa, 19\,Pup and $\eta$ Her, we determined CNO abundances and $^{12}$C/$^{13}$C using our Narval's spectra (see below). The results are summarized in Table~C.1. 

\subsection{Determination for nu3\,CMa, 19\,Pup and $\eta$ Her from Narval spectra}

\begin{table*}
\caption{Light element abundances for giants with ambiguous evolutionary status. }
\begin{tabular}{lllllllll}
\hline
HD      & Name     &$\log\varepsilon$(Li)&$\log\varepsilon$(C)&$\log\varepsilon$(N)&$\log\varepsilon$(O)& $^{12}$C/$^{13}$C & Ref.  & Branch  \\
\hline 
4128    &$\beta$ Ceti& 0.01              &                    &                    &                    &   19              & (8)   & He burning (Base RGB)\\
9746    & OP And   &  3.5                &                    &                    &                    &   24              & (2)   & RGB (He burning) \\ 
28305   &$\epsilon$ Tau & 1.2            &  8.35              & 8.40               & 8.80               &   22              &(9)(10) (6)& base RGB (He burning)\\
28307   & 77 Tau   &  0.86               &                    &                    &                    &   28 20             &(3) (4) (6)& Base RGB \\
29139   &Aldebaran &                     &  8.25$\pm 0.12$    & 8.05$\pm 0.11$     & 8.48$\pm 0.14$     &   10$\pm 2$       & (11)  & AGB (RGB) \\
31993   & V1192 Ori&  1.4                &                    &                    &                    &                   & (7)   & RGB (He burning)  \\
47442   &$\nu$3~CMa&  -0.3               &  8.31              & 8.26               & 8.73               &   22$\pm 2$       & (1)   & Base RGB (He burning) \\
62509   & Pollux   &                     &                    &                    &                    &   24              & (5)   & RGB or He burning\\
68290   & 19~Pup   &  0.9                &  8.40              & 8.30               & 8.73               &   20$\pm 2$       & (1)   & Base RGB (He burning) \\
112989  & 37 Com   &                     &                    &                    &                    &   4               & (6)   & He burning (Base RGB) \\
124897  &Arcturus  &                     & 8.06$\pm 0.09$     & 7.67$\pm 0.13$      & 8.76$\pm 0.17$    &   9$\pm 0.09$     & (11)  & RGB tip (AGB) \\
150997  & $\eta$ Her&  0.9               &  8.12              & 7.80               & 8.49               &   22$\pm 2$       & (1), Li:(9)& Base RGB (He burning) \\
\hline
\end{tabular}

\tablefoot{(1) this work, (2) Drake et al. 2002, (3) Gilroy 1989, (4) Lambert \& Ries 1981, (5) Auri\`ere et al. 2009, (6) Tomkin et al. 1976, (7) Fekel \& Balachandran 1993, (8) Tsvetkova et al. 2013, (9) Brown et al. 1989, (10) Mishenina et al. 2006, (11) Abia et al. 2012.}
\end{table*}

We now present our work for nu3\,CMa, 19\,Pup and $\eta$ Her.
Atmospheric parameters, such as effective temperature ($T_{\rm eff}$), surface gravity ($\log g$),
microturbulence ($\xi$), and metallicity, as given by 
[Fe/H]\footnote{we use the notation [X/H]=$\log(N_{\rm X}/N_{\rm
  H})_{\star} -\log(N_{\rm X}/N_{\rm H})_{\odot}$}, were taken from
Hekker \& Mel\'endez (2007)  for nu3~CMa, 
 from Takeda (2007) for 19~Pup and Massarotti et al. (2008) for $\eta$ Her. Adopted atmospheric parameters are shown 
in Table~C.2.

\begin{table}
\caption{Atmospheric parameters and metallicity.}
\label{tab:atmparam}
\begin{tabular}{lccccc}\hline
Star &$T_{\rm eff}$ &$\log g$& $\xi$& [Fe/H] & Ref\\
        &    K      & & km\,s$^{-1}$&  &                \\
\hline 
nu3~CMa &  4\,550   & 2.30   & 1.80  &  $-$0.09 & (1) \\ 
19~Pup  &  5\,028   & 2.92   & 1.21 &  $+$0.06  & (2)\\ 
$\eta$ Her& 4\,943   & 2.8    &      &  $-$0.37  & (3) \\
%37 Com  &           &        &      &           &    \\  
\hline 
\end{tabular}

\tablefoot{(1) Hekker \& Mel\'endez (2007), (2) Takeda (2007), (3) Massarotti et al. (2008)}

\end{table}

 Carbon, nitrogen, and oxygen abundances, as well as the
$^{12}$C/$^{13}$C isotopic ratio, were determined using the
spectrum synthesis technique.  Since the abundances of the CNO
elements are interdependent because of the association of carbon and
oxygen in CO molecules in the atmospheres of cool giants, the CNO
abundance determination procedure was iterated until all the
abundances of these three elements agreed.  The abundances of 
oxygen, carbon,
nitrogen, and the $^{12}$C/$^{13}$C isotopic ratio were determined
using the forbidden oxygen line  at $\lambda$6300.304~\AA\ 
and the lines of the CN and C$_2$ molecules.  The line lists
 are the same as described in Drake \& Pereira
(2008) with an exception for the dissociation energy of the 
CN molecule which in this paper was taken equal to 7.75~eV.  
The eventual contamination of the [O\,{\sc i}] $\lambda$6300.304~\AA\
line by telluric O$_2$ lines and $^{12}$CN and $^{13}$CN lines at 
$\sim\!\lambda$8004~\AA\ by telluric H$_2$O lines was checked out using 
a hot star spectrum.
The LTE model atmospheres of Kurucz (1993) and the current version (April 2010)
of the spectral analysis code {\sc moog} (Sneden 1973) were used to carry 
out the synthetic spectra calculations. Figures C.1 and C.2 present the fit of our observed and synthetic spectra for several $^{12}$C/$^{13}$C isotopic ratios. Derived light elements abundances
as well as $^{12}$C/$^{13}$C isotopic ratios are shown in Table~C.1. 

Calculations of the carbon
isotopic ratios do not depend on the uncertainties in the C and N abundances
and molecular parameters. The errors in the $^{12}$C/$^{13}$C determinations
are mainly due to uncertainties in the observed spectra, such as possible
contamination by unidentified atomic or molecular lines or uncertainties in
the continuum placement.

\begin{figure} %12C/13C
   \centering
   \includegraphics[width=9.5cm]{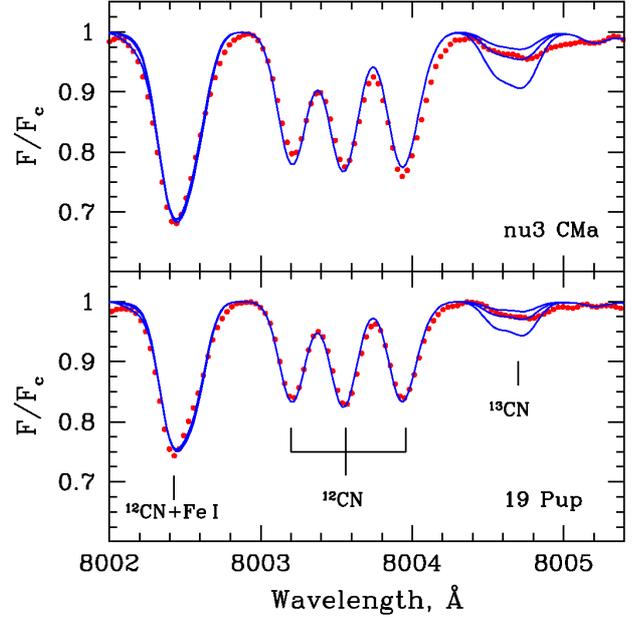}
   \caption{Observed (dotted red line) and synthetic (solid blue line)
spectra in the region around the $^{12}$CN and $^{13}$CN lines 
at $\sim$8004~\AA\, for nu3~CMa ({\it top}) and 19~Pup ({\it bottom}). Synthetic spectra 
were calculated for three $^{12}$C/$^{13}$C ratios, 10.0, 20.0, and 36.0
for nu3~CMa and 10.0, 22.0, and 36.0 for 19~Pup.}
%The green solid lines shows the spectrum of hot star used to map the telluric water lines.}
\end{figure}

\begin{figure} %12C/13C
   \centering
   \includegraphics[width=9.5cm]{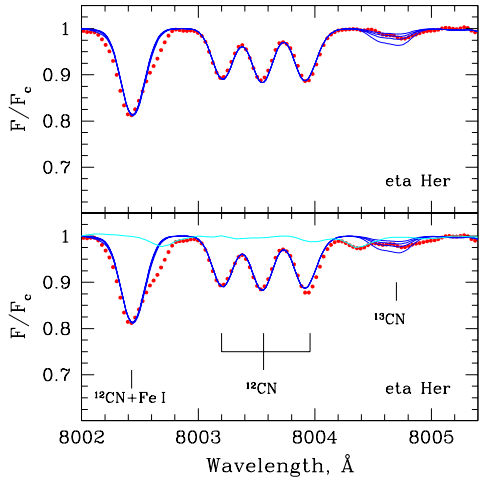}
   \caption{Observed (dotted red line) and synthetic (solid blue line)
spectra in the region around the $^{12}$CN and $^{13}$CN lines 
at $\sim$8004~\AA\, for $\eta$ Her. Synthetic spectra 
were calculated for four $^{12}$C/$^{13}$C ratios, 10, 16, 22 and 32.
Bottom part: observed spectrum of $\eta$ Her and spectrum of a hot star (cyan line) used to to ``dry"
the observed spectrum.
Upper part: observed spectrum without the contribution from the telluric H2O lines.}
\end{figure}

\end{appendix}

\end{document}